\documentclass[epj]{svjour}
\usepackage{graphics}
\usepackage[dvips]{graphicx,color}
\usepackage[body={16cm,22cm},bottom=1.5cm]{geometry}
\usepackage{amsmath}	
\usepackage{amssymb}	
\usepackage{psfrag}
\usepackage{latexsym}
\usepackage{amsfonts}

\begin{document}

\title{Plasticity and dynamical heterogeneity in driven glassy materials}
\author{Michel Tsamados}

\institute{Universit\'e de Lyon; Univ. Lyon I, Laboratoire de
Physique de la Mati\`ere Condens\'ee et Nanostructures; CNRS, UMR
5586, 69622 Villeurbanne, France}

\date{Received: date / Revised version: date}



\abstract{Many amorphous glassy materials exhibit complex spatio-temporal mechanical response and rheology, characterized by an intermittent stress-strain response and a fluctuating velocity profile. Under quasistatic and athermal deformation protocols this heterogeneous plastic flow was shown to be composed of plastic events of various sizes. In this paper, through numerical study of a 2D Lennard-Jones amorphous solid, we generalize the study of the heterogeneous dynamics of glassy materials to the finite shear-rate ($\dot\gamma\neq0$) and temperature case ($T\neq0$). The global mechanical response obtained through the use of Molecular Dynamics is shown to converge in the limit $\dot\gamma\rightarrow0$ to the quasistatic limit obtained with an energy minimization protocol. The detailed analysis of the plastic deformation at different shear rates shows that the glass follows different flow regimes. At sufficiently low shear rates the mechanical response reaches a shear-rate independent regime that exhibits all the characteristics of the quasistatic response (finite size effects, yield stress...). At intermediate shear rates the rheological properties are determined by the externally applied shear-rate. Finally at higher shear the system reaches a shear-rate independent homogeneous regime. The existence of these three regimes is also confirmed by the detailed analysis of the atomic motion. The computation of the four-point correlation function shows that the transition from the shear-rate dominated to the quasistatic regime is accompanied by the growth of a dynamical cooperativity length scale $\xi$ that is shown to diverge with shear rate as $\xi\propto\dot\gamma^{-\nu}$, with $\nu\sim0.2-0.3$. This scaling is compared with the prediction of a simple model that assumes the diffusive propagation of plastic events. 
\PACS{
      {PACS-key}{discribing text of that key}   \and
      {PACS-key}{discribing text of that key}
     } 
}

\authorrunning{M. Tsamados}
\titlerunning{Plasticity and dynamical heterogeneity in driven glassy materials}
\maketitle

\section{Introduction}

When subjected to slow driving many systems exhibit an intermittent response with the appearance of discrete and impulsive events spanning a broad range of sizes. Such a scale-invariant behavior is generally observed in driven nonlinear, dynamical systems and examples of such crackling signals are ubiquitous in nature (for a review see \cite{Sethna2001}).

Among such systems various glassy materials (granular media \cite{Losert2000,DaCruz2002}, foams \cite{Debregeas2001,Lauridsen2004}, emulsions \cite{Coussot2002,DaCruz2002}, micelles \cite{Salmon2003}, metallic glasses \cite{Bailey2007}...) were shown to exhibit, at small scales, small shear rates and for sufficiently low temperatures, such an intermittent signature in their stress-strain mechanical response along with complex spatio-temporally fluctuating velocity profiles. Such intermittent spatio-temporal features can be seen as the macroscopic manifestations of the strongly heterogeneous underlying deformation processes that take place in the driven glasses and that lead to the strain or velocity localization observed experimentally in alloys, metallic glasses \cite{Xie2008}, polymers \cite{Lu1999} , granular media \cite{DaCruz2004}, foams and colloids \cite{Debregeas2001,Kabla2003,Lauridsen2002,Varnik2004} as well as in numerous simulations, both of model systems such as Lennard-Jones glasses \cite{Varnik2003,Shi2005a,Shi2005b,Tanguy2006}, as in more realistic simulations \cite{Delogu2008a,Delogu2008b,Delogu2008c,Bailey2004b}. 

Experimental studies on disordered materials (foams, granulars), far below the glass transition temperature \cite{Dennin2004,Besseling2007,Lauridsen2002,Howell1999,Tewari1999,Majmudar2005,Kolb2004,Debregeas2001,Kabla2003}, have associated the dissipative events observed in the stress strain mechanical response (the stress drops) to the existence of a collective behavior of localized rearrangements leading to a strongly heterogeneous mechanical response as shown in the existence of shear bands in the macroscopic plasticity of such systems. Along these experimental evidences, in the last five to ten years, extensive numerical simulations of quasistatically sheared model glassy systems have confirmed this picture of a cascade mechanism but have also generated some debates first as to the validity of the potential energy minimization (PEM) methods to represent the physical reality of slowly driven systems and second as to the localized nature \cite{Lerner2009a} of the elementary rearranging plastic events associated to the macroscopic stress and energy releases of figure \ref{fig:StressStrain}.

On a theoretical level various current `mean field' models of the rheology of glasses predict reasonably well the macroscopic mechanical properties of these materials \cite{Falk1998,Sollich1997}. However to reproduce the spatial heterogeneity of their response a novel class of spatially resolved models inspired from geophysics \cite{Chen1991} have depicted the plastic flow as a result of the collective organization of interacting plastic rearrangements in an otherwise elastic medium. This type of model allows therefore a mesoscopic approach of the appearance of spatio-temporal heterogeneities based on elementary bricks, the plastic rearrangements. Nevertheless to produce a multiscale description of the mechanical response of glassy materials necessitates a microscopic (i) identification of these elementary plastically rearranging zones \cite{Tanguy2006,Maloney2004,Maloney2006}, (ii) a comprehension of their local elasto-plastic properties \cite{Tsamados2009a} and yield criteria \cite{Tsamados2008} and (iii) a study of their interaction and propagation. 

Basing our study on the detailed numerical analysis of a sheared two dimensional polydisperse Lennard-Jones glass \cite{Leonforte2004,Leonforte2005,Leonforte2006,Tanguy2006,Tsamados2008,Tsamados2009a} we discuss mainly in this paper point (iii) with an emphasis on the influence of the shear rate on the cooperative nature of the plastic response. At a microscopic level a plastic rearrangement redistributes its shear stress through a quadrupolar long-range elastic propagator \cite{Eshelby1957,Picard2004} hence favoring subsequent plastic rearrangements in its vicinity and within some prescribed directions. An estimate of the dynamical cooperativity length scale resulting from this cascading process can be computed with tools such as the four-point correlation function borrowed from the literature on dynamical heterogeneities in supercooled liquids near their glass transition \cite{Toninelli2005} or in granular media near their jamming transition \cite{Lechenault2008a}. 

In the present paper we provide an extensive study of the dynamics and plasticity in a two-dimensional model Lennard-Jones glass. To quantitatively characterize the spatio-temporal dynamics of the driven glasses we first compute various two-time observables (mean-square displacement, van Hove function, intermediate scattering function...) which inform on the typical relaxation times and their link with the external shear rate, we then proceed to a general description of the dynamical heterogeneity of the driven systems and prove the existence of a diverging dynamical cooperative length scale as the shear rate $\dot\gamma$ tends to zero. We also provide pictorial evidence of these dynamical structures for different shear rates. Finally we discuss our results in light of recent advances and compare our observations to a phenomenological mesoscopic yield stress model.

\section{Rheological response of the system}
\label{Rheology}
%

\subsection{Rheological and mechanical characteristics.} 
\begin{figure}[!hbtp]
\begin{center}
\includegraphics[width=7cm]{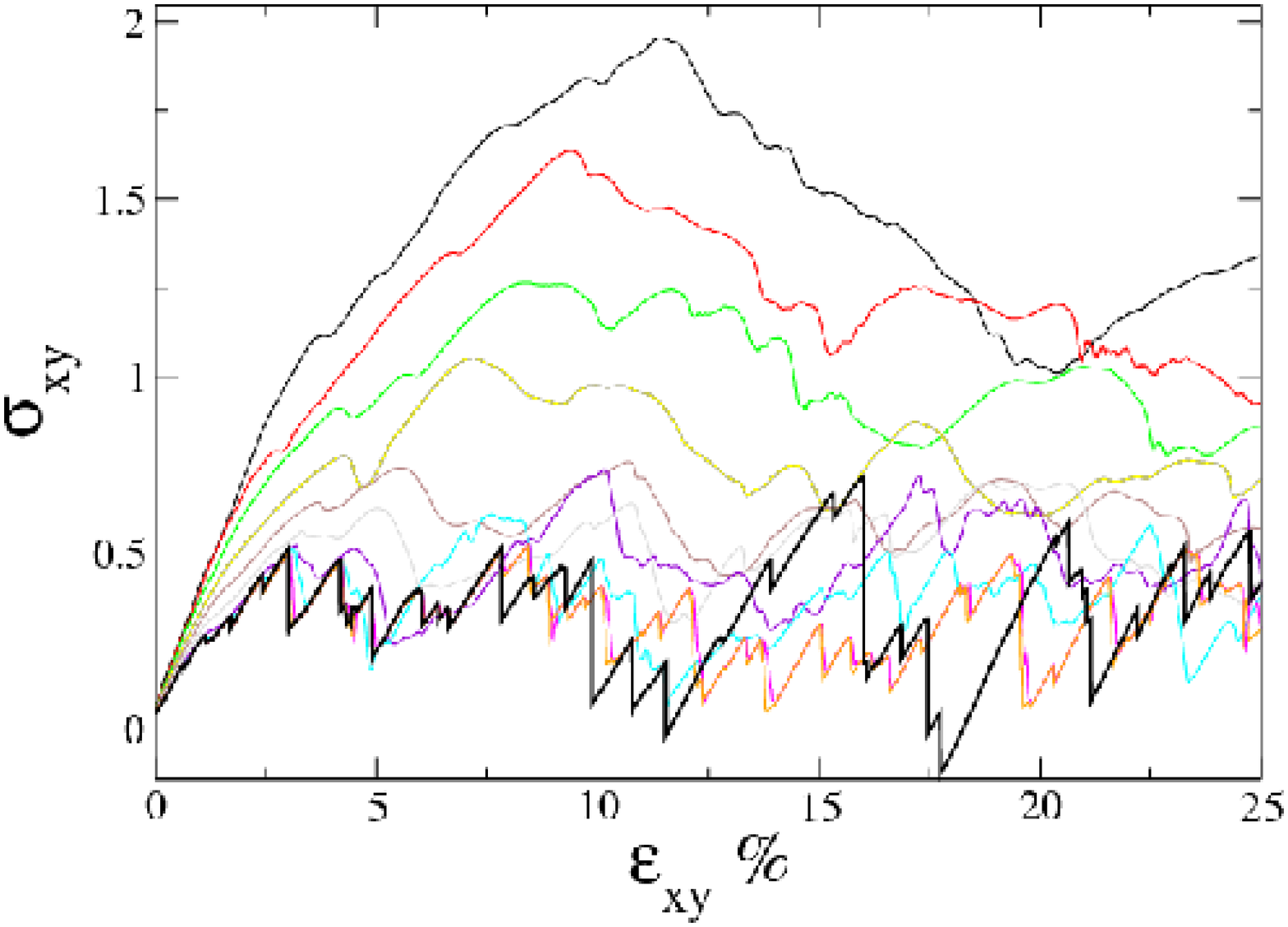}
\includegraphics[width=7cm]{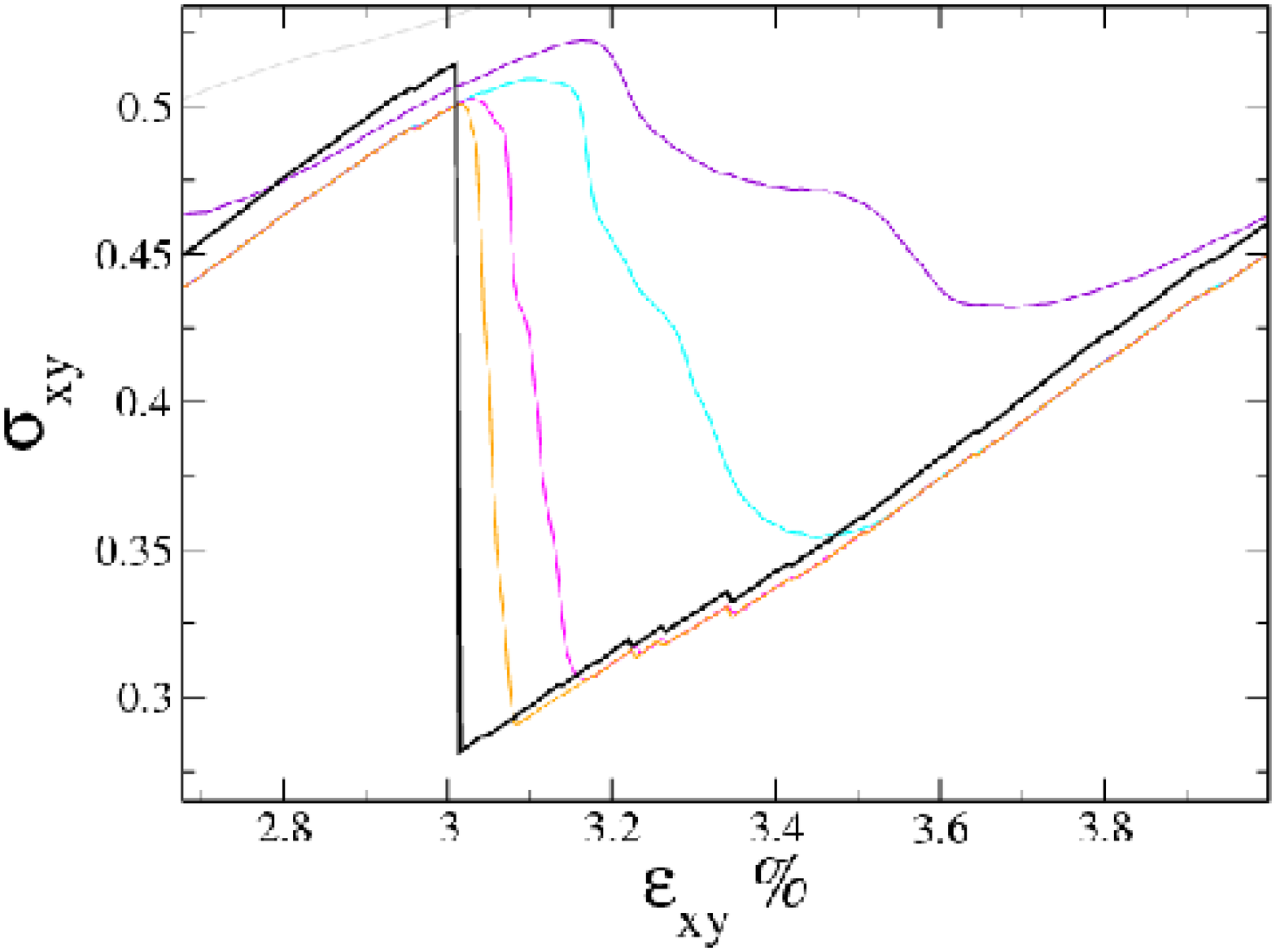}
\caption{\textbf{ Left :} stress strain mechanical response of a glass sample containing 625 particles sheared under RWBCs at shear rates ranging from $\dot\gamma=10^{-2}$ to $\dot\gamma=10^{-5}$ (thin colored lines from top to bottom, $\dot\gamma=10^{-2}$ (black), $\dot\gamma=5\cdot10^{-3}$ (red), $\dot\gamma=2.5\cdot10^{-3}$ (green), $\dot\gamma=10^{-3}$ (dark green), $\dot\gamma=5\cdot10^{-4}$ (brown), $\dot\gamma=2.5\cdot10^{-4}$ (gray), $\dot\gamma=10^{-4}$ (violet), $\dot\gamma=5\cdot10^{-5}$ (cyan), $\dot\gamma=2.5\cdot10^{-5}$ (magenta) and $\dot\gamma=10^{-5}$ (orange)). The thick black line corresponds to the quasistatic shear protocol. \textbf{Right :} Zoom in a portion of the total mechanical response, illustrating the typical relaxation time associated to a plastic rearrangement in the glass signaled in the quasistatic protocol by an abrupt stress drop.}
\label{fig:StressStrain}
\end{center}
\end{figure}

\begin{figure}[!hbtp]
\begin{center}
\includegraphics[width=7cm]{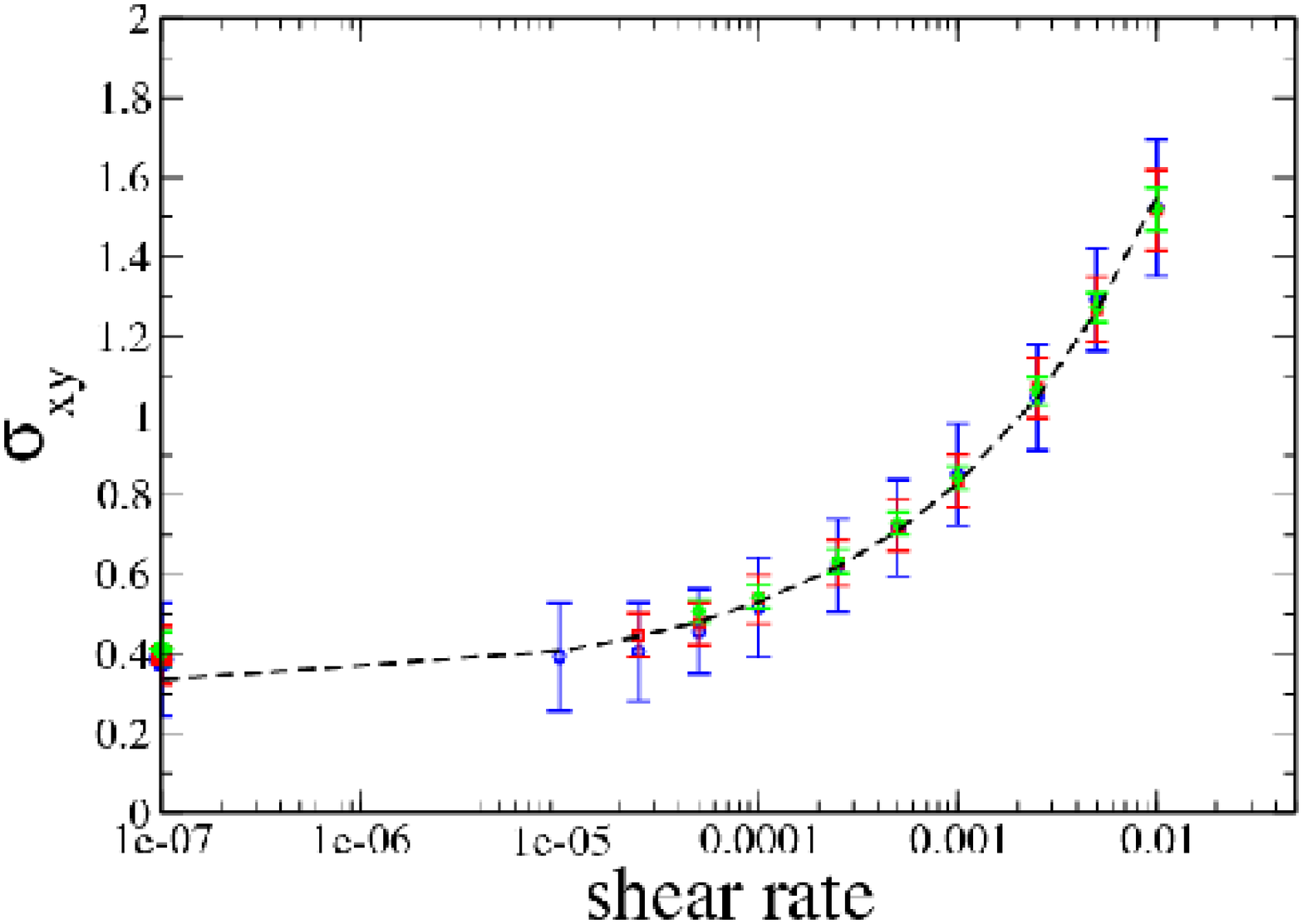}
\includegraphics[width=7cm]{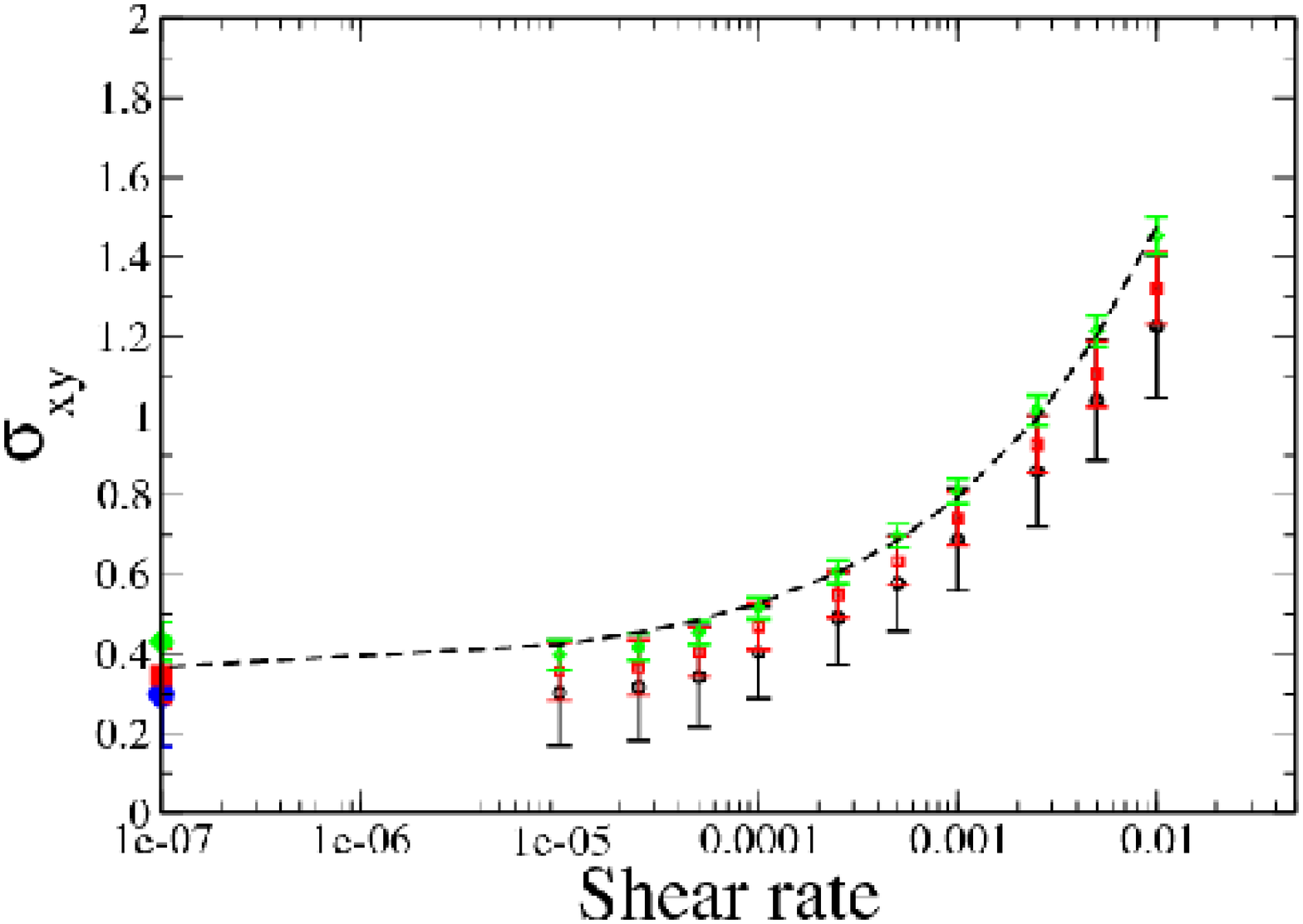}
\caption{Flow curves associated with the different system sizes. Small open symbols correspond to finite shear rate values ranging from $10^{-5}$ to $10^{-2}$. The larger solid points on the vertical axis correspond to the quasistatic protocol $\dot\gamma=0$. \textbf{ Left :} Shear under LEBCs with fit parameter to the Herscel-Bulkley rheological law, $\tau_{Y}\simeq0.32$, $c_{1}=7.2$ and $\beta=0.38$. \textbf{Right :} Shear under RWBCs, with the dashed line representing a Herschel-Bulkley fit $\tau = \tau_{Y} + c_{1} \dot\gamma^{\beta}$, with $\tau_{Y}\simeq0.36$, $c_{1}=7.4$ and $\beta=0.4$.}\label{fig:StressStrainrate}
\end{center}
\end{figure}

We do not reproduce here the details of the quasistatic simulation procedures that we use in this paper as they were presented already in details in \cite{Leonforte2004,Leonforte2005,Leonforte2006,Tanguy2006}. Here we extend the results obtained on the quasistatically and athermally sheared two dimensional polydisperse glasses presented in our earlier work to finite shear rates by the use of molecular dynamics while the temperature is maintained well below the glass transition temperature. As in \cite{Varnik2004} we impose the transverse temperature ($T_{y}=5\cdot10^{-8}$) through the use of a simple velocity rescaling thermostat on the transverse component of the particle velocities. For the applied shear an equivalent amount of runs were produced under two boundary condition protocols, namely rigid walls boundary conditions (simplified notation RWBCs) at $y = \pm H/2$, where H is the height of the sample, and Lees-Edwards boundary conditions (noted LEBCs). In total we have analyzed here 24 samples of (625 particles, Lx=25.9938, Ly=25.9938), 8 (2500,51.9875,51.9875) and 8 (10000,103.975,103.975) all corresponding to a density $\rho=0.925$. We have sheared at finite shear rates also one larger sample (40000,206.950,206.950) that was not sheared with the quasi-static protocol. For each simulation, we collect data over four strain units ($\epsilon=4$) and store all the positions of the particles at a regular strain interval of $\delta\epsilon=10^{-3}$. In figure \ref{fig:StressStrain} we have reported the mechanical response of the smallest sample for RWBCs. We observe the characteristic flow behavior associated with glassy materials with a convergence of the response to the quasistatic limit as the shear rate is progressively reduced and a global non linear flow curve of the Herschel-Bulkley type : $\tau = \tau_{Y} + c_{1} \dot\gamma^{\beta}$, where $\tau_{Y}$ is the yield stress, and where one can define the viscous stress $\tau_{V}=\tau - \tau_{Y}$. In figure \ref{fig:StressStrainrate} we draw these flow curves for both boundary conditions and for all system sizes. The stress-strain rate curves of figure \ref{fig:StressStrainrate} are obtained by averaging the macroscopic stress values of the sheared glass obtained in figures \ref{fig:StressStrain} for strains larger than $\epsilon>25\%$, i.e. deep in the plastic flowing regime once the stationary flowing regime is established. Indeed we have checked that a linear velocity profile is established in the different samples for typical strains of the order of $\epsilon\sim2.5\%$.
 

\subsection{Convergence to the quasistatic limit.} Importantly the results presented above bridge the gap between the two types of approaches used in the literature, namely quasistatic energy minimization protocols and finite shear rates molecular dynamics methods, and resolve the controversy relative to the validity of the quasistatic protocols. Indeed as is evident from the mechanical response shown in figures \ref{fig:StressStrain} the quasistatic stress-strain curve appears as the limiting curve of the finite shear rates procedures. The superposition of the quasistatic response with the $\dot\gamma=10^{-5}$ shear rate response is in fact almost perfect in the early parts of the curves before small differences are amplified irremediably. One sees for example on the right hand side of figure \ref{fig:StressStrain} how even very small relaxations at around $\epsilon\sim3.2-3.4\%$ in the quasistatic response (black thick line) are also visible in the lowest shear rate curves (yellow and magenta).  Note that these small features of the mechanical response of the glass would not be visible if the temperature was higher and therefore inducing a noisier stress signal. The good convergence to the quasistatic protocol is also apparent in figures \ref{fig:StressStrainrate} were one can indeed observe that the values of the lowest finite shear rates are in good agreement with the $\dot\gamma=0$ shear rate method (under both boundary conditions $\sigma_{xy}(\dot\gamma<10^{-5})\sim0.4$), again confirming the physical relevance of the quasistatic method.

\subsection{Decomposing the plastic events in elementary units.} 
\begin{figure}[!hbtp]
\begin{center}
\includegraphics[width=7cm]{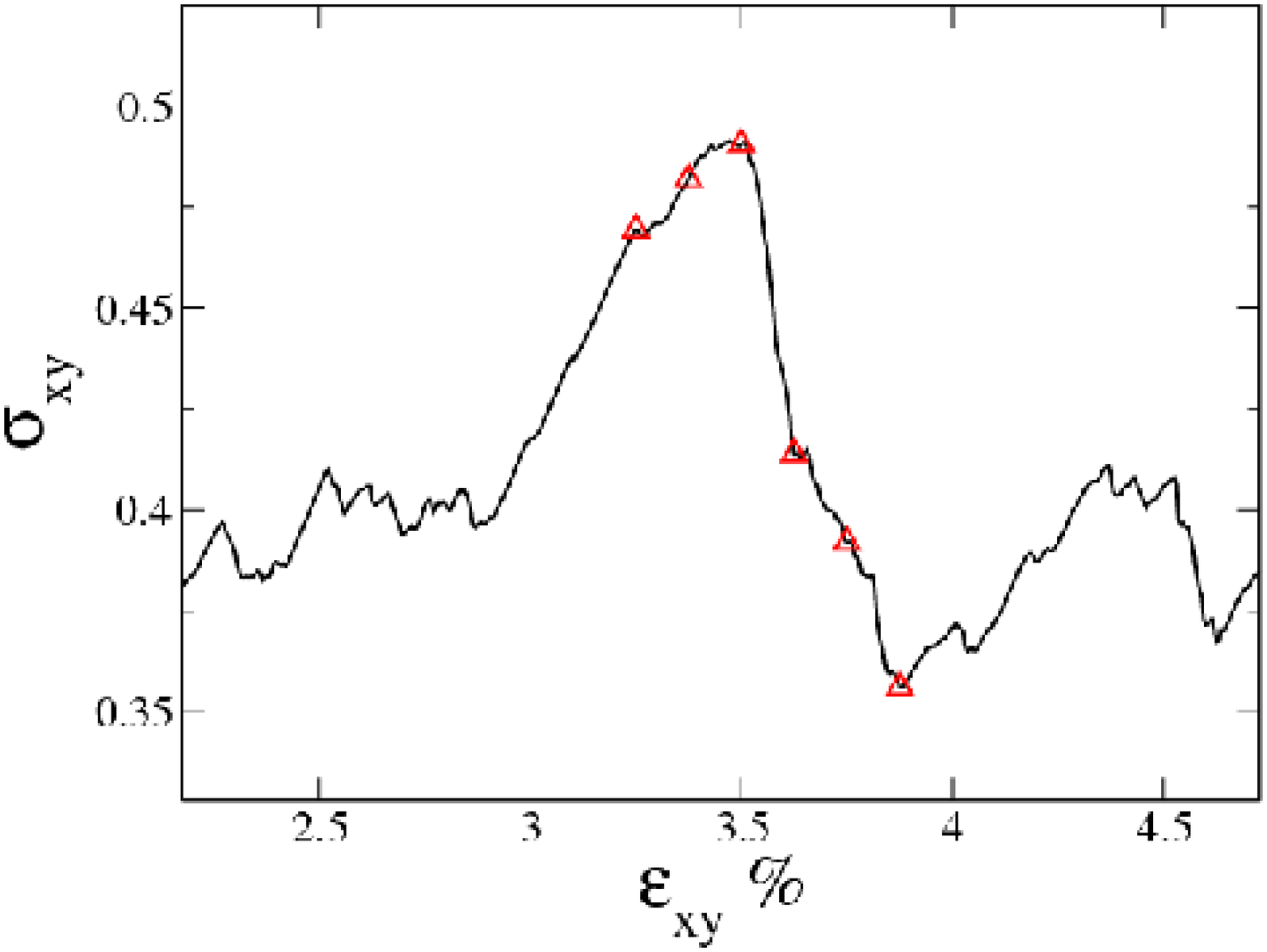}
\includegraphics[width=7cm]{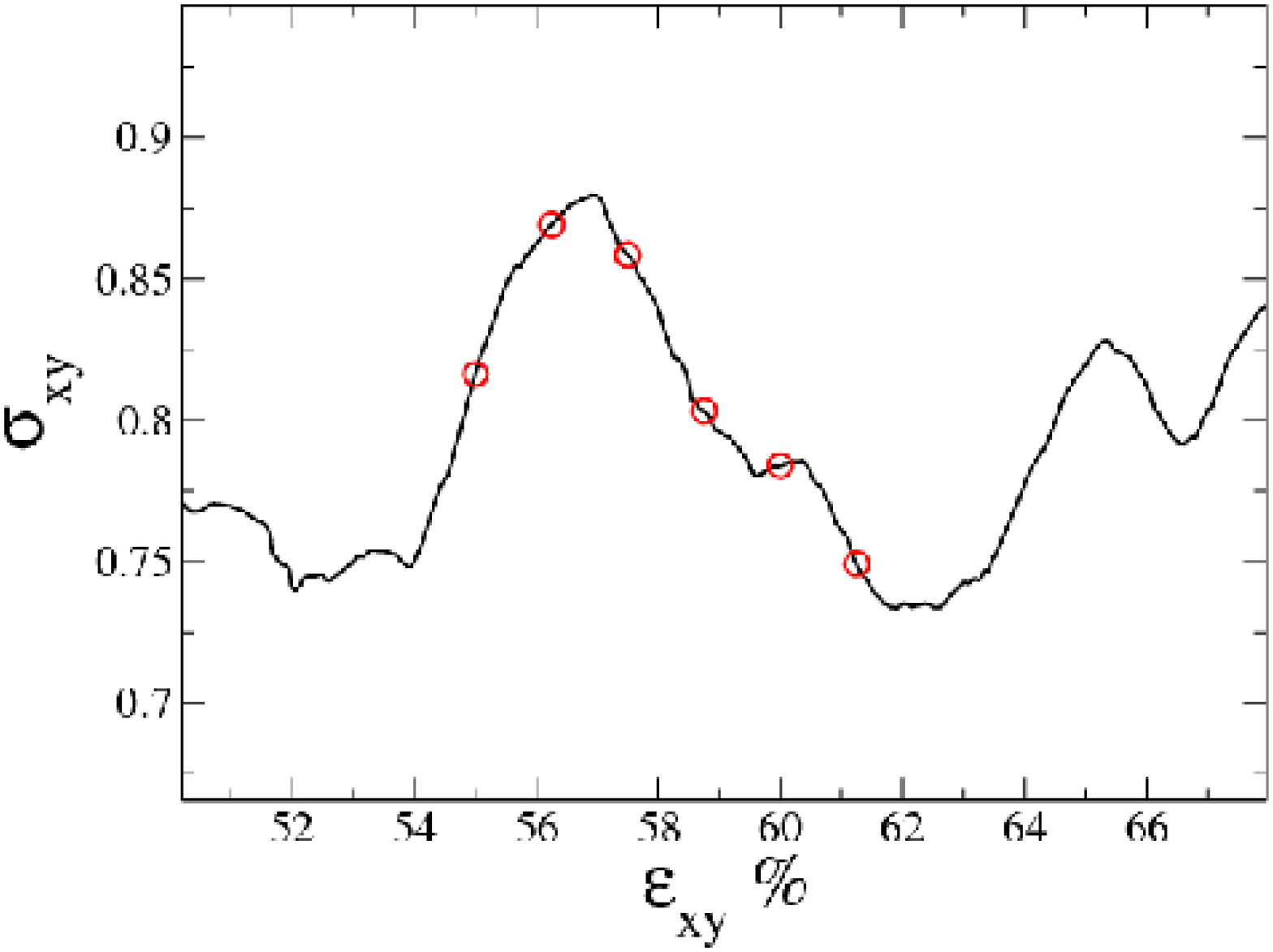}
\caption{Portion of the mechanical stress-strain response of a Lennard-Jones glass containing 10000 particles and sheared under LEBCs for two shear-rates. \textbf{Left :} $\dot\gamma=10^{-5}$. \textbf{Right : } $\dot\gamma=10^{-3}$.}\label{fig:RelaxEvent}
\end{center}
\end{figure}

\begin{figure}[!hbtp]
\begin{center}
\includegraphics[width=7cm]{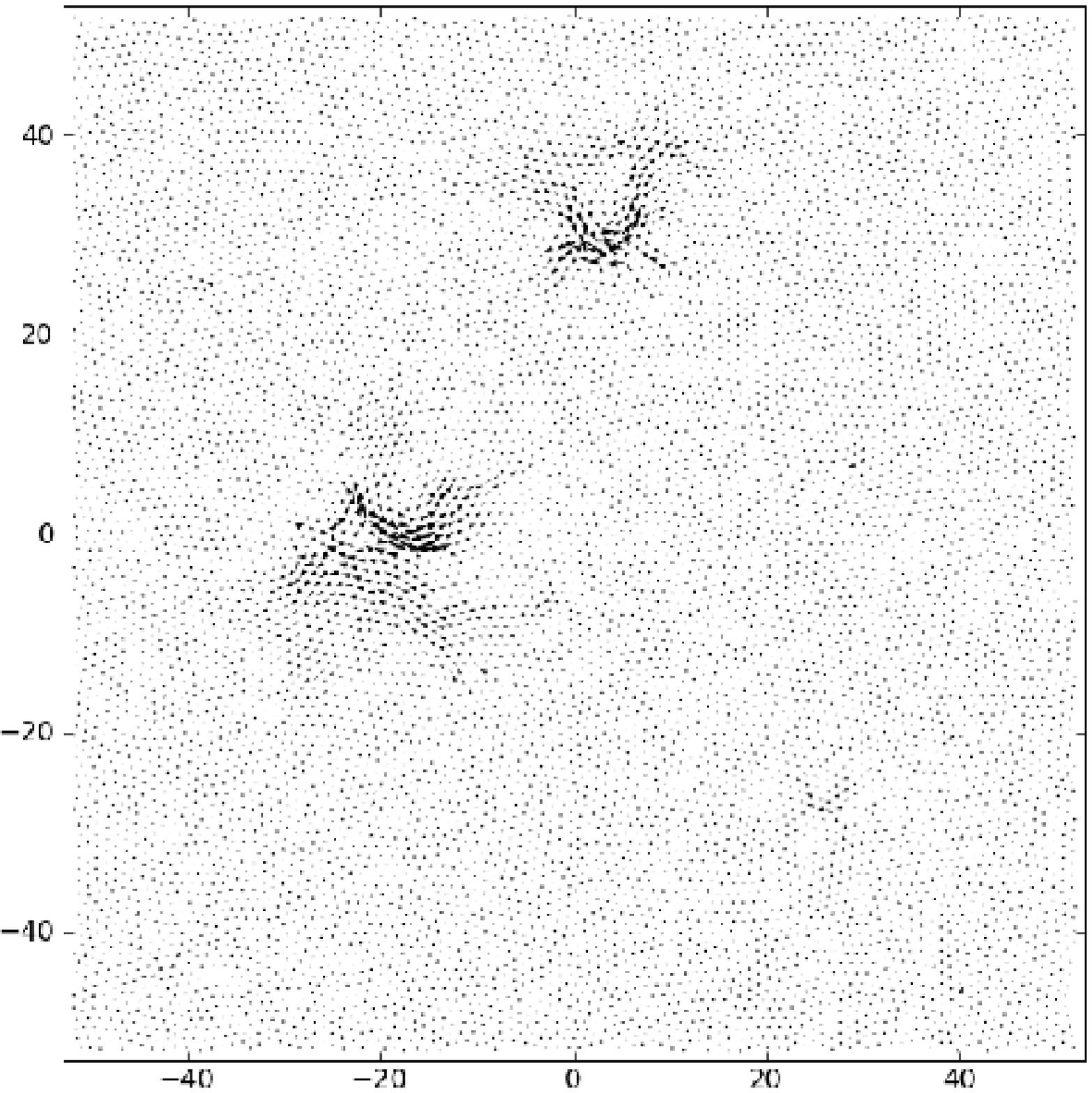}
\includegraphics[width=7cm]{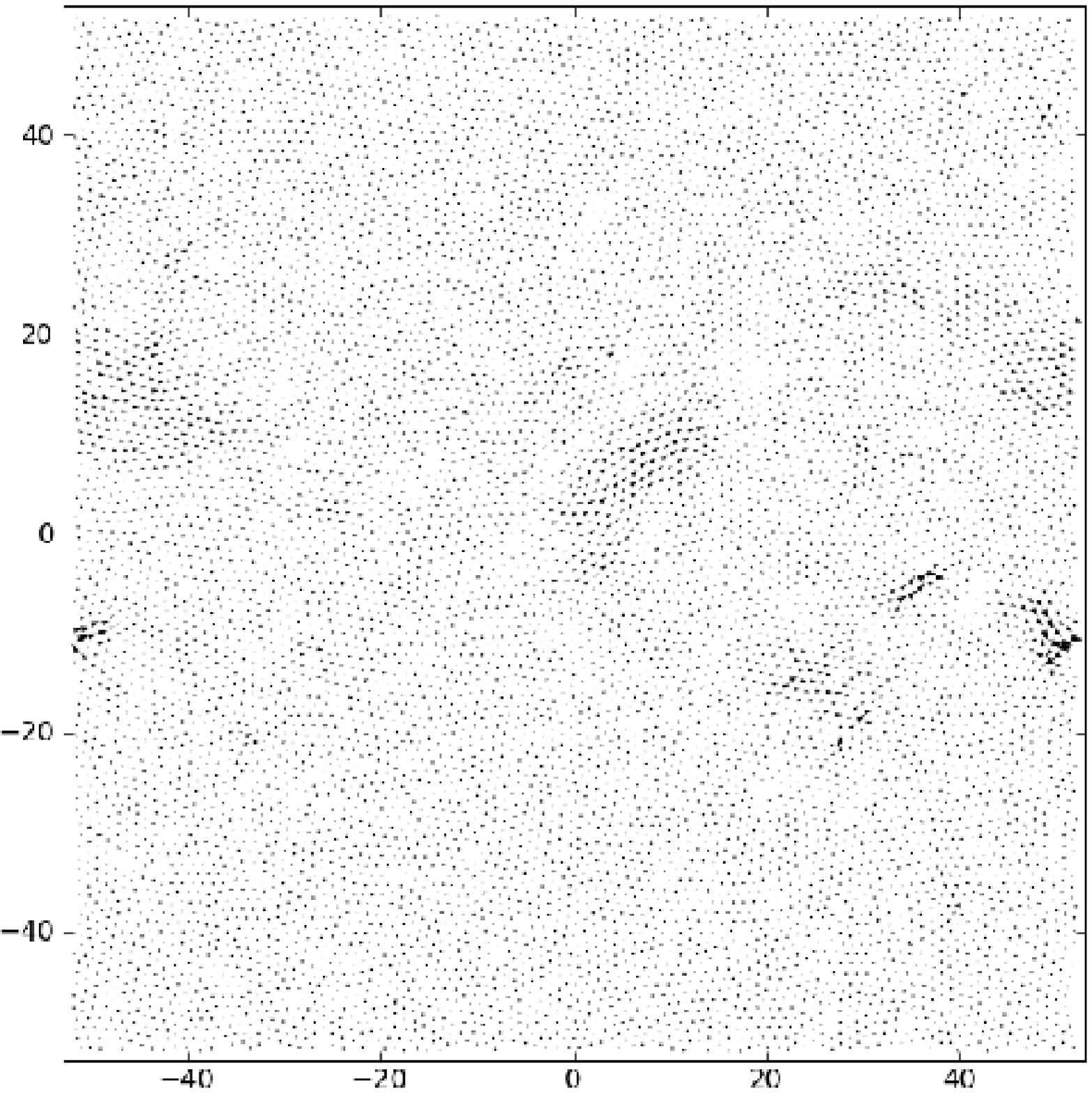}
\includegraphics[width=7cm]{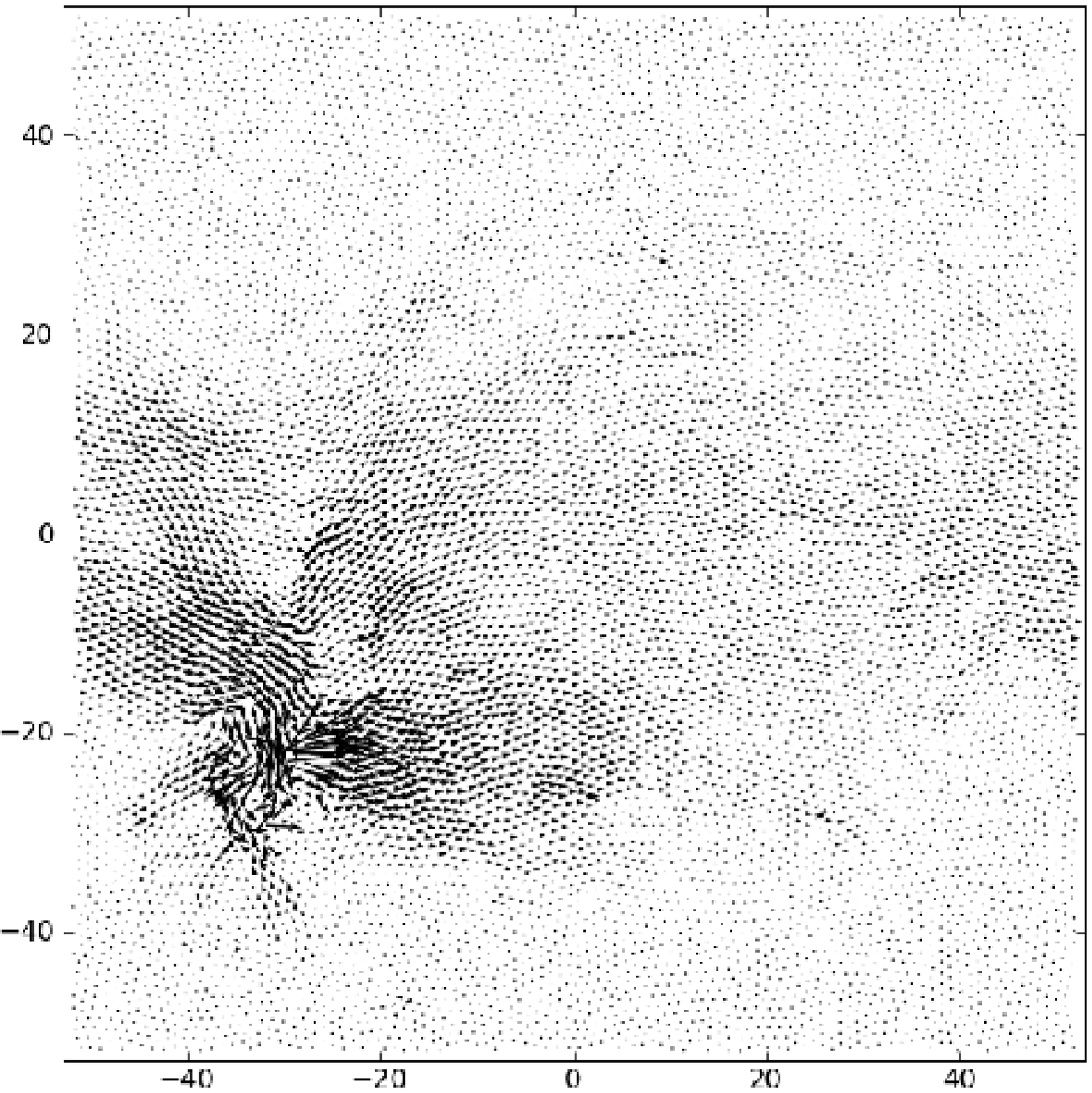}
\includegraphics[width=7cm]{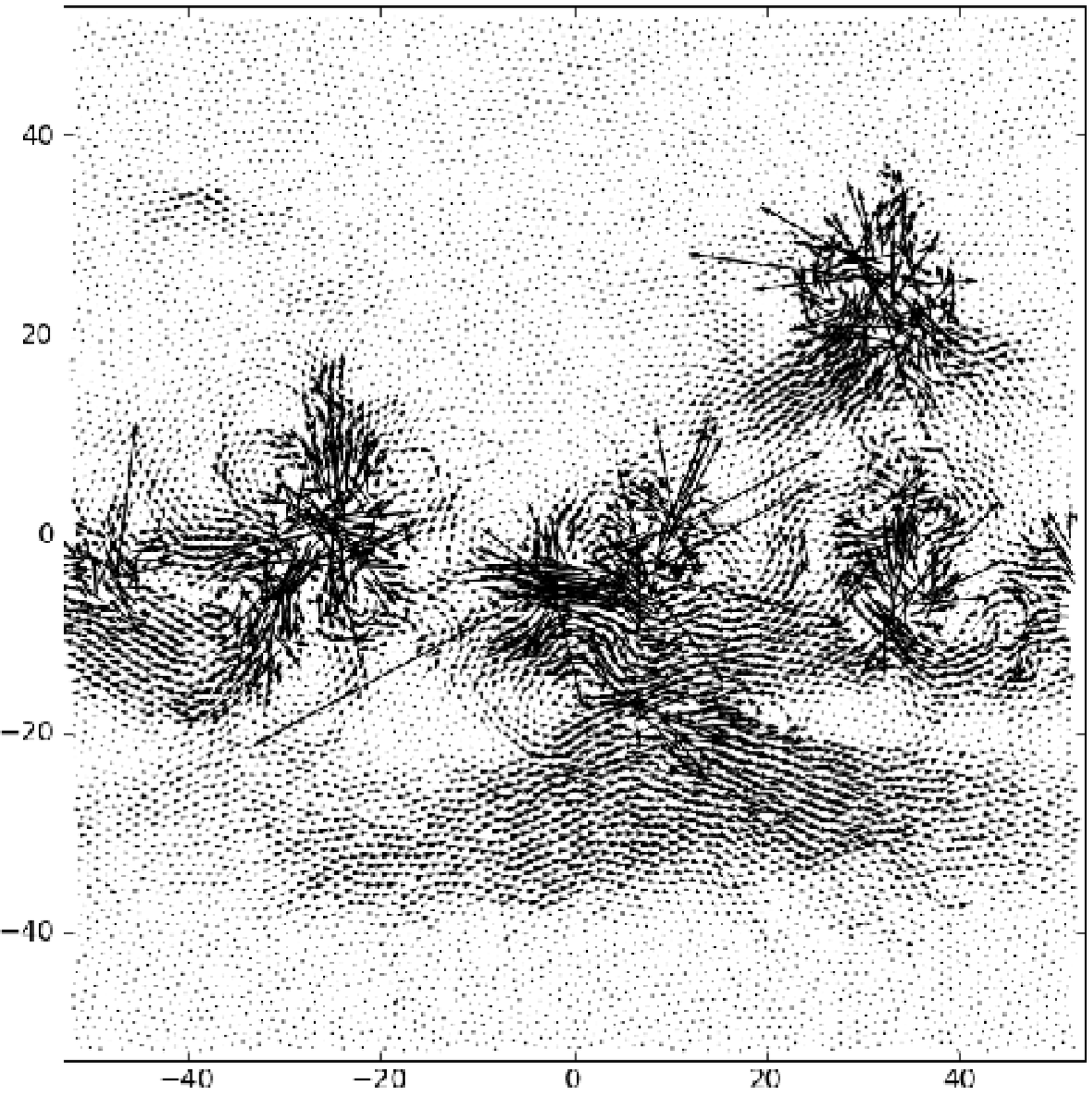}
\includegraphics[width=7cm]{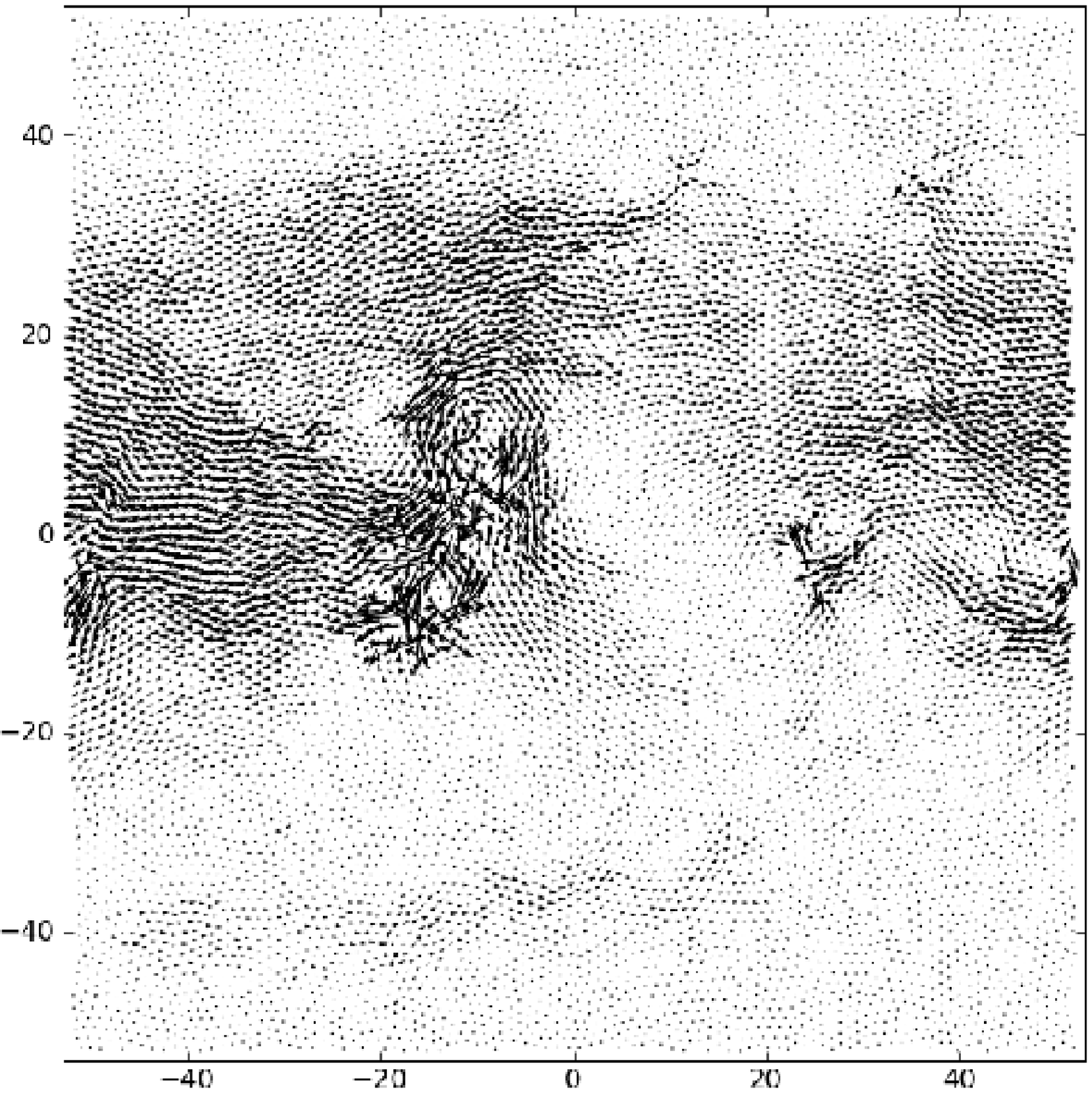}
\includegraphics[width=7cm]{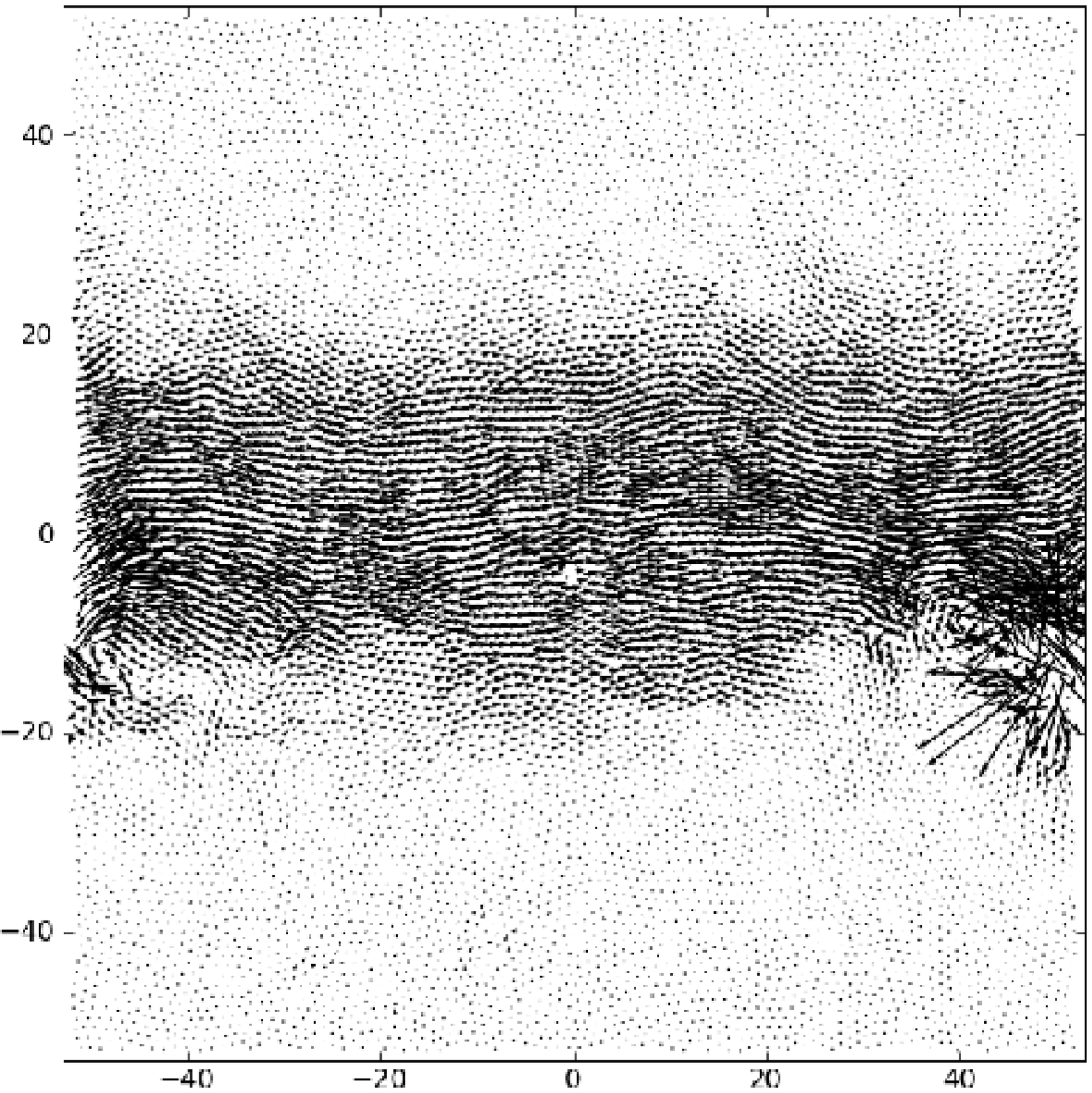}
\caption{Snapshots of the instantaneous displacement field (magnified ten times) during 1 LJU and corresponding to the triangular symbols of figure \ref{fig:RelaxEvent}. The shear-rate is here $\dot\gamma=10^{-5}$.}\label{fig:Frames1e-5}
\end{center}
\end{figure}

\begin{figure}[!hbtp]
\begin{center}
\includegraphics[width=7cm]{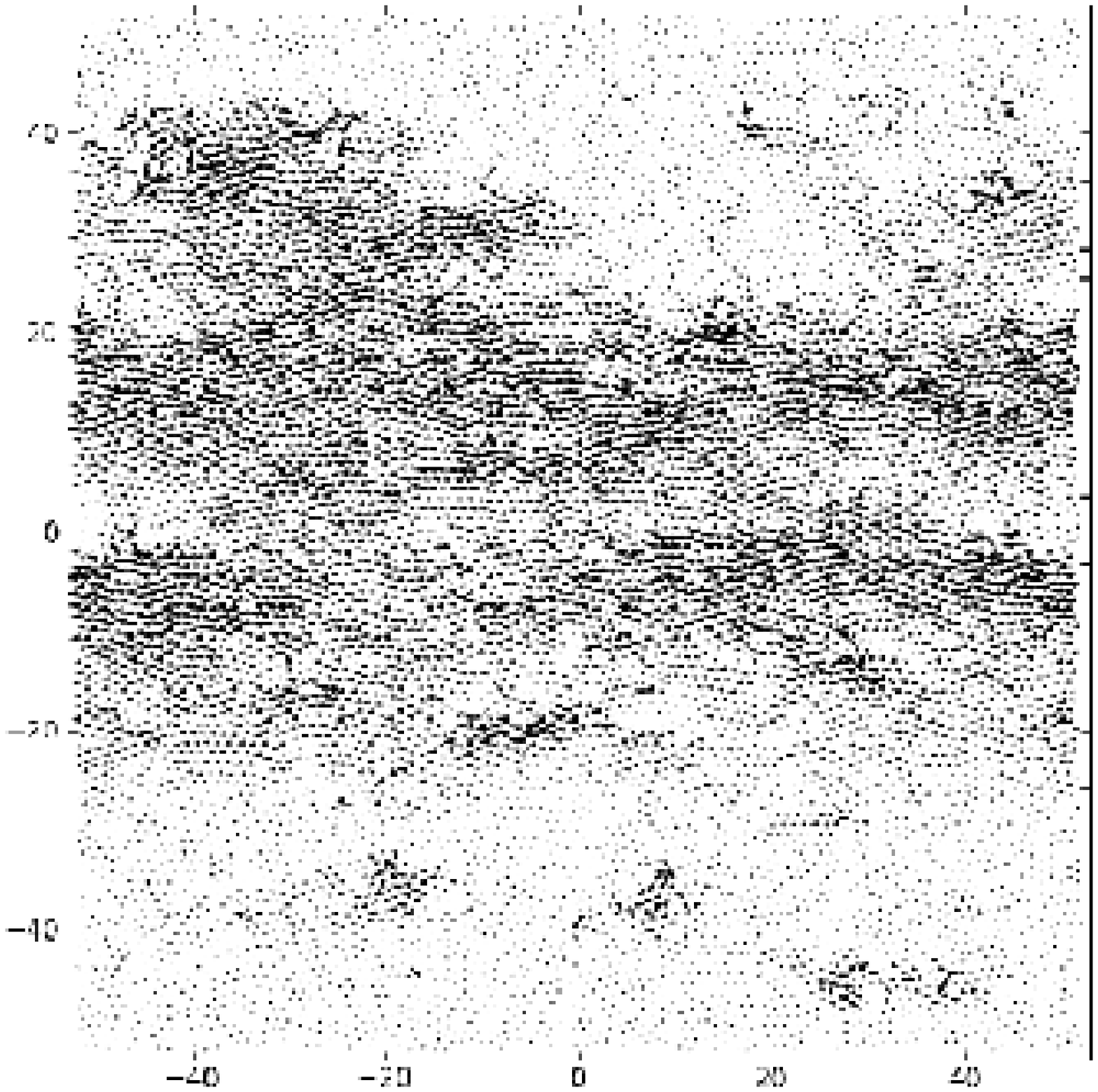}
\includegraphics[width=7cm]{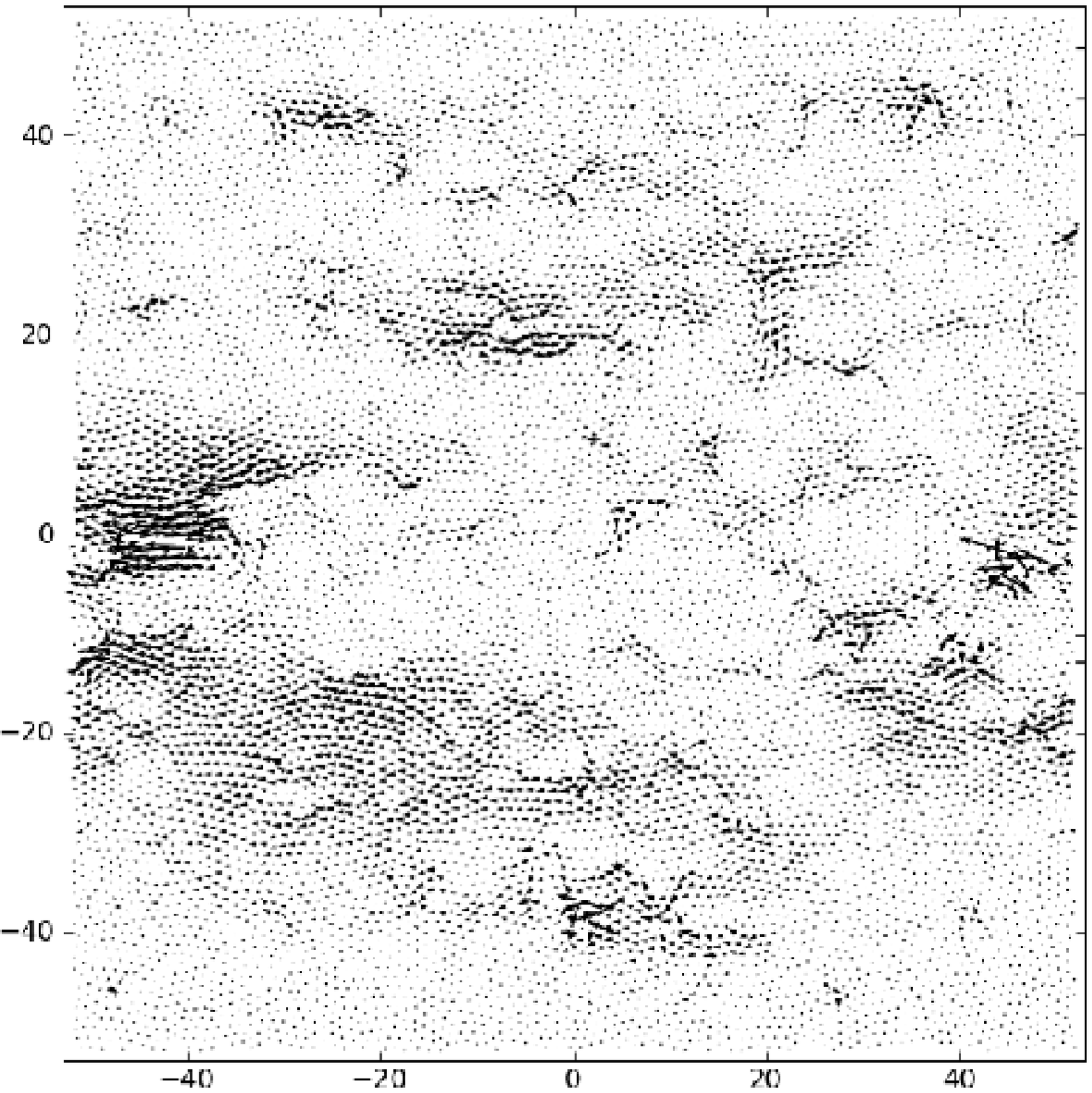}
\includegraphics[width=7cm]{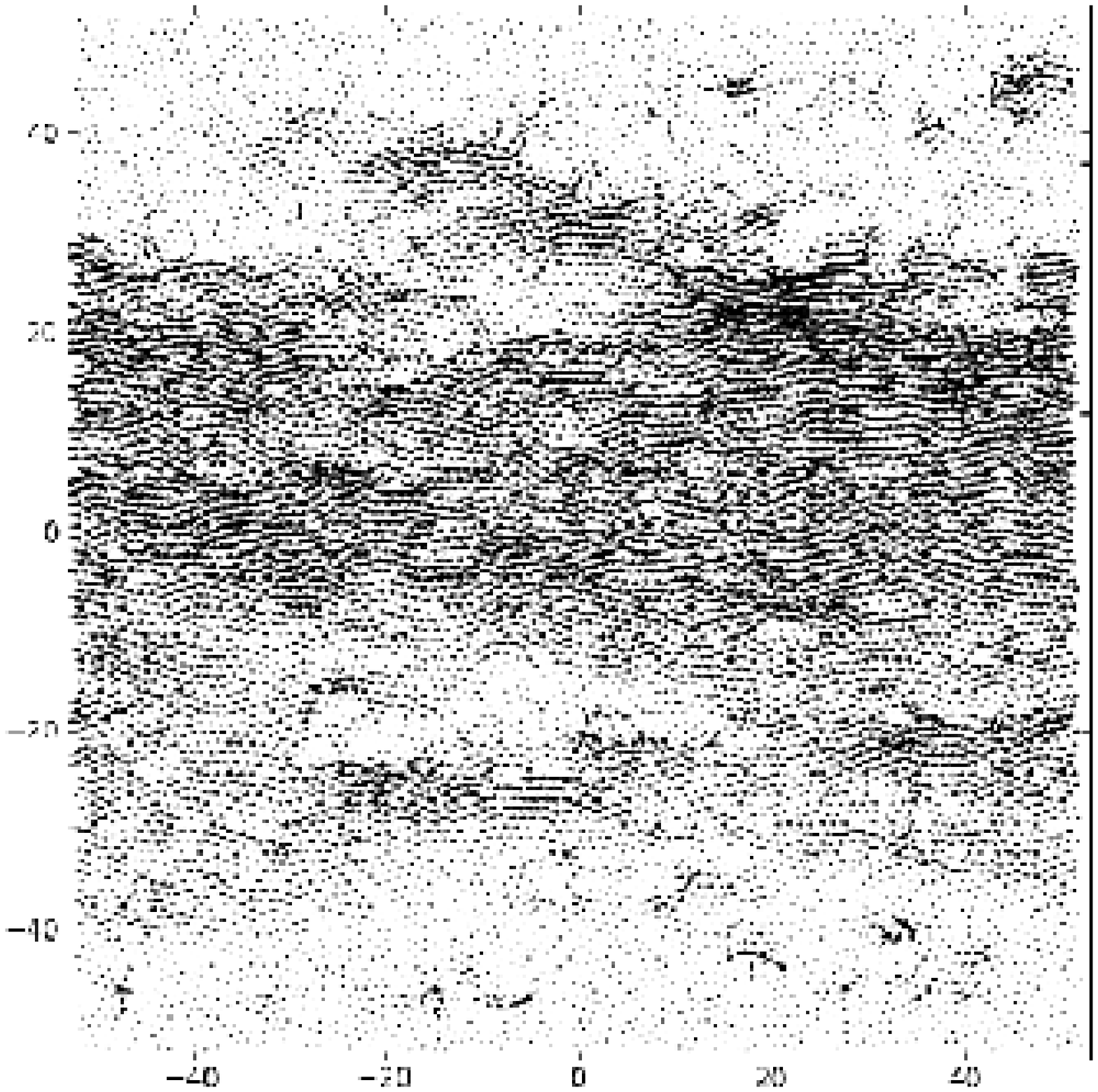}
\includegraphics[width=7cm]{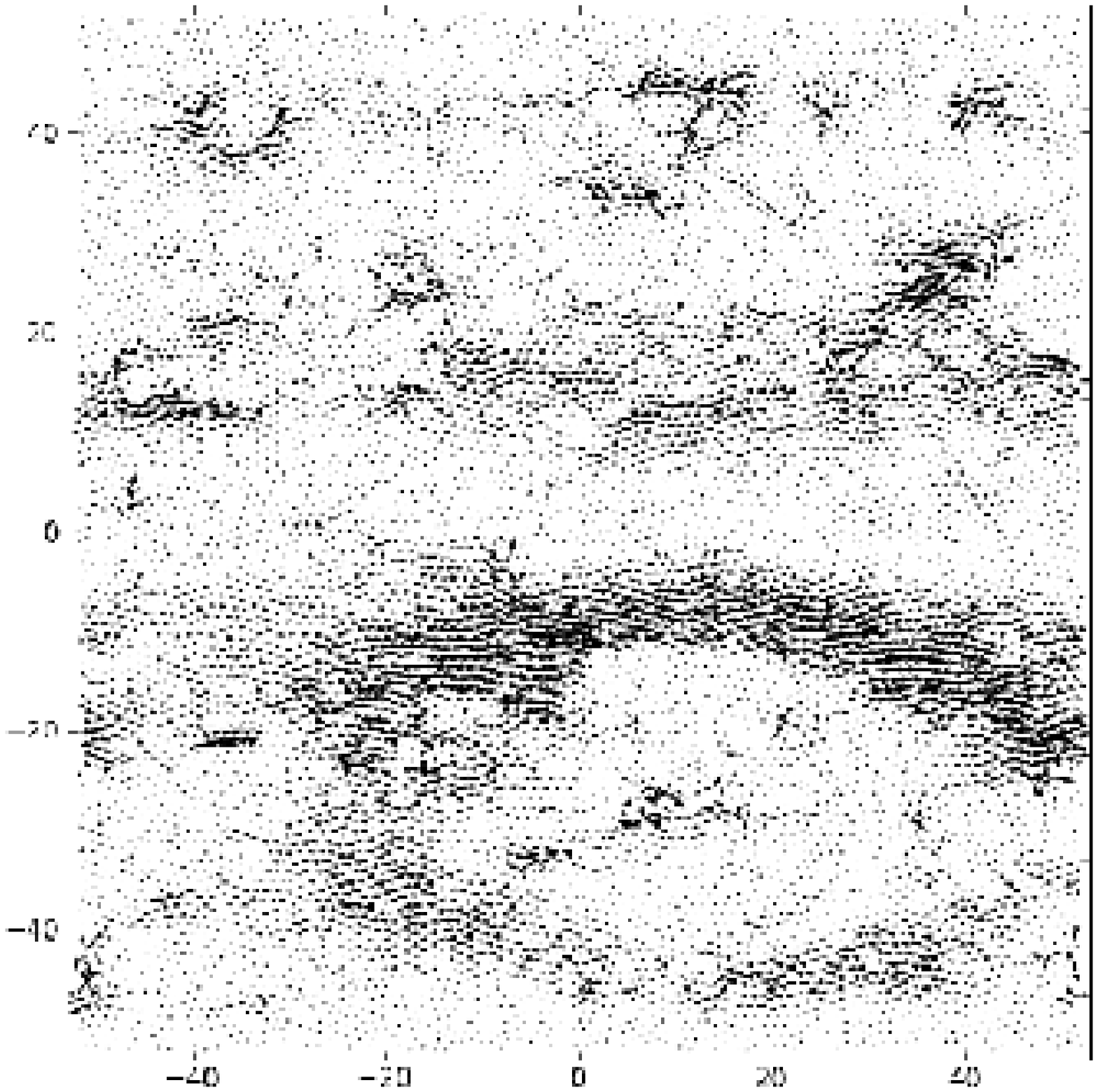}
\includegraphics[width=7cm]{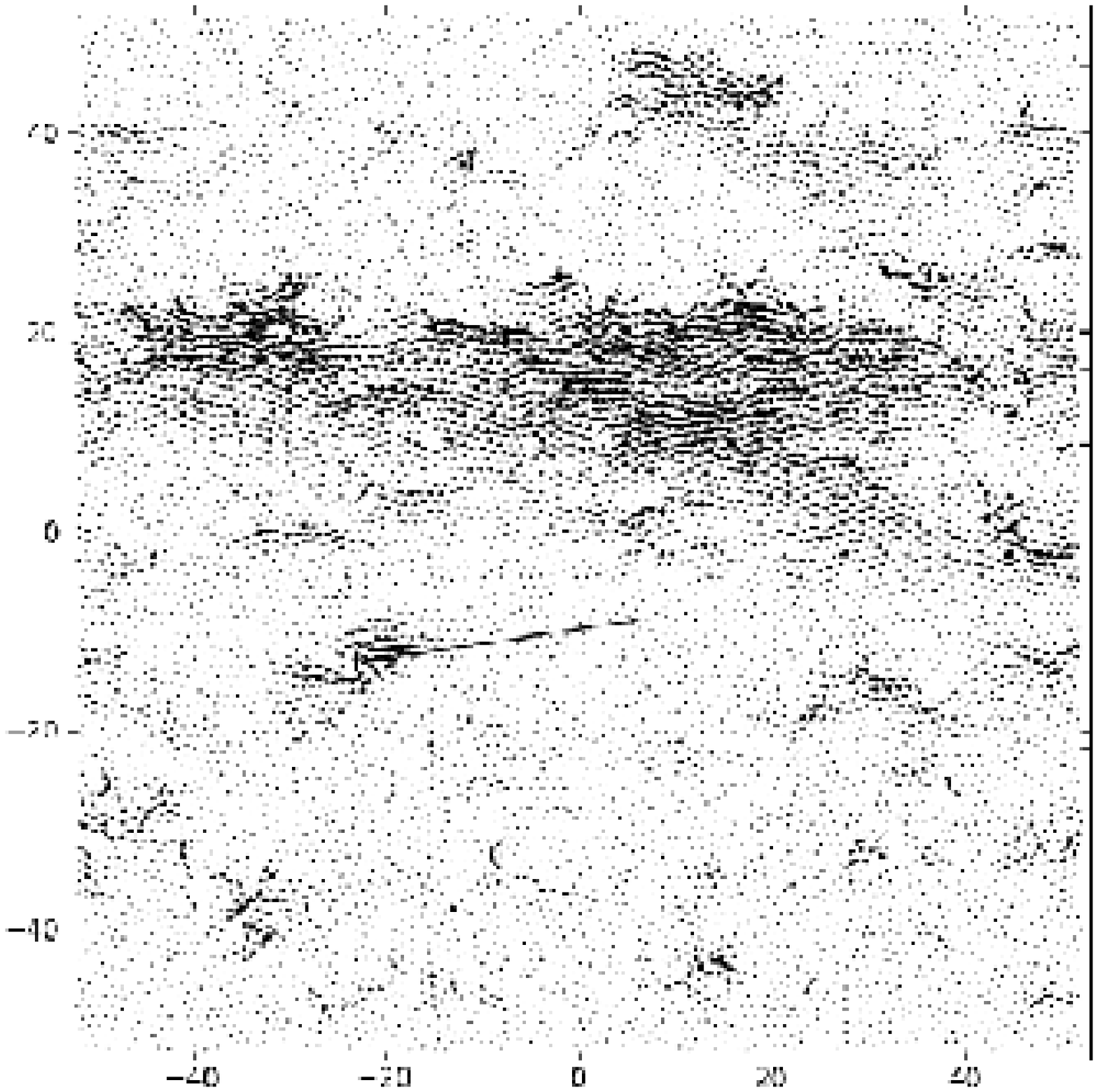}
\includegraphics[width=7cm]{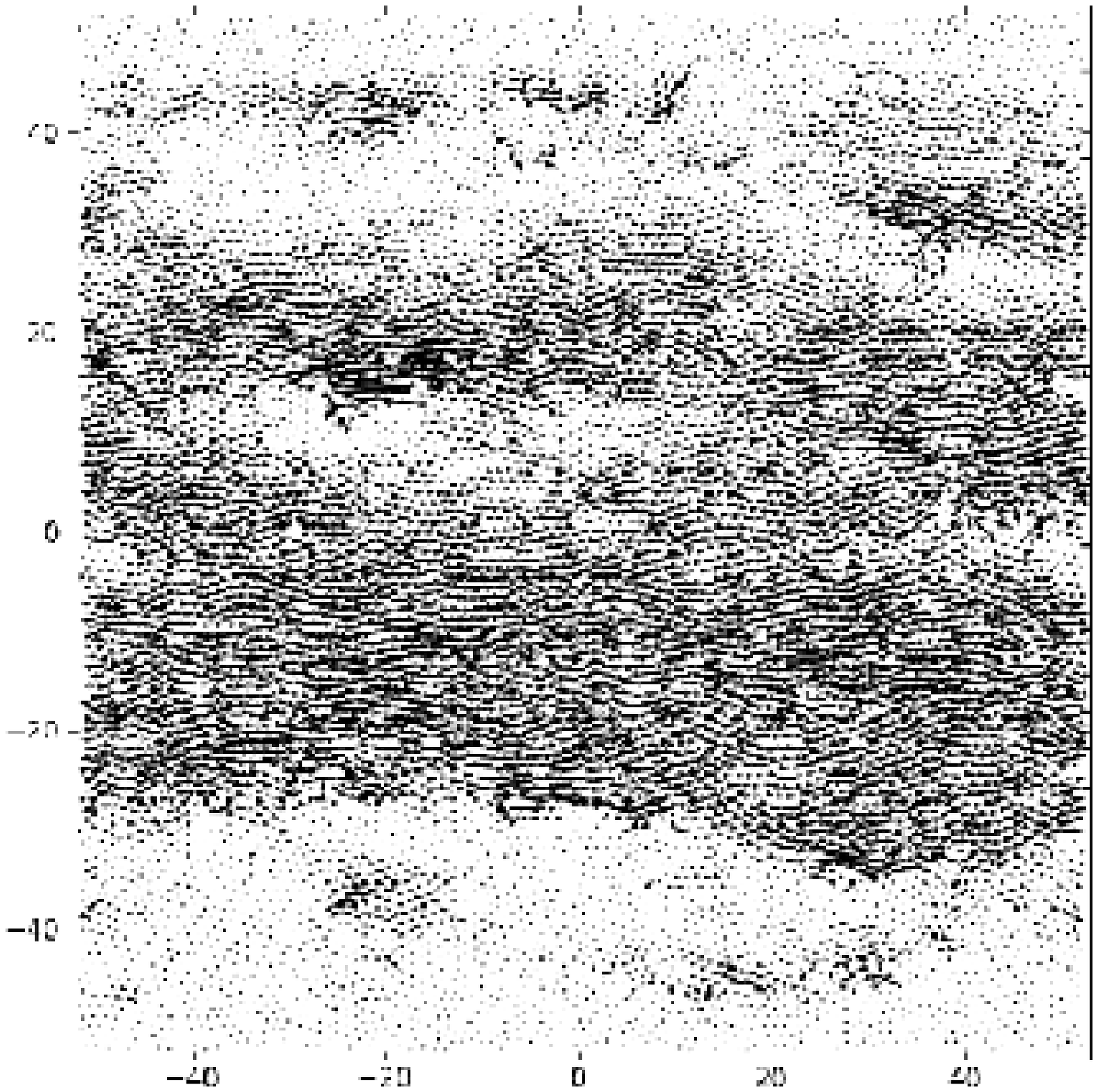}
\caption{Snapshots of the instantaneous displacement field during 1 LJU and corresponding to the circular symbols of figure \ref{fig:RelaxEvent}.  The shear-rate is here $\dot\gamma=10^{-3}$.}\label{fig:Frames1e-3}
\end{center}
\end{figure}

In an important series of papers \cite{Maloney2004,Maloney2006} it was recently shown by Lema\^itre and Maloney that the relaxation in mechanically driven glasses occurs through the formation of cascades of quadrupolar elementary units. In the quasistatic protocol to decompose the cascade in its subunits one needs to study in detail the evolution of the positions of the particles during the minimization procedure, with the inherent limitation of the minimization algorithms that one can not associate a time scale to the successive elementary rearranging units. This limitation is automatically overcome in the molecular dynamics simulation where time appears explicitly in the algorithm and where one can follow the evolution of the cascade in time. As is apparent at finite shear rates in figures \ref{fig:StressStrain} the plastic relaxation is not instantaneous during a stress drop and this typical lifetime of the plastic events is transposed in the stress-strain mechanical response in the downward portions with the slopes that increase with increasing strain rate. To understand intuitively the mechanisms involved in the mechanical response of the sheared glasses and the different time scales that are relevant it is highly instructive to visualize movies of the instantaneous non-affine displacement field during the deformation. The two panels of figures \ref{fig:Frames1e-5} and \ref{fig:Frames1e-3} show snapshots of such movies taken at regularly spaced strain intervals corresponding to the red symbols on figure \ref{fig:RelaxEvent}. As can be seen in this last figure, for the slowest strain rate $\dot\gamma=10^{-5}$ the strain interval between snapshots is $\delta\epsilon=2.5\times 10^{-3}$ (red triangles) and for the fastest strain rate $\dot\gamma=10^{-3}$ it is $\delta\epsilon=1.25\times 10^{-2}$ (red circles). When looking at the two panels one must bare in mind that while in the slowly sheared case the total strain applied between the first figure and the last is less than  $1\%$ in the fast case it is more than $5\%$. Also the non-affine displacement field represented on each snapshots correspond to the displacement during a time interval of $\delta t=1 LJU$ and therefore the associated strain $\delta\gamma = \dot\gamma \times \delta t$ is one hundred times larger for the fast shear than for the slow shear. If one assumes that the density of weak triggering zones of plasticity is homogeneously distributed through the sample with a shear rate independent density per unit strain $\Omega$ then one expects also to see 100 times more such nuclei of plasticity in the fast case. This explains the visual impression that there exists a higher density of local plastic displacements (black arrows) on panel \ref{fig:Frames1e-3} than on panel \ref{fig:Frames1e-5} where only a few local plastic rearrangements are observed. Another difference between the slowly and fast driven regimes is that while in the fast case the nuclei do not seem to merge or percolate or evolve in a correlated manner in the slowly driven case the local quadrupoles show a cooperative and correlated avalanche dynamics. This is particularly visible on the fourth panel of figure \ref{fig:Frames1e-5} where one can see about 5 such elementary plastic units forming a L-shape with four events horizontally aligned. Interestingly the typical distance between quadrupoles on this figure is about $\xi\sim20LJU$ which is reminiscent of the length scale that emerged for example in the autocorrelation function of the non-affine field in recent studies of similar Lennard-Jones glasses \cite{Leonforte2005}. These findings are in line with observations of similar weak zones that grow and trigger the flip of neighbouring zones as depicted in fig. 1 of \cite{Lemaitre2009} confirming the validity of the picture of the dynamics of slowly driven glassy materials as dominated by the accumulation and cascading of plastic events. Beyond this general mechanism a careful study of the spatio-temporal signal associated to the slow drive ($\dot\gamma=10^{-5}$) shows different time scales $\tau$ associated with an entire zoology of typical sequences of plastic events. First one observes short lived local quadrupoles, typically visible only during one snapshot $\tau_{e}<1LJU$, and that do not trigger a cascade. Some local rearrangements seem to be locked and to survive for longer time intervals of the order of $\tau_{e}\sim10-100LJU$. In general this type of rearrangement triggers in its vicinity (vertically or horizontally) subsequent similar events. Sometimes as is the case on the forth snapshot of figure \ref{fig:Frames1e-5} this cascade leeds to the formation of a system spanning shear band. Finally one can associate also a timescale to the global relaxation process of figure \ref{fig:RelaxEvent} which is here for the slow shear rate $\tau_{c}=\gamma_{c}/\dot\gamma\sim1000$ LJU ($\gamma_{c}$ corresponds here to the duration of the relaxation event, i.e. the total strain associated to each downward slopes in the stress-strain response) and for the fastest shear rate  $\tau_{c}\sim 100$ LJU. Note that these values of the typical duration of an entire relaxation process are in line with the values that one can compute from figure \ref{fig:StressStrain}. In this figure one also sees that for low enough shear rates $\dot\gamma<10^{-4}$ there is an intrinsic lifetime associated to a plastic rearrangement process which is proven by the fact that the slope of the stress curves is proportional to the shear rate. For shear rates larger than $\dot\gamma\sim10^{-4}$ the relaxation strain becomes larger than the typical strain between relaxation events and therefore one can see this value of the shear rate as a mark of a transition to a different type of rheology also characterized by an important increase of the average yield stress as can be seen from figure \ref{fig:StressStrainrate}. It is very striking that the avalanche like behavior seems to be somehow screened when the shear rate is increased. This result has been reported also elsewhere in atomic scale simulation \cite{Lemaitre2009} but also in mesoscopic yield stress models \cite{Picard2005} and \cite{BocquetPrivate}. Only rarely studied in driven glassy material the growth of a cooperativity length scale near the glass transition is well known in the supercooled liquids literature (see for example \cite{Dalle-Ferrier2007}) or in simple lattice gas models \cite{Ashton2008} and have been interpreted in the framework of facilitated dynamics (for a review see \cite{Ritort2003}). Of course the shear that one applies on glassy systems breaks the symmetry of supercooled liquids (this is apparent for example in the existence of preferred orientations for the local quadrupoles and for the system spanning shear bands along the neutral axis of the external applied strain. In supercooled liquids the directions of the rearrangements are isotropic.) but nevertheless it appears tempting to find, in line with supercooled liquids, a mapping between the dynamics of the sheared glass and a simpler facilitated model. The detailed description of the elementary rearranging processes that we propose here should help to devise reasonable ingredients for these models. A first approach in this direction was proposed by Picard \textit{et al} \cite {Picard2005} and we will briefly compare our results to this model at the end of this paper. To conclude with the description of the panels \ref{fig:Frames1e-5} and \ref{fig:Frames1e-3} let us mention that these plastic rearrangements independently of the shear rate emit a transverse sound wave propagating at a typical transverse sound speed characteristic of the Lennard-Jones glass ($c_{s}\sim\sqrt{\mu/\rho}\sim3-4$ LJU) and appear in general as dark regions on the snapshots of panel \ref{fig:Frames1e-5} and \ref{fig:Frames1e-3}. This allows to introduce a new timescale $\tau_{s}=L/c_{s}\sim10$ LJU, i.e. comparable to the life time $\tau_{e}$ of the elementary plastic rearrangements for a system of size $L=50$. We now turn to the detailed study of the dynamics in the driven systems at the atomic level.

\section{Motion of particles}
\label{Motion}

\subsection{Two time correlation functions.} 
\begin{figure}[!hbtp]
\begin{center}
\includegraphics[width=7cm]{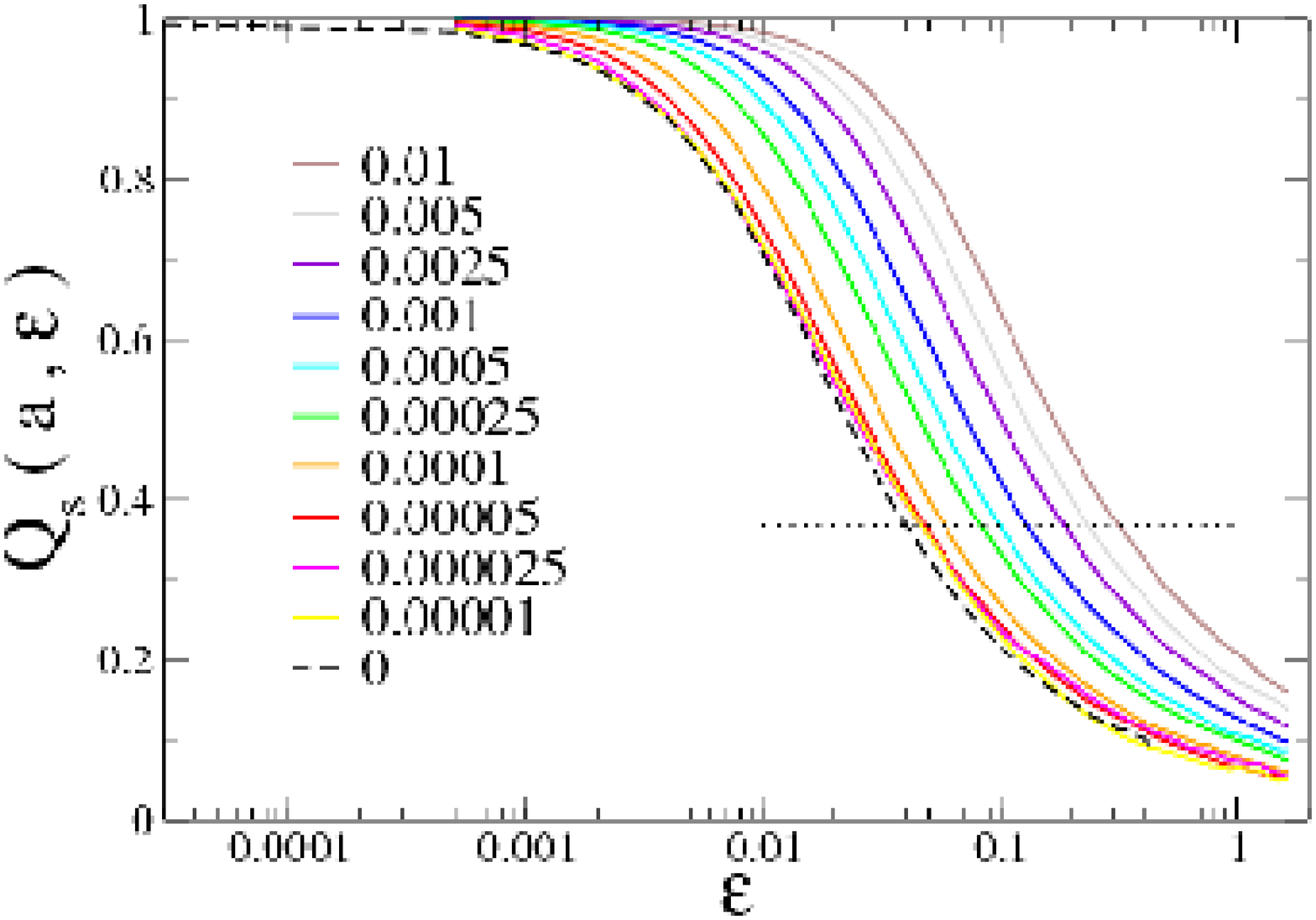}
\includegraphics[width=7cm]{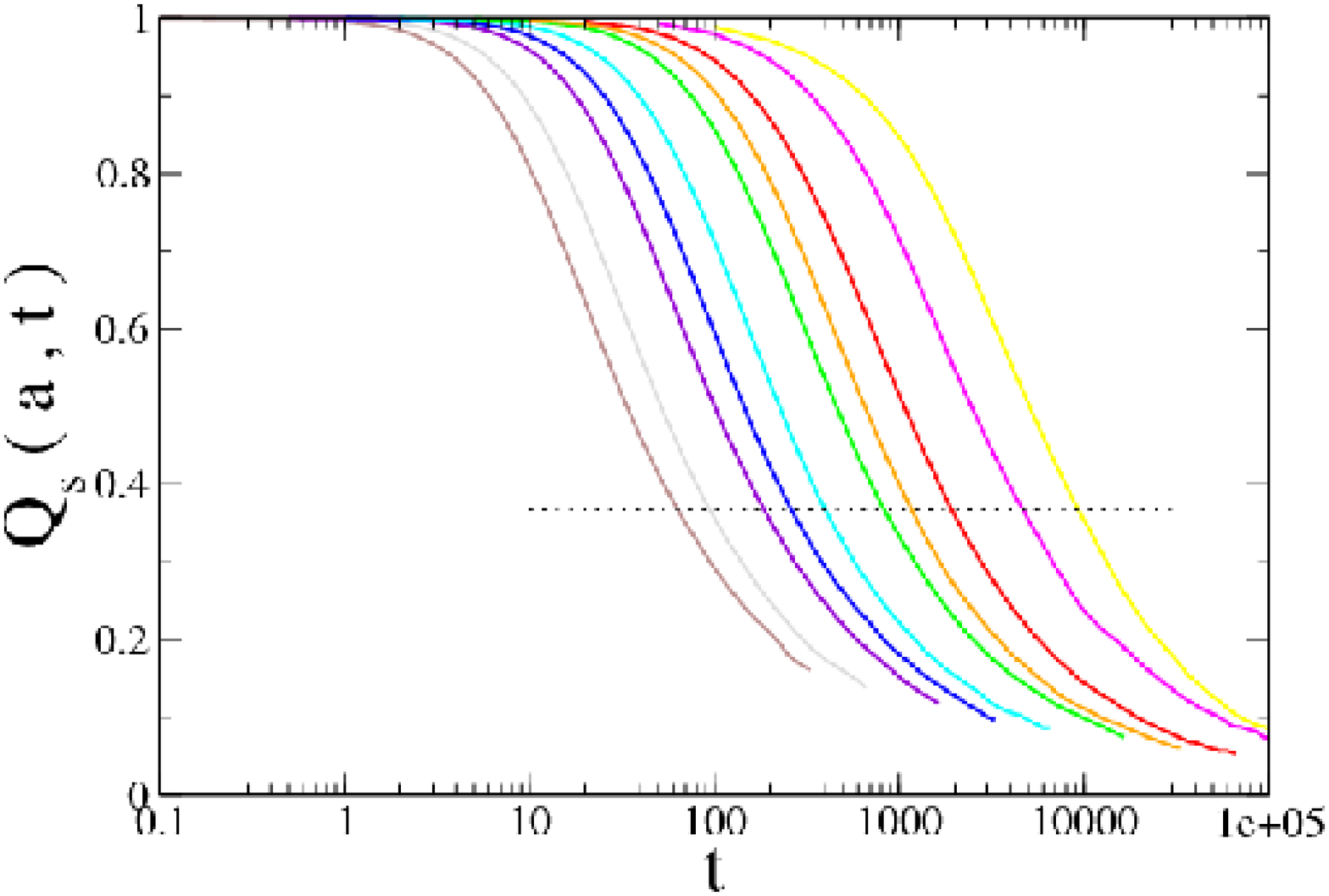}
\caption{\textbf{Left} : 2-point correlation function $Q_{s}(a,\epsilon)=\sum_{i}\exp(-\frac{(y_{i}(\epsilon)-y_{i}(0))^2}{2a^2})$ against strain $\epsilon$ for a system with LEBC. Thin lines (color on pdf version) : from top to bottom shear rate values are 0.01,0.005,0.0025,0.001,0.0005,0.00025,0.0001,0.00005 and 0.000025. Dashed line : $Q_{s}(a,\epsilon)$ vs $\epsilon$ for quasistatic shear. Doted line corresponds to the value 1/e for which we calculate the 1/e-relaxation strains and times. \textbf{Right}: $Q_{s}(a,t)$ against time $t$. In both figures the parameter $a$ is chosen equal to $0.1$.}\label{fig:Qs}
\end{center}
\end{figure}

The study of two time correlation function, besides its direct importance to quantify the characteristic relaxation times, enables to make comparisons with simple models of the rheology. In what follows, to examine the dynamics of the local density and the relaxation associated times, we compute therefore, on an equal foot, the self intermediate scattering function $F_{s}(\textbf{k},t)$,
\begin{equation}
\label{eqn:sisfbis}
F_{s}(\textbf{k},t)=\frac{1}{N}\sum_{i}\cos\left[\textbf{k}\cdot\left(\Delta \textbf{r}_{i}(t)\right)\right]\mbox{  ,}
\end{equation}
and the self correlation function (as in \cite{Lechenault2008a}) $Q_{s}(a,t)$ defined by,

\begin{equation}
\label{eqn:Qs} 
Q_{s}(a,t)=\frac{1}{N}\sum_{i}\exp(-\frac{\Delta \textbf{r}_{i}(t)^{2}}{2a^{2}})\mbox{  ,}
\end{equation}
where $\Delta \textbf{r}_{i}(t)=\textbf{r}_{i}(t'+t)-\textbf{r}_{i}(t')$ is the displacement vector. In what follows it is the spatial and time average of these two function that we compute and in general in what follows we replace the displacement vector $\Delta \textbf{r}_{i}(t)$ by its non-affine contribution $\Delta \textbf{r}^{na}_{i}(t)$ (or even simply by the transverse displacement) defined rigorously by,
\begin{equation}
\Delta \textbf{r}^{na}_{i}(t)=\textbf{r}_{i}(t'+t)-\dot\gamma\int_{t}^{t'+t}dt''y_{i}(t'')\textbf{e}_{x}-\textbf{r}_{i}(t) \mbox{  ,}
\end{equation}
for a shear in the x direction. In practice we find that within a good degree of accuracy $\Delta \textbf{r}^{na}_{i}(t)=\textbf{r}_{i}(t'+t)-\dot\gamma y_{i}(t')\textbf{e}_{x}-\textbf{r}_{i}(t)$ and we use this expression. Figure \ref{fig:Qs} represent the function $Q_{s}$ both in function of time and strain. After a slow decrease at small times/strains the function $Q_{s}$ exhibits a power law decay as a function of strain $\epsilon$, $Q_{s}(a,\epsilon)\propto\epsilon^{-\beta}$ with $\beta\gtrsim 0.5$ signaling shear induced structural relaxation. This power law decay is in contrast with the more `classic' self intermediate scattering function that exhibits (not shown here) a compressed exponential decay $F_{s}(k_{y},\epsilon)\propto\exp(-(\epsilon/\epsilon_{c})^{\beta})$, with $\beta\gtrsim1.0$. Note also that the $Q_{s}(a,\epsilon)\propto\epsilon^{-\beta}$ decay is compatible with a Gaussian distribution function of the transverse displacements $P(\Delta y,\epsilon)$ and with a diffusive transverse mean square displacement $\langle\Delta y^{2}\rangle$. As shown in figure \ref{fig:Qs} the convergence to the quasistatic curve is verified when $Q_{s}$ is plotted against strain for values of the shear rate $\dot\gamma\lesssim10^{-4}$. 

%

\begin{figure}[!hbtp]
\begin{center}
\includegraphics[width=7cm]{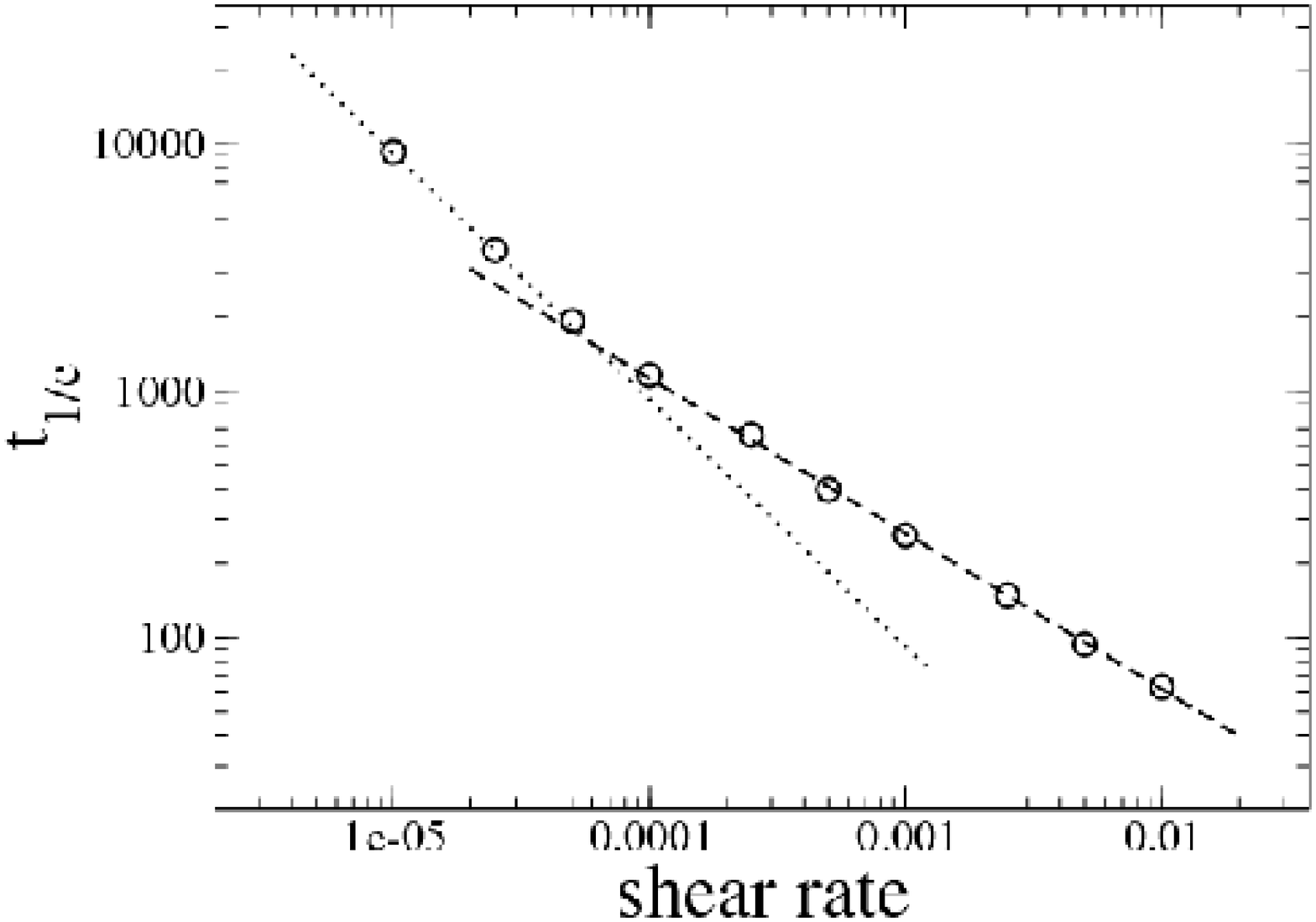}
\includegraphics[width=7cm]{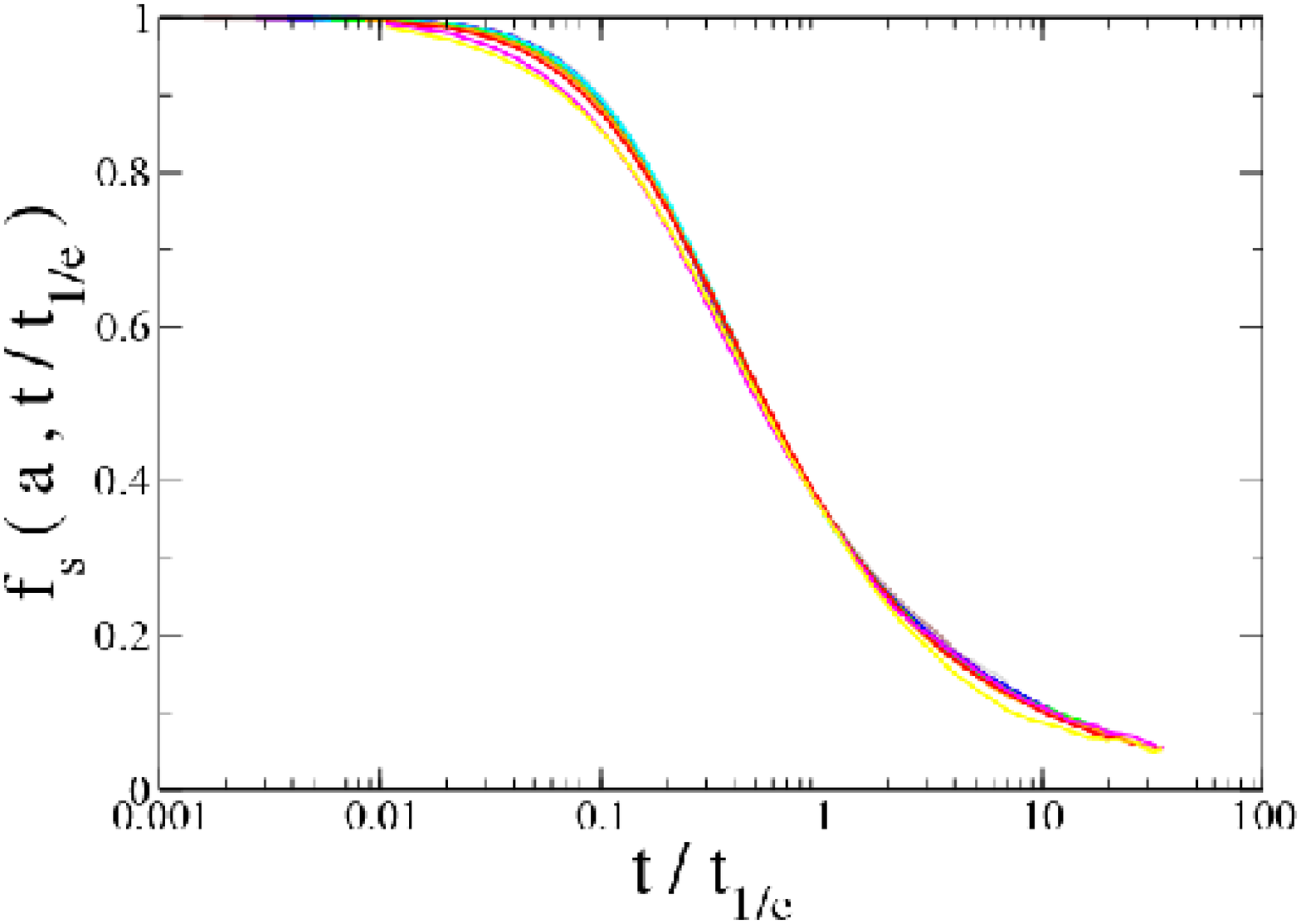}
\caption{\textbf{Left} : Relaxation time $t_{1/e}$ vs shear rate $\dot\gamma$ for the system of figure \ref{fig:Qs}. Dotted line $t_{1/e}\propto\dot\gamma^{-1}$, dashed line $t_{1/e}\propto\dot\gamma^{-0.63}$. \textbf{Right}: Rescaled self correlation function $Q_{s}(a,t)$ of figure \ref{fig:Qs} when time is rescaled by the structural relaxation time $t_{1/e}$. All curves superimpose rather well on a master curve $f_{s}(a,t/t_{1/e})$.}\label{fig:relax1overe}
\end{center}
\end{figure}

In figure \ref{fig:relax1overe} we have reported the relaxation times $t_{1/e}$ - the points of intersection of the dotted line with the colored curves in figure \ref{fig:Qs} verifying $Q_{s}(a,t_{1/e})=1/e$ - for different shear rates and different system sizes. Of course the relaxation strains and relaxation times are related through the simple relation $\epsilon_{1/e}=t_{1/e}*\dot\gamma$ and we therefore only report the relaxation times for simplicity. In \cite{Mobius2009,Besseling2007} similar analysis were reported in experiments respectively on foams and colloids. Looking at the relaxation time dependence on shear rate in these studies the authors found scalings of the form $t_{1/e}\propto\dot\gamma^{-\nu}$ with $\nu\simeq0.66$ in \cite{Mobius2009} and $\nu\simeq0.8$ in \cite{Besseling2007}. Similarly in extensive numerical studies \cite{Varnik2006a,VarnikHabilit} of sheared Lennard-Jones glassy materials the authors have computed these relaxation times without explicitly writing to our knowledge the functional form of the dependence of the relaxation time on shear rate. 


Here as shown in figure \ref{fig:relax1overe} we find two regimes : for high shear rates,  $\dot\gamma\gtrsim10^{-4}$, the structural relaxation time of the sheared glass scales with the global shear rate as $t_{1/e}\propto\dot\gamma^{-\nu}$ with $\nu\simeq0.63$, while for lower shear rates,$\dot\gamma\lesssim10^{-4}$, the relaxation functions $Q_{s}(a,\epsilon)$ reach a quasistatic limit strain limit $\epsilon_{1/e}\sim0.04$ and therefore the associated relaxation times scale as $t_{1/e}\propto\dot\gamma^{-\mu}$ with $\mu\simeq1$. The crossover between the quasistatic and shear rate dominated regimes is size dependent with the transition shear rate $\dot\gamma_{c}$ being lowered as the size of the system is increased (not shown here). Our data confirm the theoretically predicted `time-shear superposition principle' \cite{Berthier2003,Fuchs2002} : when time is scaled by $t_{1/e}$ the relaxation follows a master curve $f_{s}(a,t/t_{1/e})$ as shown in figure \ref{fig:relax1overe}. It is tempting to try and relate as in \cite{Varnik2006a,Mobius2009,Besseling2007} the scaling exponent $\nu$ of the structural relaxation time to the scaling exponent $\beta$ that appears in the Herschel-Bulkley type macroscopic rheological flow curve of the material where $\sigma-\sigma_{Y}\propto\dot\gamma^{\beta}$ (see figure \ref{fig:StressStrainrate}). Taking, as is often assumed and verified numerically \cite{VarnikHabilit}, the structural relaxation time $t_{1/e}$ as proportional to viscosity provides an expression of an effective stress $\sigma_{eff}=\mu t_{1/e}\dot\gamma$, where $\mu$ is the macroscopic shear modulus. Surprisingly reporting the scaling of the relaxation time $t_{1/e}$ in this expression we see that for high enough shear rates $\dot\gamma\gtrsim10^{-4}$ the effective stress scales with shear rate as $\sigma_{eff}\propto\dot\gamma^{1-\nu}$. This is in good agreement with the global mechanical response of the material and one finds a posteriori that the two coefficients $\beta$ and $\nu$ are compatible with the hypothesis made above and one has indeed to a good approximation $\beta=1-\nu$ for high shear rates. This relation breaks down for lower shear rates in a regime where shear banding becomes the dominant relaxation mechanism, as can be seen for example in the panels \ref{fig:Frames1e-5} and \ref{fig:Frames1e-3} corresponding respectively to the shear rates $\dot\gamma=10^{-5}$ and $\dot\gamma=10^{-3}$.

\subsection{Mean square displacement (MSD).} 
\begin{figure}[!hbtp]
\begin{center}
\includegraphics[width=7cm]{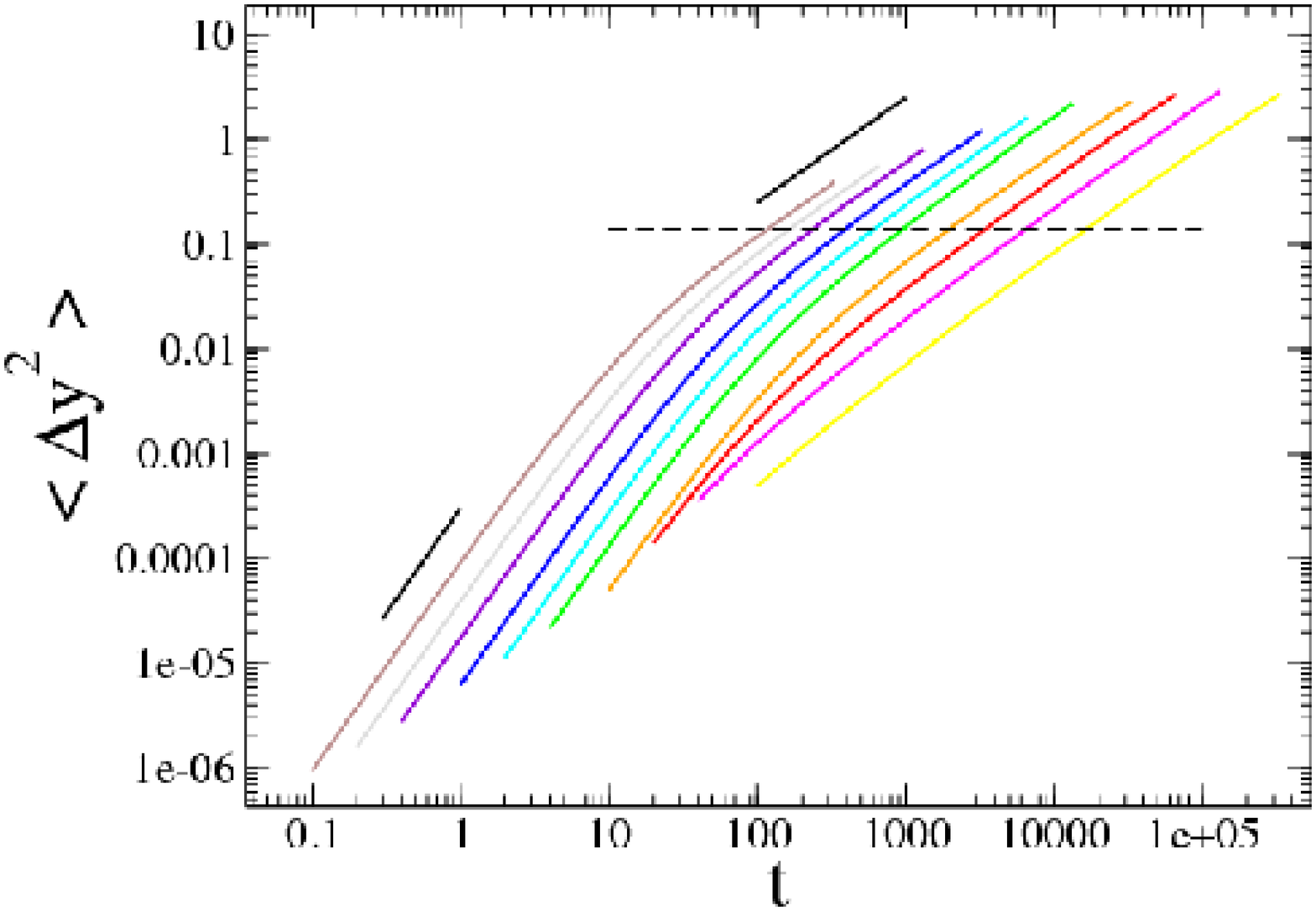}
\includegraphics[width=7cm]{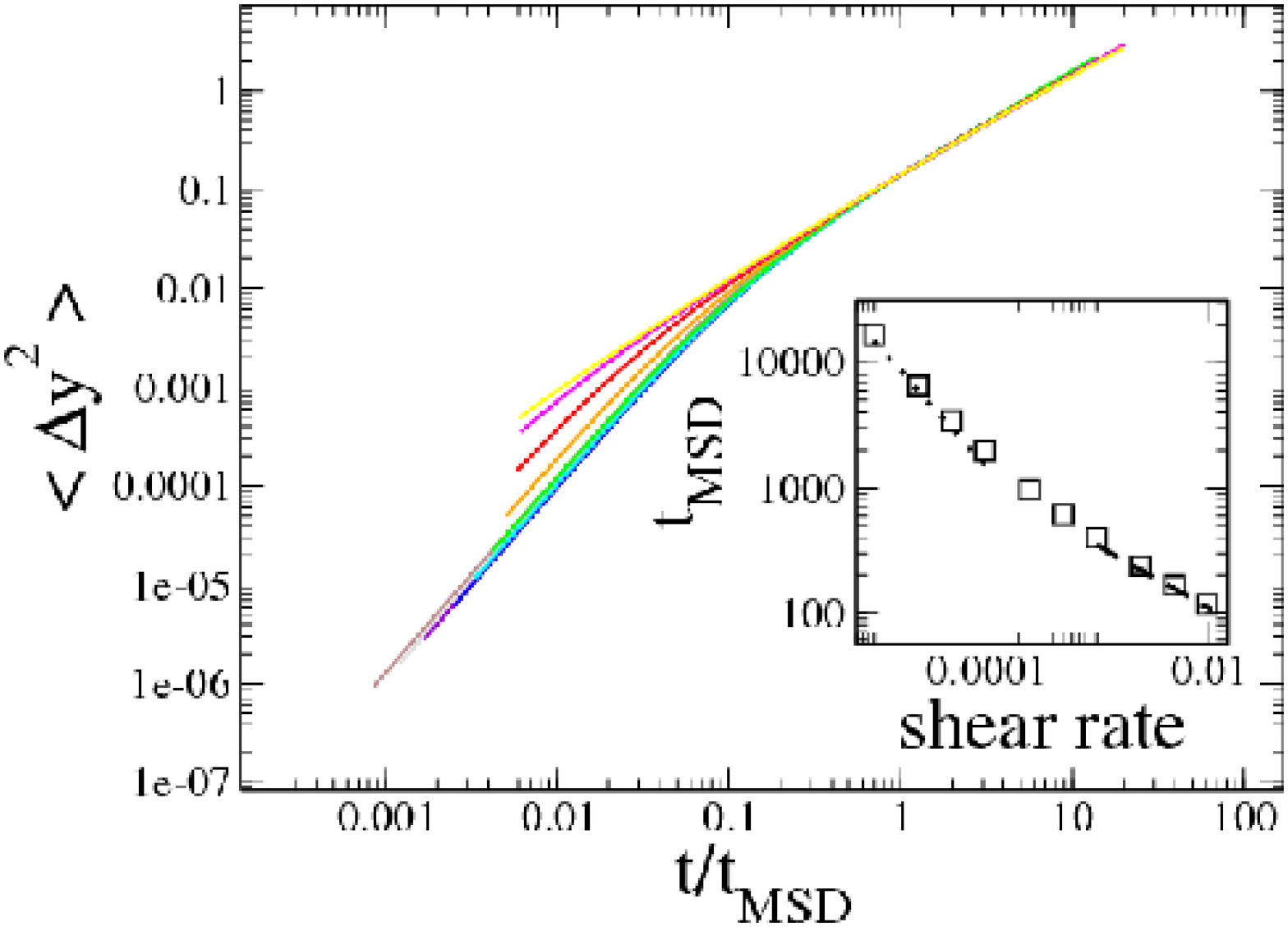}
\caption{\textbf{Left} : Transverse mean square displacement (MSD) $\langle\Delta y^{2}\rangle$ versus time for different shear rates $\dot\gamma$ (same color code as in previous figures). The dashed line marks the arbitrarily chosen criterion distance a verifying $a^{2}=0.14$ corresponding to a `Lindemann' criterion for  the transverse MSD of the particles. The intersection of this line with the colored curves marks the times $t_{MSD}$. The two thick black lines correspond to the power laws $t^{2}$ and $t^{1}$, i.e. respectively to the ballistic and diffusive regime.\textbf{Right}: same figure where time is rescaled by the times $t_{MSD}$. These plots are obtained in configurations with RWBCs and the average are computed in a central region of the samples therefore avoiding direct influence of the walls.}\label{fig:MSDy}
\end{center}
\end{figure}

Usually to quantify the dynamics at a particle level one also calculates the MSD. In our two dimensional simulations the diffusion along the x and y directions are not equivalent. Indeed while the diffusion in the shear direction (in our case the x axis) is enhanced by the shear, the diffusion in the transverse shear-gradient direction ( y axis) is unaffected. Here we therefore present the MSD $\langle\Delta y(t)^{2}\rangle$ in the transverse direction. Of course in the presence of rigid boundary conditions the diffusion along the y axis is limited by the presence of walls and one must be cautious to compute the average MSD sufficiently far away from the boundaries. Figure \ref{fig:MSDy} shows the typical transverse MSD for a system containing 625 particles under RWBCs and averaged over a total cumulative strain of $200\%$ for each of 24 glass samples. Moreover in order to avoid boundary effects the average is computed over one third of the sample in the central region. Larger samples as well as LEBCs yield similar results and we have not reproduced these here for clarity. From figures \ref{fig:MSDy} we see that MSD exhibits a transition from ballistic motion at short times ($\langle\Delta y^{2}\rangle\propto t^{2}$) to diffusive motion ($\langle\Delta y^{2}\rangle\propto t^{1}$) for larger times. For high shear rate values ($\dot\gamma\gtrsim10^{-4}$ one can rescale the entire MSD curves on a master curve $g(t/t_{MSD})$, while for smaller shear rates the scaling doesn't hold for small times. The times $t_{MSD}$ are defined here as the intersection of the MSD curves with a `Lindemann' like criterion defined at $\langle\Delta y^{2}\rangle=0.14$ as in \cite{Mobius2009}. The time $t_{MSD}$ is a nonlinear function of shear rate and follows the same trend as $t_{1/e}$ defined earlier (see figure \ref{fig:Qs}), but with a slightly different exponent $t_{MSD}\propto\dot\gamma^{-\nu_{2}}$, with $\nu_{2}\sim0.5$ (see doted line in the inset of figure \ref{fig:MSDy}).


\subsection{Diffusion coefficient.} 
\begin{figure}[!hbtp]
\begin{center}
\includegraphics[width=7cm]{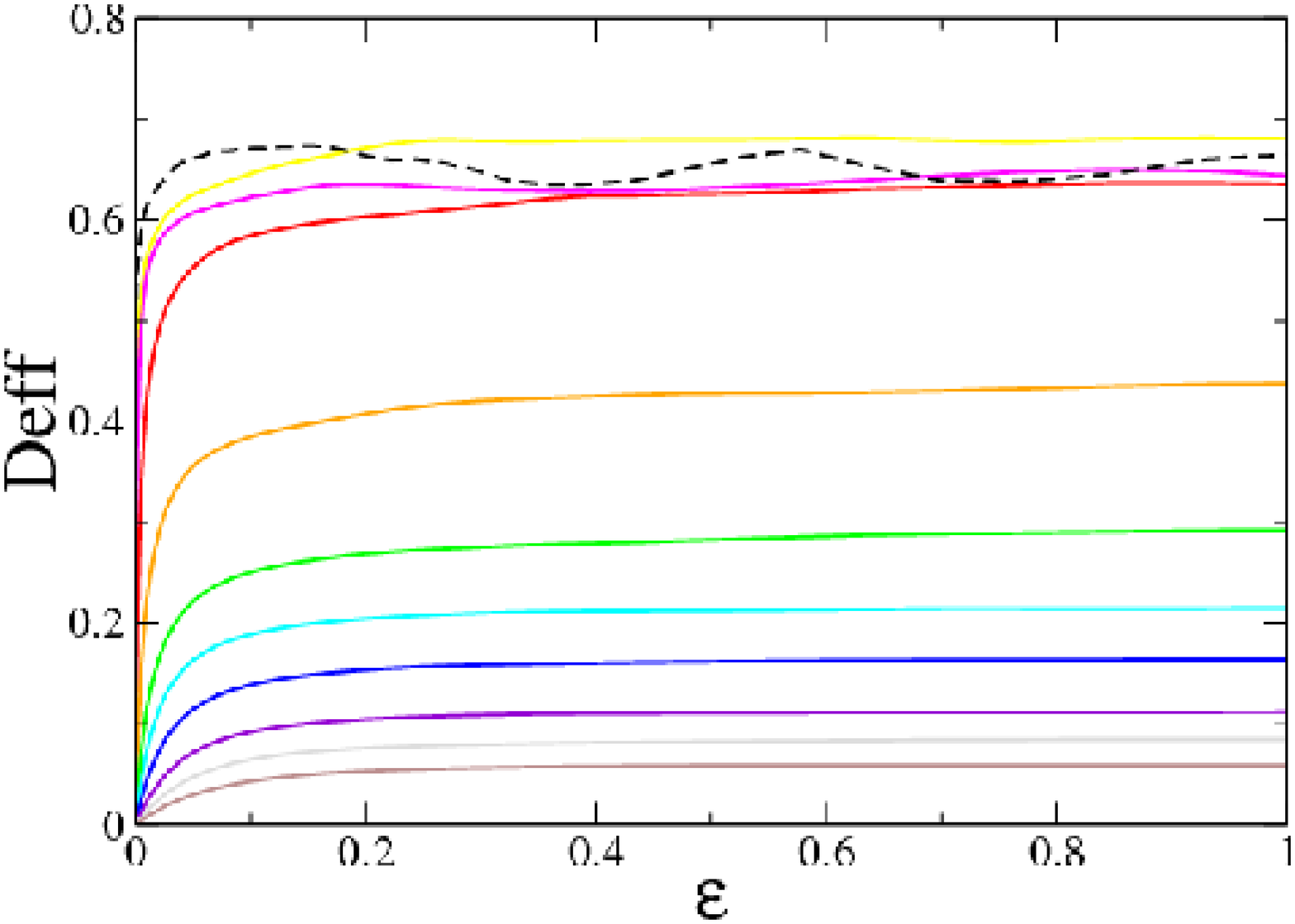}
\includegraphics[width=7cm]{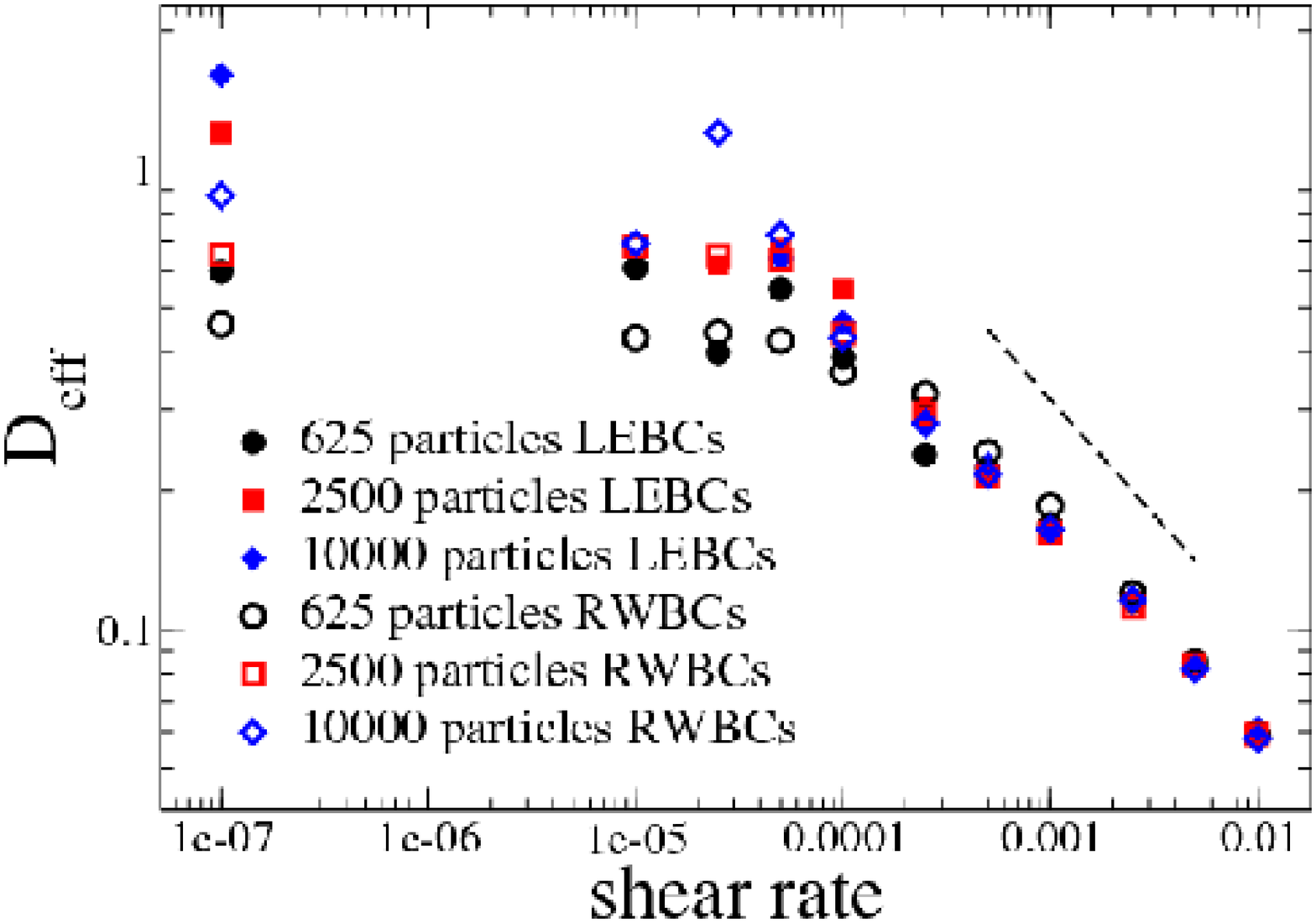}
\caption{\textbf{Left} : Effective diffusion coefficients defined as $D_{eff}(\Delta\gamma)=\langle\Delta y^{2}\rangle/2\Delta\gamma$ for different shear rates ranging from bottom to top from $\dot\gamma=10^{-5}$ to $10^{-5}$ (same color code as previous figures) and for the quasistatic shear protocol (black dashed line). $D_{eff}$ is computed here for sample containing 2500 particles under RWBCs. The spatial averaging is performed sufficiently far from the rigid walls to avoid boundary effects. \textbf{Right}: The asymptotic values $D_{eff}$ are plotted for various shear rates, system sizes and boundary conditions. The dashed line marks the power law $D_{eff}\propto\dot\gamma^{-0.5}$ as a guide to the eyes.}\label{fig:Deff} 
\end{center}
\end{figure}

The scaling of the MSD curves at different shear rates allows to identify the dependence of the transverse diffusion coefficient, defined by $\langle\Delta y^{2}\rangle=2 D_{y} t$, with shear rate $\dot\gamma$, $D_{y}\propto1/t_{MSD}\propto\dot\gamma^{\nu_{2}}$. In order to allow comparison with diffusion in quasistatically sheared glasses we follow Lema\^itre \cite{Lemaitre2007} and compute the quantity $D_{eff}(\Delta\gamma)=\langle\Delta y^{2}\rangle/2\Delta\gamma$ which is related to the usual transverse diffusion coefficient through $D_{eff}=D_{y}\dot\gamma$. In figure \ref{fig:Deff} we plot the effective diffusion coefficient $D_{eff}(\Delta\gamma)$ for the various system sizes and for various finite shear rates as well as under a quasistatic protocol. For all system sizes and shear rates we see that $D_{eff}(\Delta\gamma)$ increases from a finite initial value (that increases with decreasing shear rate) to an asymptotic value for large strain. The transient strain interval appears not strongly shear rate or size dependant and is of the order of $\epsilon_{transient}=0.25$. The dependence of the asymptotic values of $D_{eff}$ (we will call this asymptote D from now on) on shear rate is shown in figure \ref{fig:Deff} and displays for high shear rates ($\dot\gamma\gtrsim10^{-4}$), as expected from the relation $D_{eff}=D_{y}\dot\gamma$, the scaling $D\propto\dot\gamma^{\nu_{2}}$ ($\nu_{2}\sim 0.5$). 



Lema\^itre \cite{Lemaitre2009} finds the following two limiting scaling behaviors of the effective diffusion coefficient $D_{eff}$. First in the high shear rate limit one gets uncorrelated localized plastic events and 

\begin{equation}
\label{eqn:Deff1}
D_{eff}\propto\ln(L)
\end{equation}

whilst at low shear rates Lema\^itre predicts a linear scaling

\begin{equation}
\label{eqn:Deff1bis}
D_{eff}\propto L.
\end{equation}

This linear scaling was also obtained numerically in \cite{Maloney2008}. In this article the linear scaling of $D_{eff}$ is recovered by the authors with a simple argument if one assumes that the mechanical response of the material is dominated by uncorrelated system spanning slip lines. One then simply has $D_{eff}=\langle\Delta r^{2}/\Delta \gamma^{2}\rangle \approx \left(\Delta\gamma/(a/L)\right)a^{2}/12$, where a is the slip amplitude, $\Delta\gamma$ is the applied strain increment and $a^{2}/12$ is the average mean square displacement associated to an individual slip line and accumulated during a strain of $\sim a/L$. Note that a is assumed to be size independent.

The difference between Maloney's and Lema\^itre's approach is in the fact that while the later assumes the elementary constituents of the response to be the avalanches observed in deformed glassy materials the former believes that one must take into account the entire system spanning slip lines that are formed by a cluster of avalanches as elementary constituents of the rheology.

Due to the large fluctuations of the effective diffusion our results (figure \ref{fig:Deff}) do not allow us to resolve clearly these questions.  Therefore while we can not rule out the three main observations made by Lema\^itre we can not either confirm them. First pertaining to the system size dependence of D at low shear rates we indeed observe that D grows with system size but our number box sizes and the uncertainty for each measure of D stops us from descriminating between a $D_{eff}\propto\ln(L)$ or a $D_{eff}\propto L$ scaling. Second, at higher shear rates, with a simple argument based on the long range elastic propagator of the local quadrupoles Lema\^itre predicts $D_{eff}\propto\ln(L)$. This seems in contrast with our findings where $D_{eff}$ at high shear rates seems system size independent (see figure \ref{fig:Deff} and compare with figure 5 in \cite{Lemaitre2009}). Finally due to the very long simulation time required to produce runs for shear rates below $\dot\gamma\simeq10^{-5}$\footnote{A run at $\dot\gamma=10^{-5}$ for a system of size $L=100$ takes of the order of a few days for $400\%$ strain.} we can not extract clearly the system dependence of the critical shear rate $\dot\gamma_{c}$ separating the system size dominated regime from the shear rate dominated regime. In line with previous numerical studies (for example \cite{Lemaitre2009,Ono2003,Varnik2006a}) this change of behavior for the three system sizes presented here is located in the range $10^{-5}\lesssim\dot\gamma\lesssim10^{-4}$. While these results convincingly illustrated the influence of shear rate on the atomic motion in a sheared model glass they also call for extended simulation runs. In the next section we focus on what is thought to be associated with this change of behavior namely the existence of a growing dynamical heterogeneity length scale as the shear is reduced. 

\section{Dynamical heterogeneity}
\label{DynamicalHeterogeneity}

\subsection{General study}

As discussed in the introduction the dynamical heterogeneity is quantified in supercooled liquids near the glass transition and more recently in jammed systems near the jamming point by the so-called four-point correlation function. Here we propose to extend these approaches to the case of sheared glasses where instead of $T$, the temperature, in the case of supercooled liquids or $\phi$, the volume fraction, in granular materials we consider here $\dot\gamma$ as the new control parameter. The analytical framework allowing to quantify the dynamical heterogeneity remains nevertheless identical and we therefore use these tools in our present analysis, in particular our analysis parallels the experimental study by Lechenault \textit{et al} in \cite{Lechenault2008a} on the critical scalings and heterogeneous dynamics near the jamming/rigidity transition of a granular material. 

\begin{figure}[!hbtp]
\begin{center}
\includegraphics[width=7cm]{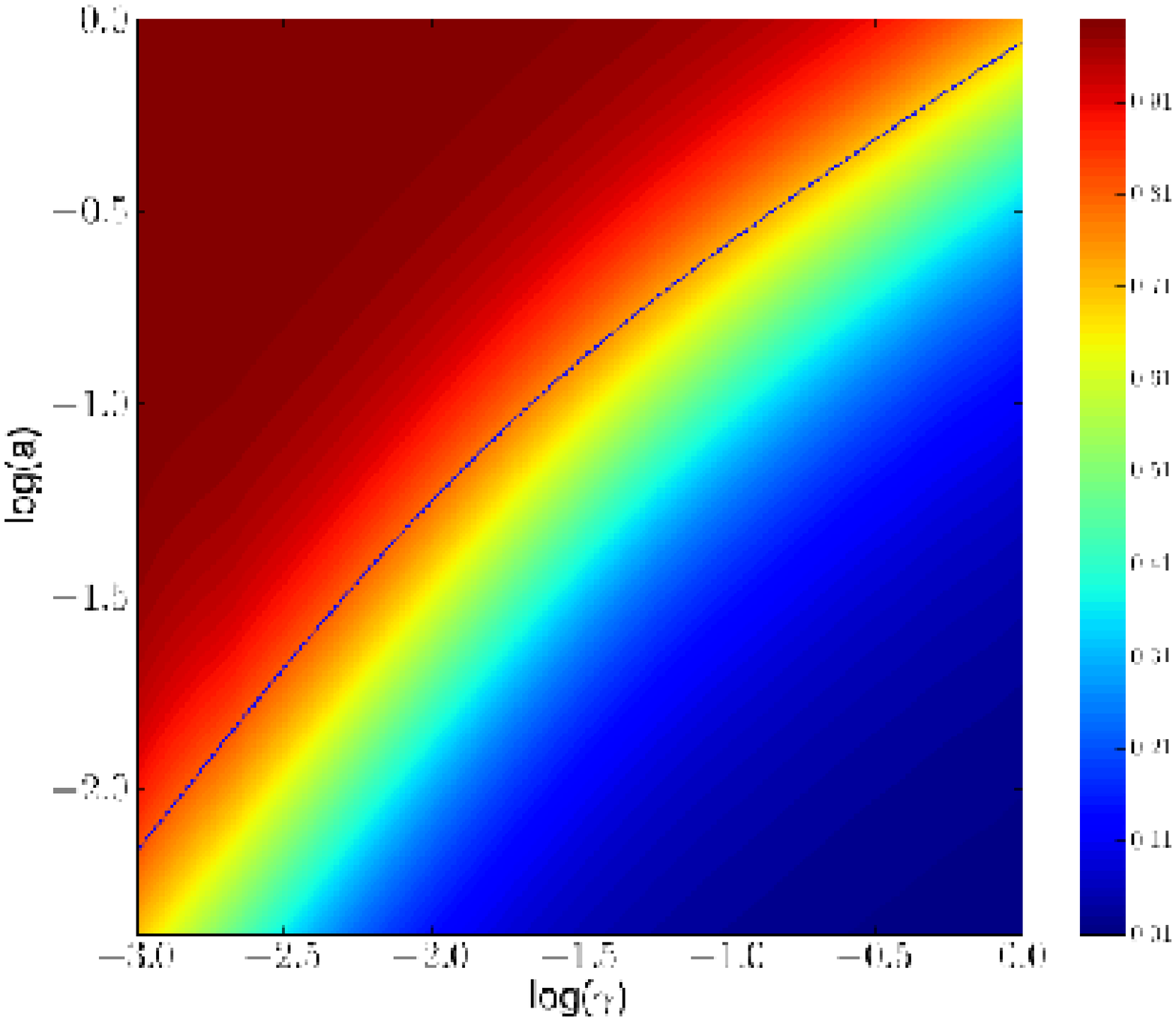}
\includegraphics[width=7cm]{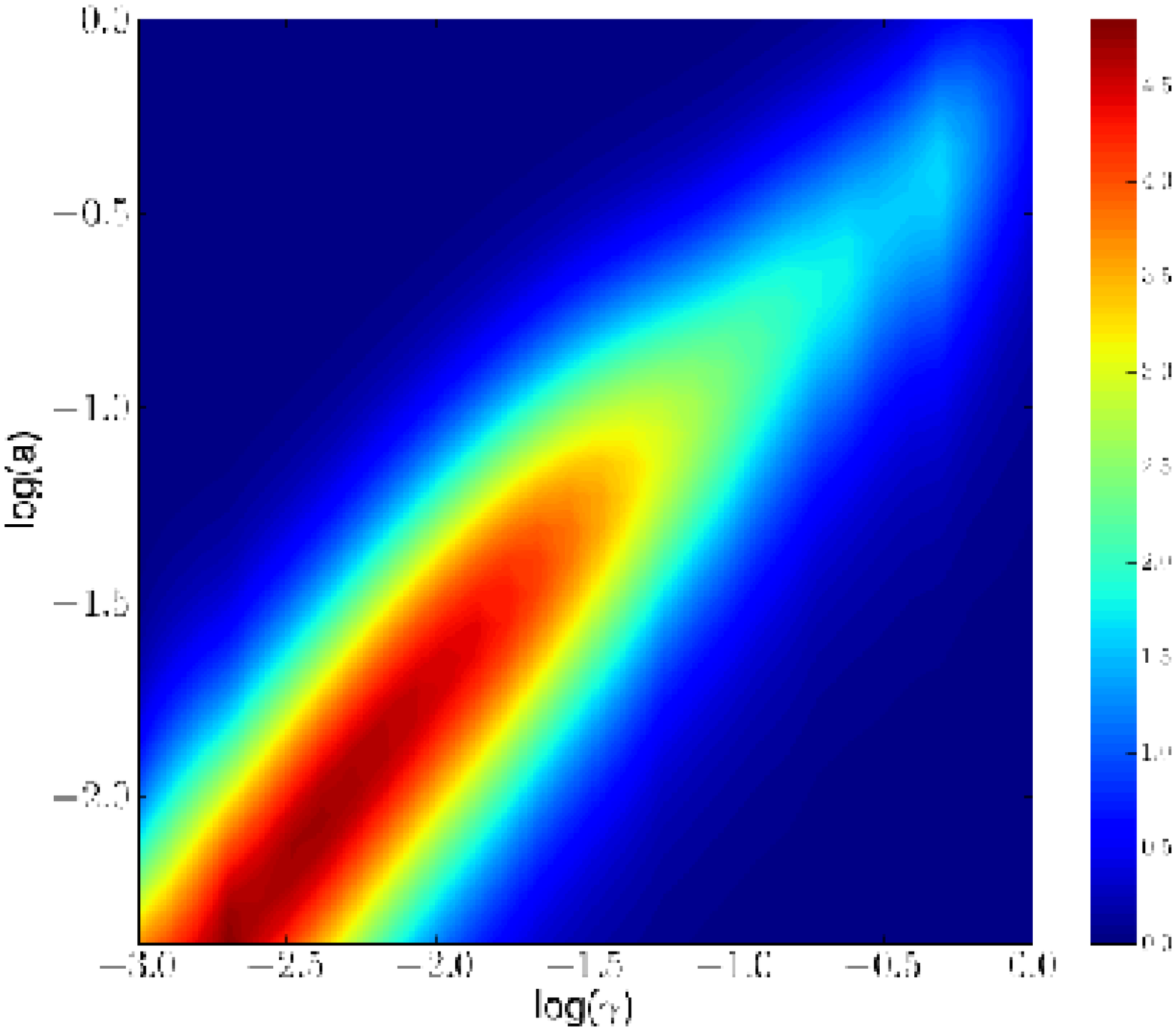}
\caption{Dynamical correlation functions computed over the particles of sample containing 2500 particles and sheared at $\dot\gamma=10^{-4}$ under RWBCs for a total strain $\epsilon_{tot}=200\%$.\textbf{Left} : Correlation function $Q_{s}(a,\gamma)$ as a function of the probing length a and the strain $\gamma$ in a log-log colormap. \textbf{Right}: Four-point correlation function $\chi_{4}(a,\gamma)$ in a log-log colormap.}\label{fig:Qsvsavsg}
\end{center}
\end{figure}

The dynamical cooperativity is quantified as the fluctuations of a two-point correlation function. Here we choose as a two point correlation function the transverse part self correlation function (as in equation \ref{eqn:Qs}) that we express here rather than in function of time in terms of the strain interval $\epsilon$ as,

\begin{equation}
\label{eqn:Qsepsilon} 
Q_{s}(a,\epsilon)=\frac{1}{N}\sum_{i}\exp(-\frac{\Delta y_{i}(\epsilon)^{2}}{2a^{2}})\mbox{  ,}
\end{equation}
where a is a characteristic length scale over which the dynamics is probed. In figure \ref{fig:Qsvsavsg} we represent the dependence of the spatial (index i for each particle) and temporal/strain (index $\epsilon$) average $Q(a,\epsilon)=\langle Q_{s}(a,\epsilon)\rangle_{i,\epsilon}$ as a function of both the parameter a and the strain interval $\epsilon$, or rather $\gamma=\epsilon/2$. The function is here plotted for a shear rate of $\dot\gamma=10^{-4}$ and for a sample containing 2500 particles under RWBCs. $Q(a,\epsilon)$ takes values in the range [0,1], with $Q\sim1$ typically when the transverse motion is small relatively to a, $\Delta y\ll a$, and $Q\sim0$ in the opposite situation when $\Delta y \gg a$. Following \cite{Lechenault2008a}, we superimpose in figure \ref{fig:Qsvsavsg} on the colormap of Q the root of the transverse MSD, $\delta(\gamma,\dot\gamma)=\sqrt{\Delta y^{2}}(\gamma)$. Interestingly, as for granular materials around the jamming volume fraction we see that the MSD follows very well the decay of $Q(a,\epsilon)$ and that here also the function $Q(a,\epsilon)$ can be rescaled for all shear rates as $Q(a,\epsilon)=Q'(\delta(\epsilon,\dot\gamma)/a)$ showing that the MSD defines the only microscopical relevant distance for a given strain and shear rate. 


Turning now to the fluctuations of the self correlation function $Q_{s}(a,\epsilon)$ we
define the four-point correlation function $\chi_{4}(a,\epsilon)$ as,

\begin{equation}
\label{eqn:chi4epsilon}
\chi_{4}(a,\epsilon)=N\left[\langle Q_{s}(a,\epsilon)^{2}\rangle_{i,\epsilon}-\langle Q_{s}(a,\epsilon)\rangle_{i,\epsilon}^{2}\right]\mbox{  .}
\end{equation}

Again we remind the reader that $\chi_{4}(a,\epsilon)$ gives an estimate of the number of particles that move cooperatively when the sample is subjected to global strain $\epsilon$. We produce an example of this function for a shear rate of $\dot\gamma=10^{-4}$ in figure \ref{fig:Qsvsavsg} where we see that at this shear rate the maximum cooperativity is of the order of 5 particles. Again in line with \cite{Lechenault2008a} we obtain the same scaling with the mean square displacement of the four-point correlation function $\chi_{4}(a,\epsilon)$ that can be rescaled as $\chi_{4}(a,\epsilon)=h(\dot\gamma,\epsilon)\chi_{4}'(\delta(\epsilon,\dot\gamma)/a)$ where the amplitude $h(\dot\gamma,\epsilon)$ depends both on shear rate and strain interval. From this figure we can not determine if there is a point (log(a),log($\epsilon$)) with finite values in this map corresponding to an absolute maximum of the function  $\chi_{4}(a,\epsilon)$ as is observed by Lechenault \textit{et al.} \cite{Lechenault2008a} or if the maximum is pushed at non-finite values.
%

\begin{figure}[!hbtp]
\begin{center}
\includegraphics[width=15cm]{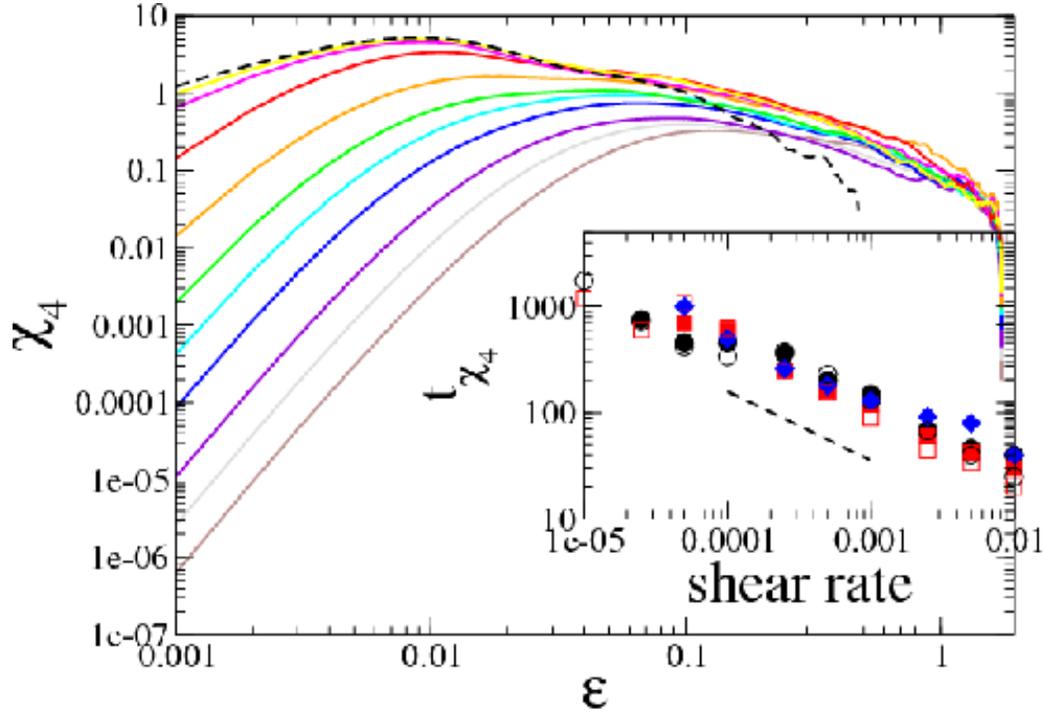}
\caption{Four-point correlation function as a function of strain $\epsilon$ for different shear rates (same color code as previous figures) computed on samples containing 2500 particles and sheared under RWBCs. The averaging is done over 15 samples over a total strain of $\epsilon=200\%$ on each configuration. \textbf{Insert :} Shear rate dependence of the time $t_{\chi_{4}}^{max}$ associated to the peak of $\chi_{4}$. The different symbols correspond to different system sizes and boundary conditions, see figure \ref{fig:MSDy} for the legend. The dashed line corresponds to the power law $\dot\gamma^{-0.5}$ and is shown as a guide to the eyes.}\label{fig:khi4s}
\end{center}
\end{figure}

We now turn to the influence of the shear rate on the dynamical cooperativity in the driven glasses.
Figure \ref{fig:khi4s} shows for different shear rates the build up of cooperativity in a glass sample containing 2500 particles. The curves start from a low value of $\chi_{4}(a,\epsilon)$ at low strain go through a maximum $\chi_{4}^{max}(\dot\gamma)$ at a corresponding strain $\epsilon_{\chi_{4}}^{max}(\dot\gamma)$ ($t_{\chi_{4}}^{max}=\epsilon_{\chi_{4}}^{max}/\dot\gamma$) and decay to zero for larger strains. Here the log-log representations allows to see that the growth with strain is of the form $\chi_{4}(\epsilon)\propto\epsilon^{4}$ for high shear rate values while towards the quasistatic limit the behavior changes toward a $\chi_{4}(\epsilon)\propto\epsilon^{1}$ growth. Note that the $\epsilon^{4}$ behavior is consistent with a ballistic regime while $\epsilon^{1}$ is consistent with a regime dominated by collectively rearranging regions (see \cite{Toninelli2005}). 
The curves of figure \ref{fig:khi4s} allow us to extract the dependence of the time scale $t_{chi_{4}}^{max}$ with shear rate $\dot\gamma$ and we display this in the inset of figure \ref{fig:khi4s}. We find a scaling of the form $t_{chi_{4}}^{max}\propto\dot\gamma^{-\nu_{3}}$ with $\nu_{3}\sim0.65$, hence very close to the coefficient $\nu_{1}\sim0.63$ observed for the relaxation time $t_{1/e}$ associated to the correlation functions $Q_{s}$. This time dependence differs on the other hand slightly from the time $t_{MSD}$ extracted from the Lindemann criterion on the transverse mean square displacement $\nu_{2}\sim0.5$ (see figure \ref{fig:MSDy}). 

%

\begin{figure}[!hbtp]
\begin{center}
\includegraphics[width=15cm]{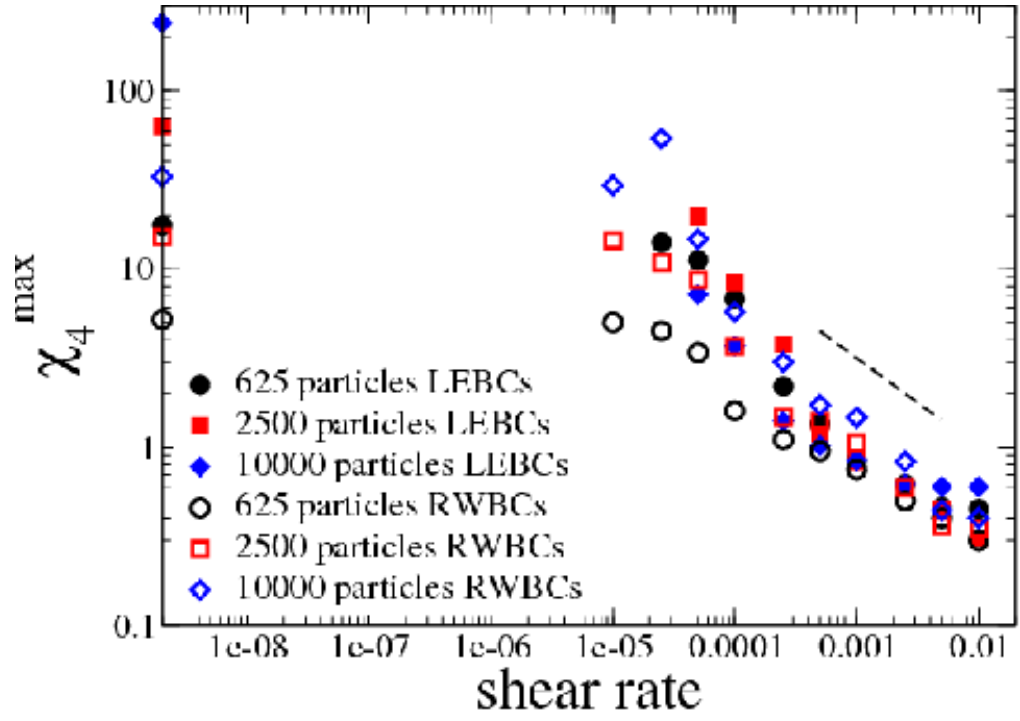}
\caption{maximum values $\chi_{4}^{max}$ of the four-point correlation function as a function of shear rate $\dot\gamma$ for various system sizes and boundary conditions. The dashed line shows the power law $\propto\dot\gamma^{-\mu}$, with $\mu=0.6$.}\label{fig:maxkhi4s}
\end{center}
\end{figure}

Again it is satisfactory to observe, in figure \ref{fig:khi4s}, the convergence of the $\chi_{4}(a,\epsilon)$ associated to the finite shear rate deformation runs towards the quasistatic data as the two sets of simulations are produced from completely independent codes and procedures. For the system size analyzed in this figure (L=50) we see that the number of particles moving in a cooperative manner is of the order of $\chi_{4}^{max}\sim18$ in the quasistatic regime. In figure \ref{fig:maxkhi4s} we have collected all the values of $\chi_{4}^{max}$ for the various system sizes, boundary conditions and shear rates that we have analyzed. The result shows strikingly the growth of cooperative length scale with decreasing shear rate for all system sizes. This plot illustrates again (as was the case for the flow curve or for the diffusion properties) two regimes, namely a high shear rate regime (above a critical shear rate $\dot\gamma\gtrsim\dot\gamma_{c}$) where $\chi_{4}^{max}$ grows with decreasing $\dot\gamma$ and a plateau regime where $\chi_{4}^{max}$ saturates to a system size limited value. While the data are still quite scattered they allow to extract a typical scaling coefficient $\mu$ for the dependence of $\chi_{4}^{max}$ on $\dot\gamma$ as $\chi_{4}^{max}\propto\dot\gamma^{-\mu}$, with $\mu\sim0.4-0.6$. The dependence of $\chi_{4}^{max}$ on system size in the quasistatic regime is too noisy to be quantified precisely at the present stage of our study. It is obvious nevertheless that the dynamical cooperativity $\chi_{4}^{max}$ grows in the quasistatic regime with system size in a way that indicates finite size effects. We will come back to the scaling exponent $\mu$ in the last section of this paper and relate our findings with previous observations. But let us try first to visualize the spatial structures associated to this build up of a growing dynamical heterogeneity length scale as $\dot\gamma\rightarrow0$.


\subsection{Spatio-temporal structures and a simple model for the cooperativity length} 
\begin{figure}[!hbtp]
\begin{center}
\includegraphics[width=7cm]{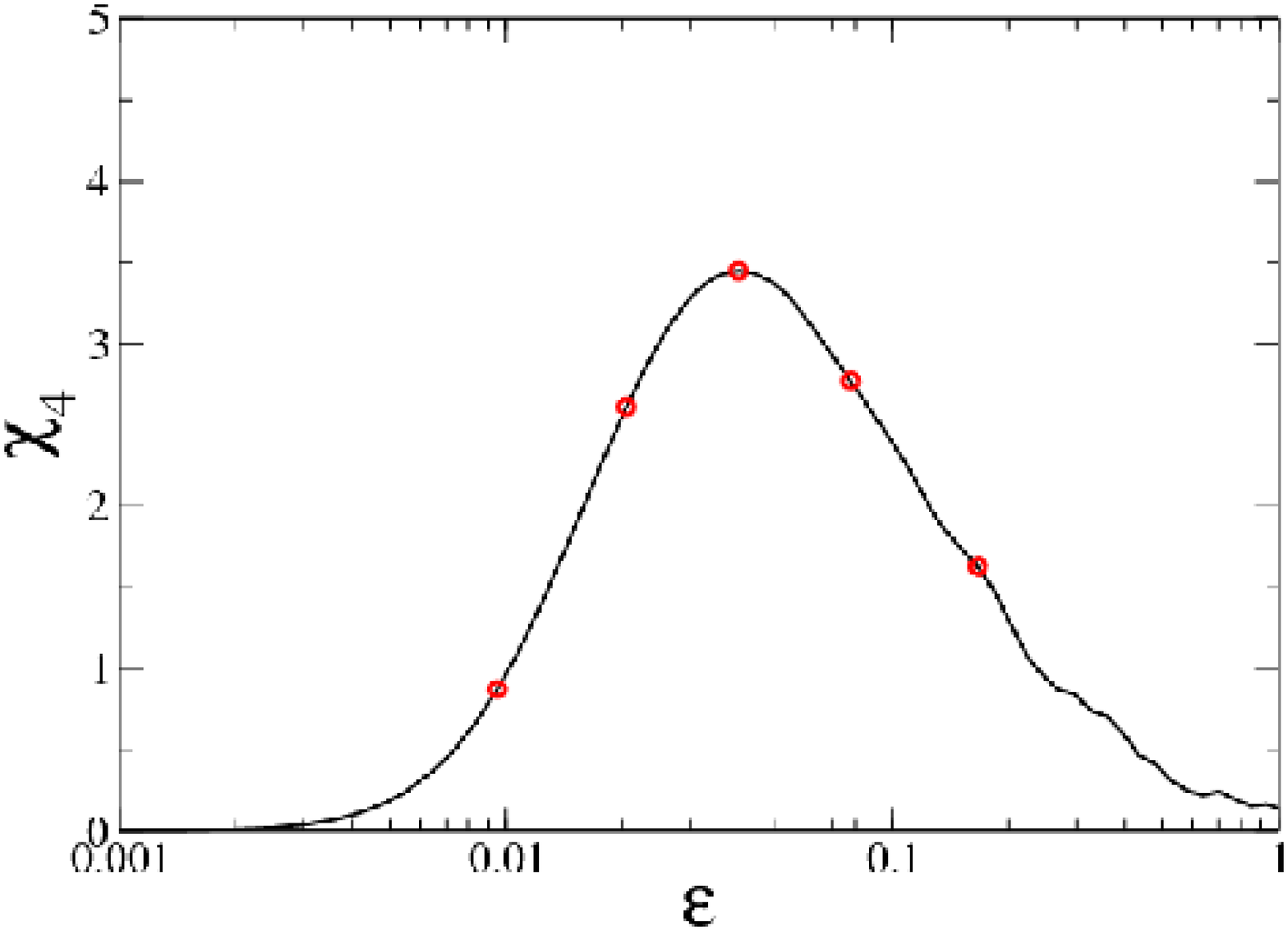}
\includegraphics[width=7cm]{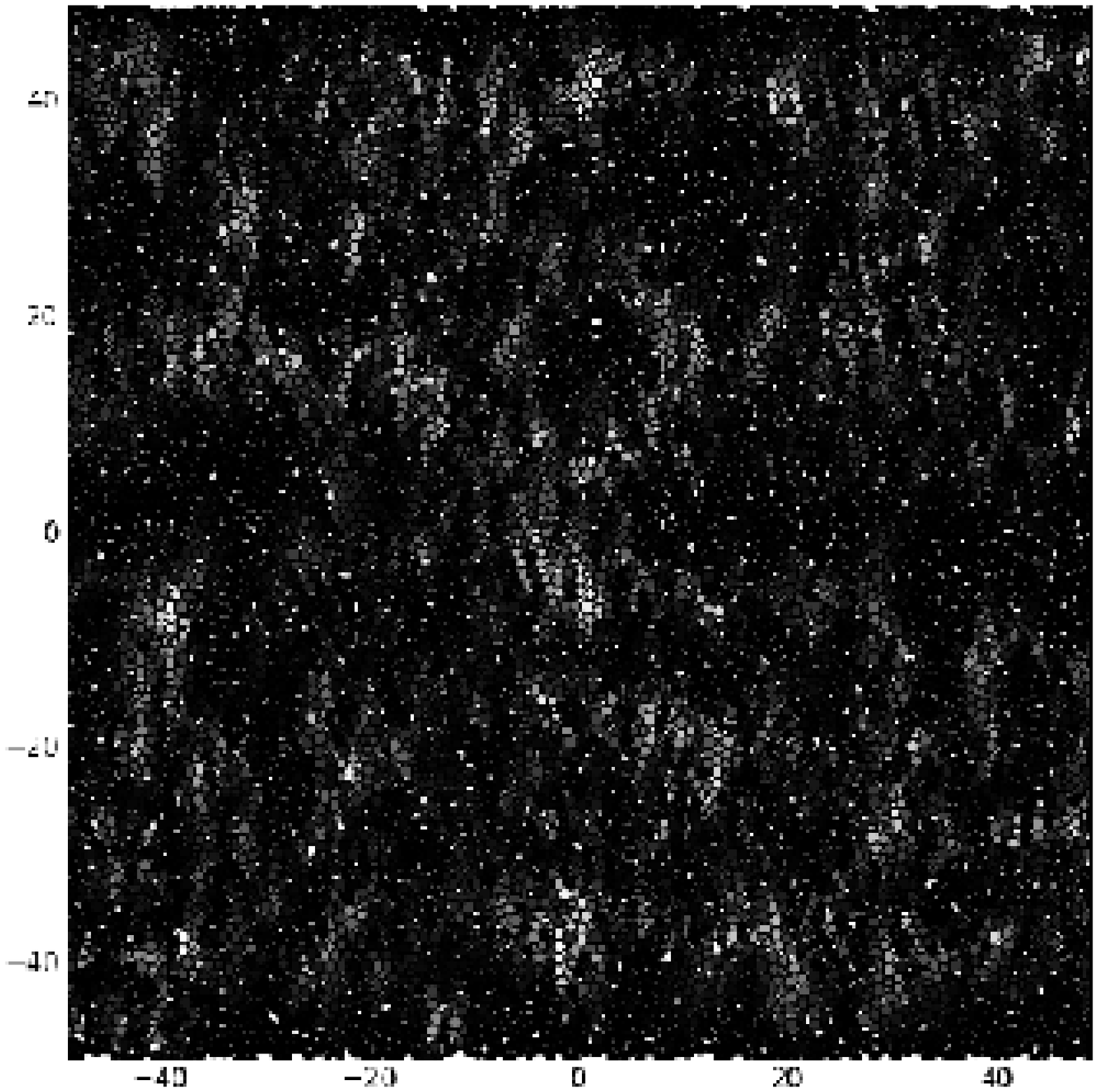}
\includegraphics[width=7cm]{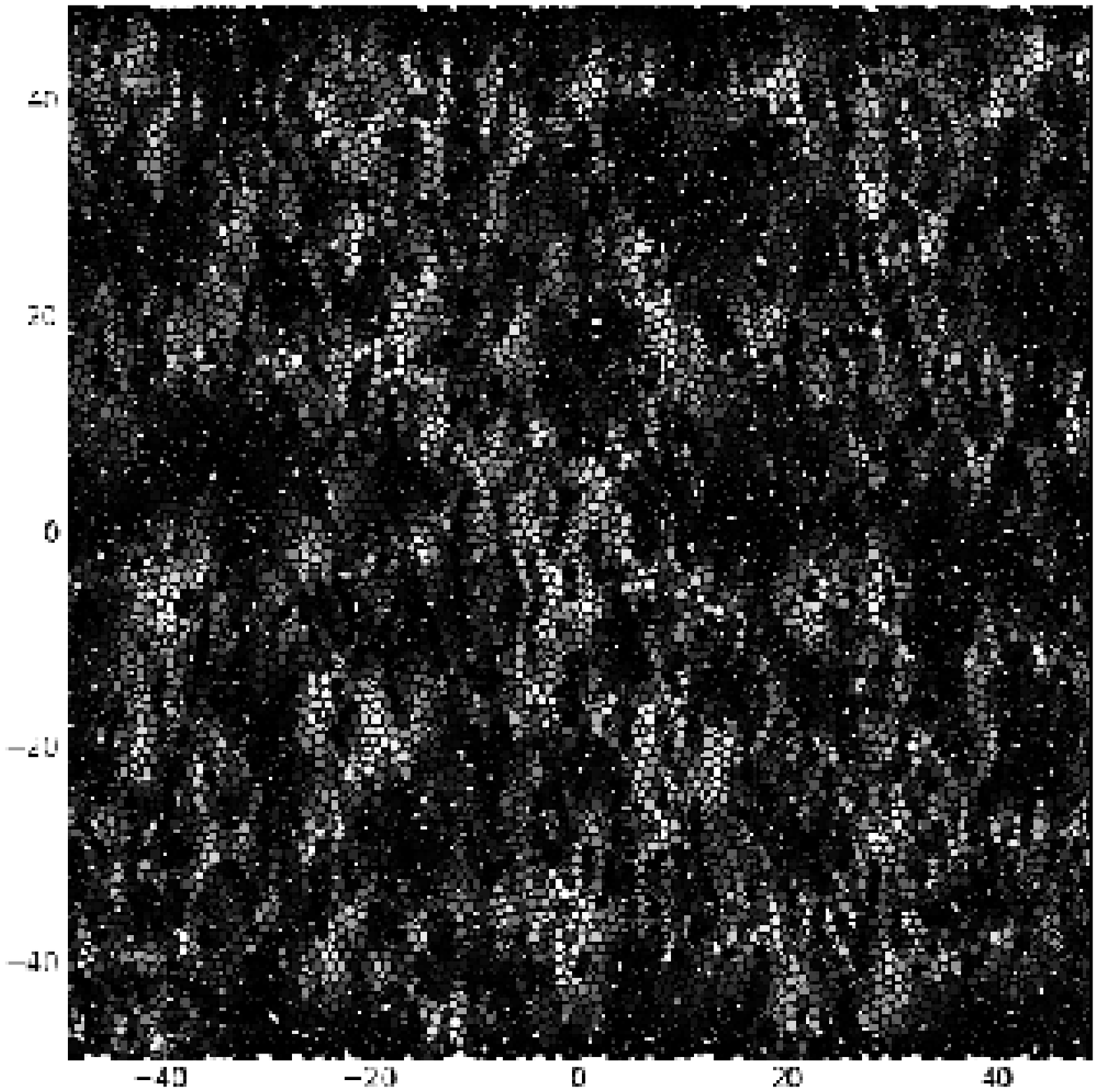}
\includegraphics[width=7cm]{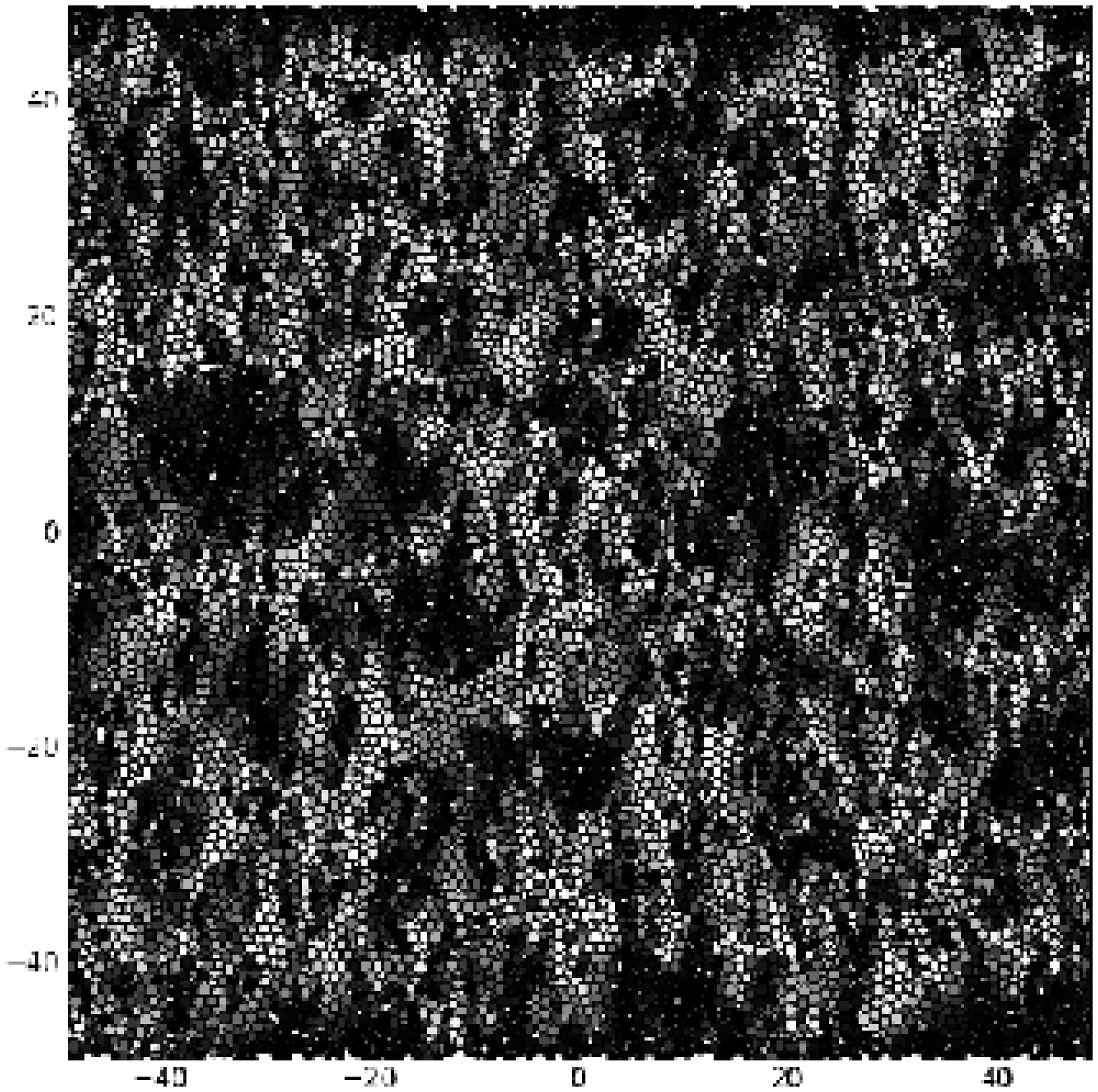}
\includegraphics[width=7cm]{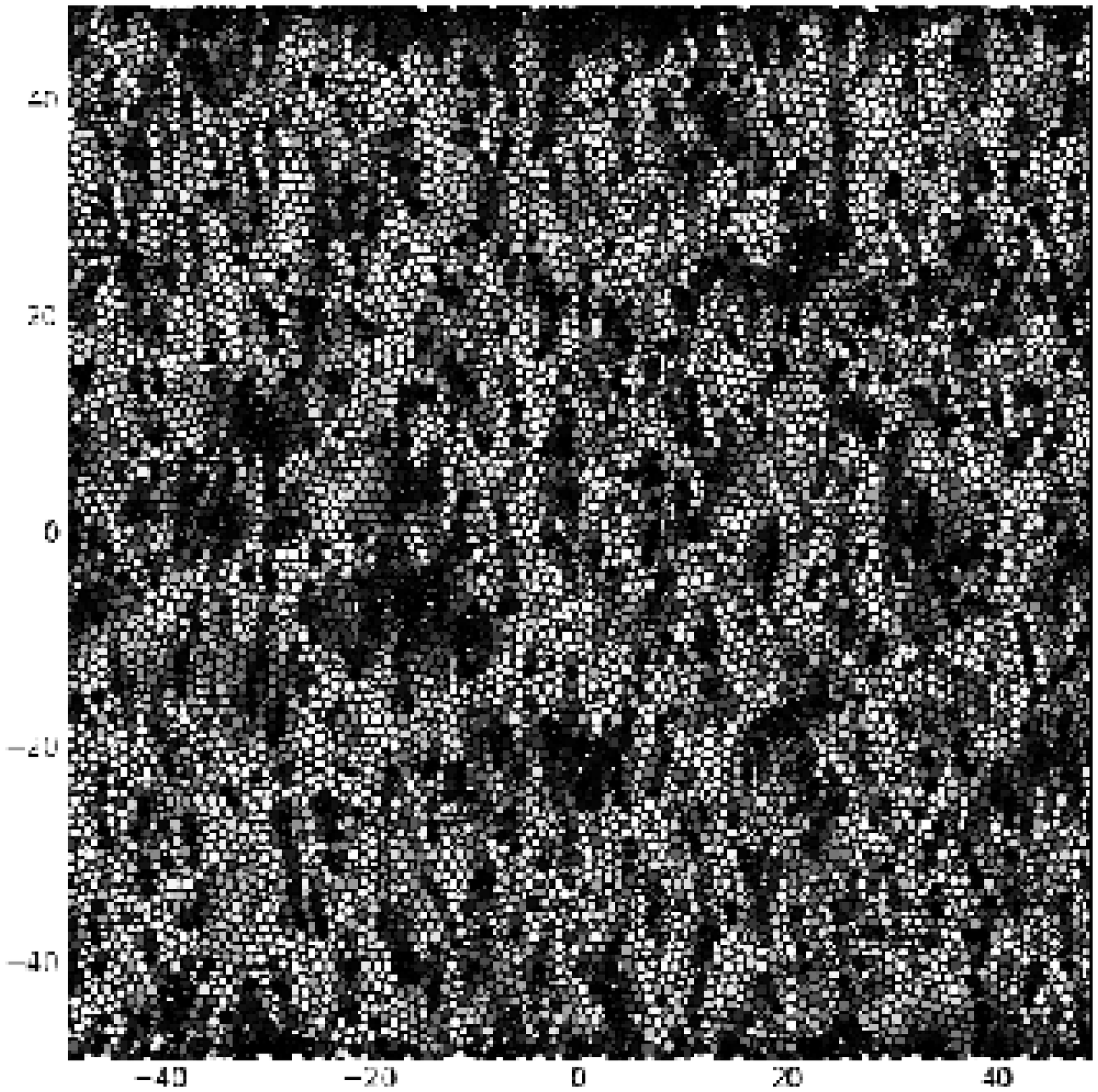}
\includegraphics[width=7cm]{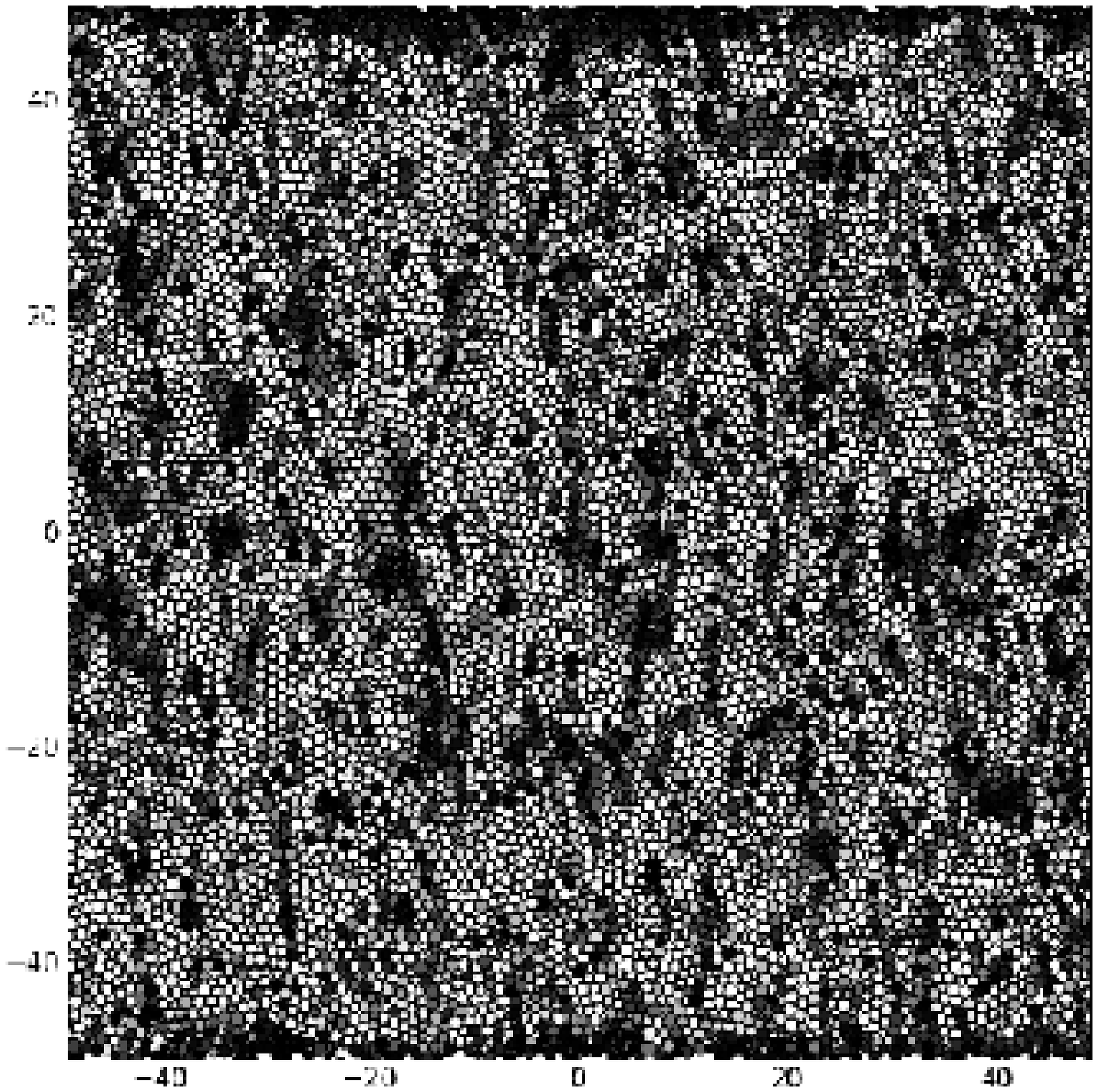}
\caption{\textbf{Top left corner :} $\chi_{4}(\epsilon)$. The red marks correspond to the strain intervals at which the five spatial maps of the self correlation function $Q_{s}^{i}(\epsilon)=\exp{\left(\frac{\Delta y_{i}(\epsilon)^{2}}{2a^{2}}\right)}$ are computed. From top to bottom and from left to right, $\epsilon=10^{-2}$,$\epsilon=2\cdot10^{-2}$,$\epsilon=4\cdot10^{-2}$,$\epsilon=8\cdot10^{-2}$,$\epsilon=16\cdot10^{-2}$. All figures are obtained on a sample containing 10000 particles under RWBCs and at a shear rate of $\dot\gamma=5\cdot10^{-4}$.}\label{fig:QyMapsvsStrain}
\end{center}
\end{figure}

\begin{figure}[!hbtp]
\begin{center}
\includegraphics[width=7cm]{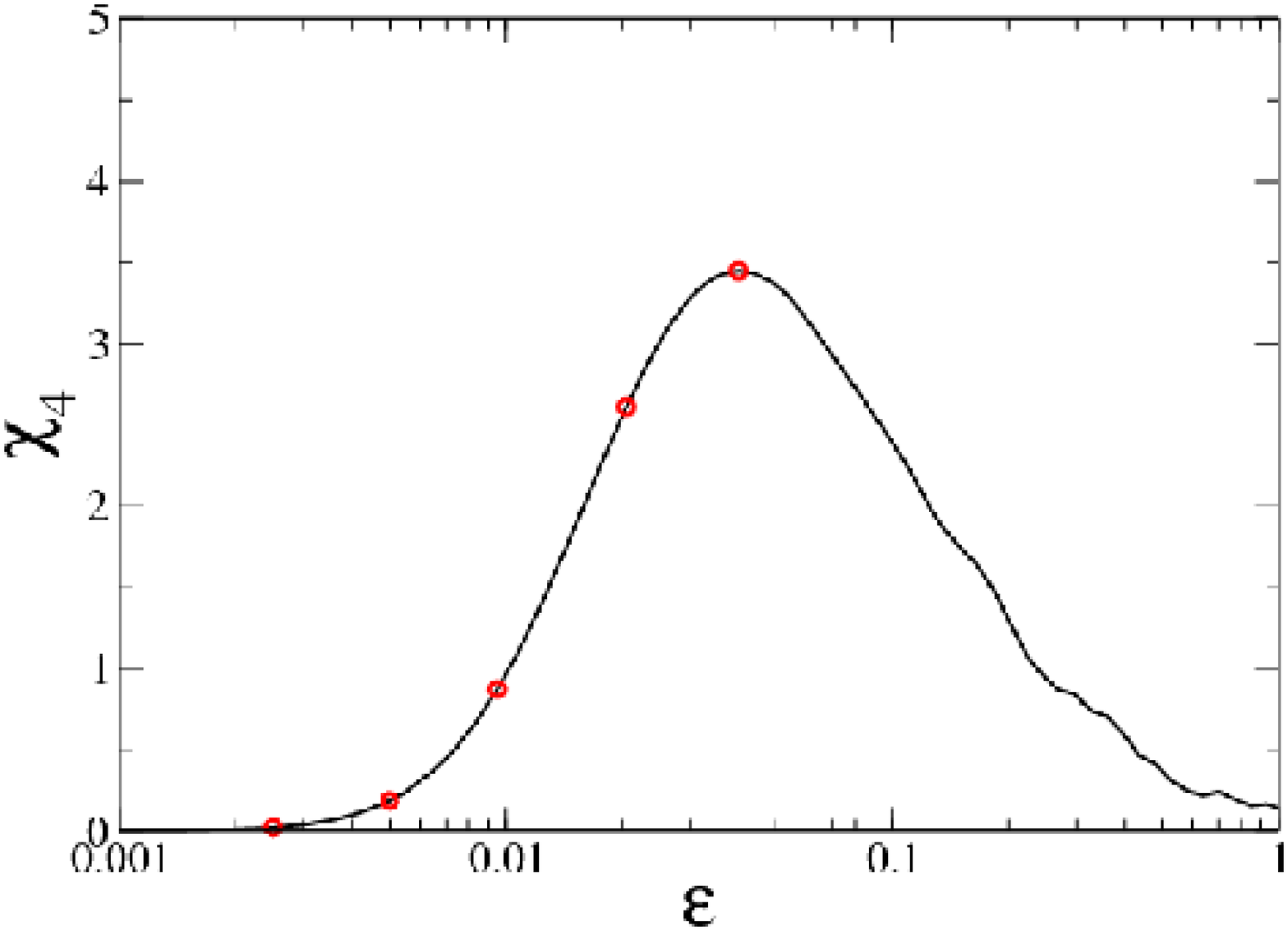}
\includegraphics[width=7cm]{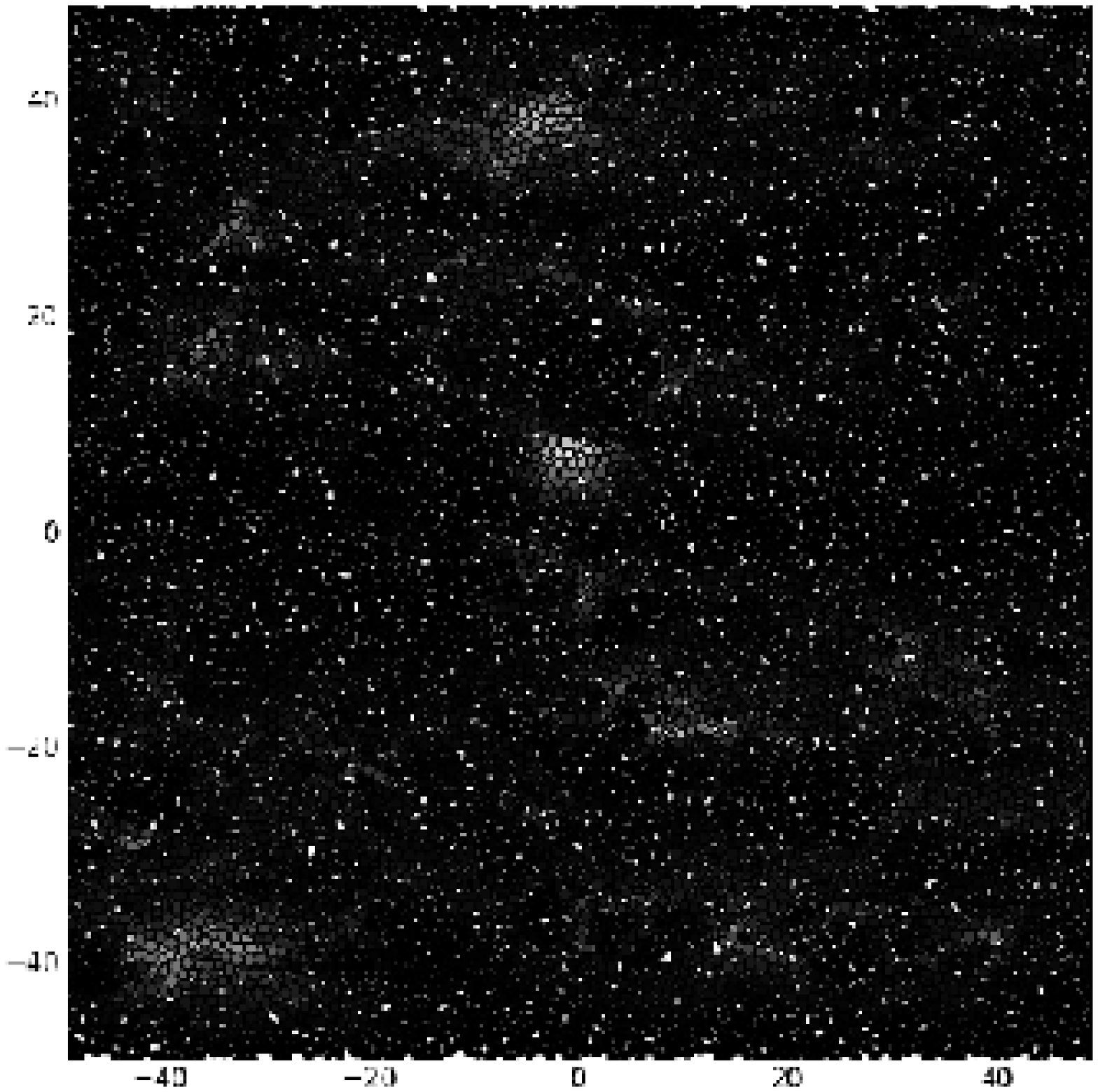}
\includegraphics[width=7cm]{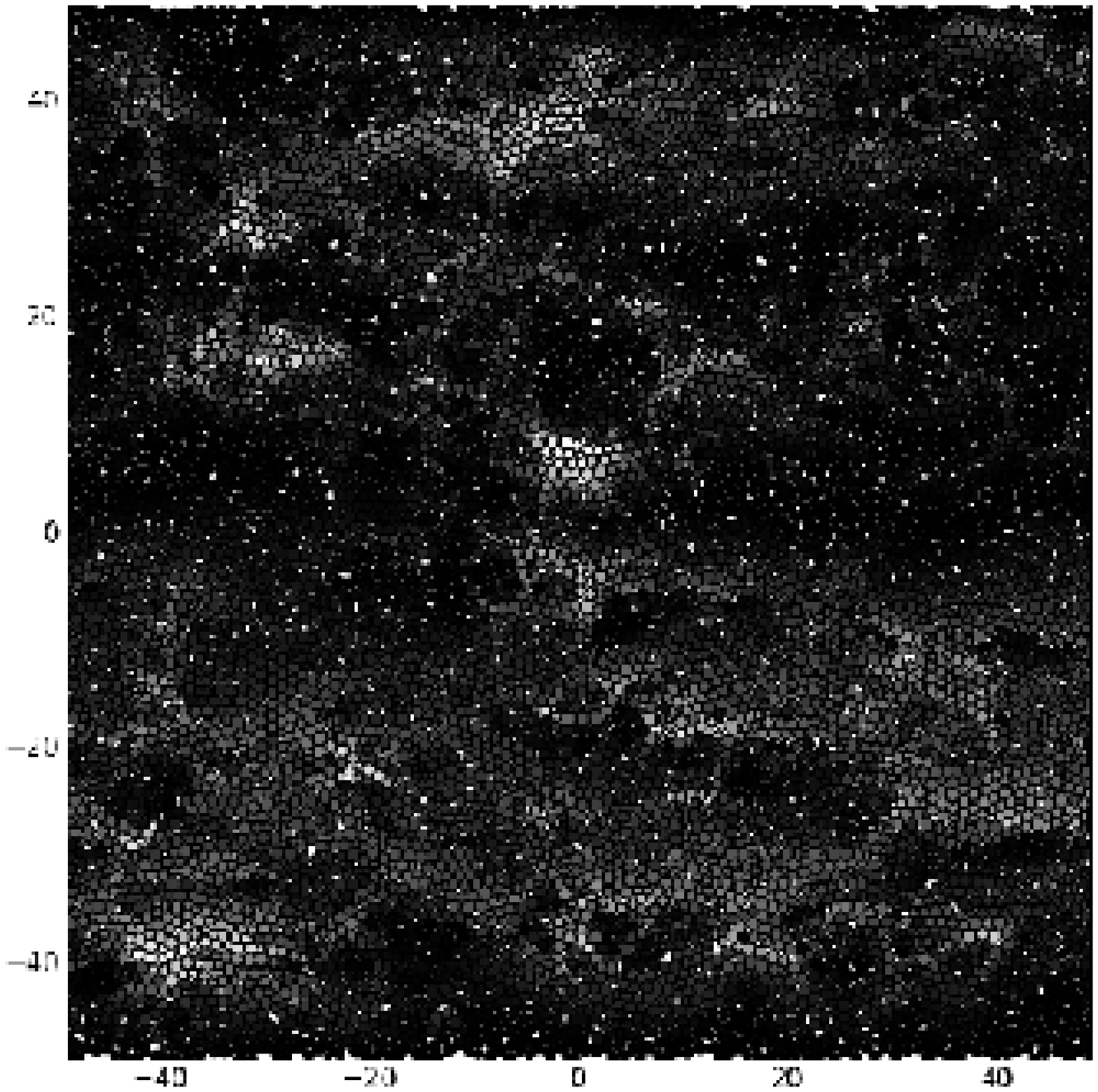}
\includegraphics[width=7cm]{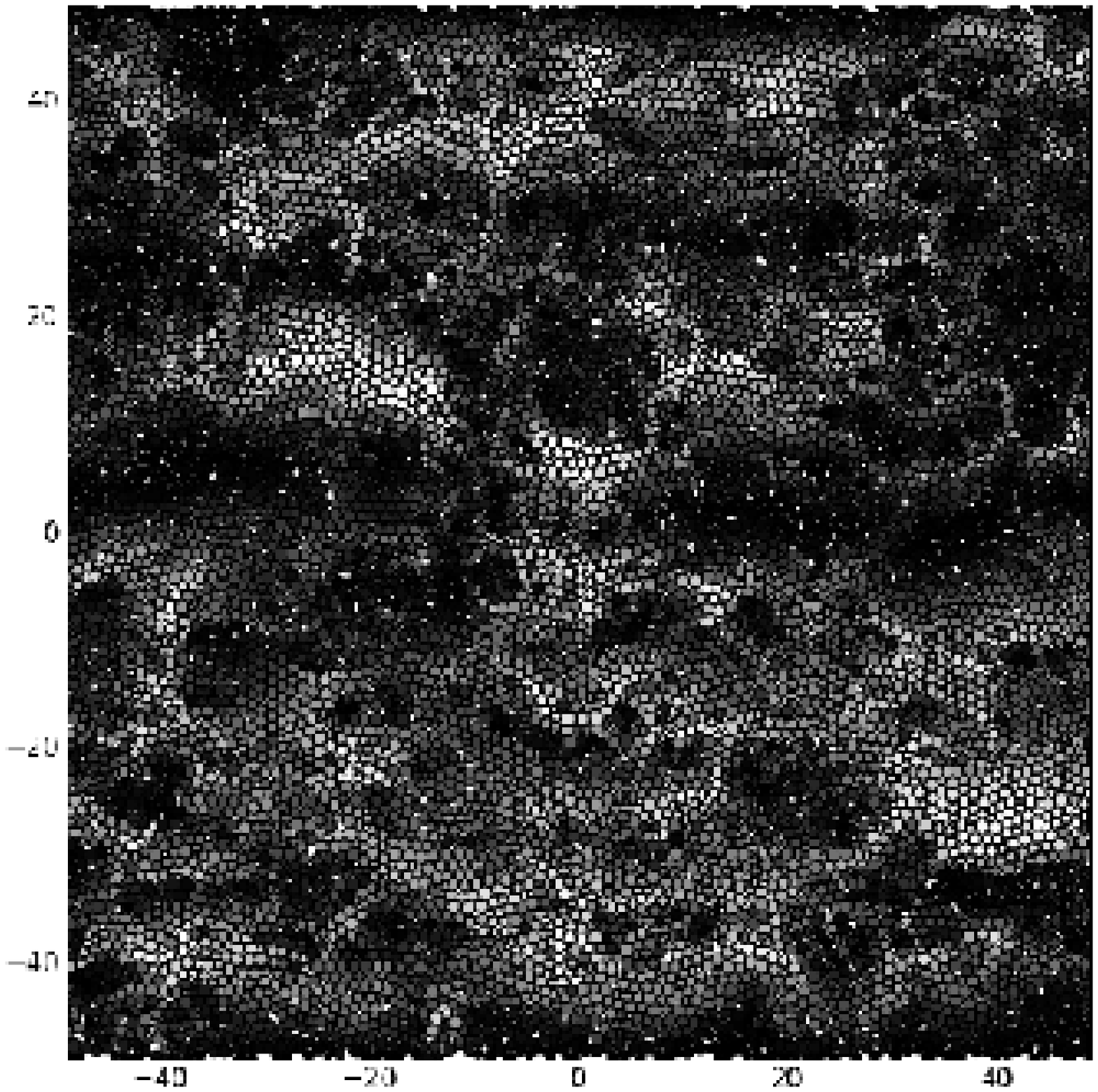}
\includegraphics[width=7cm]{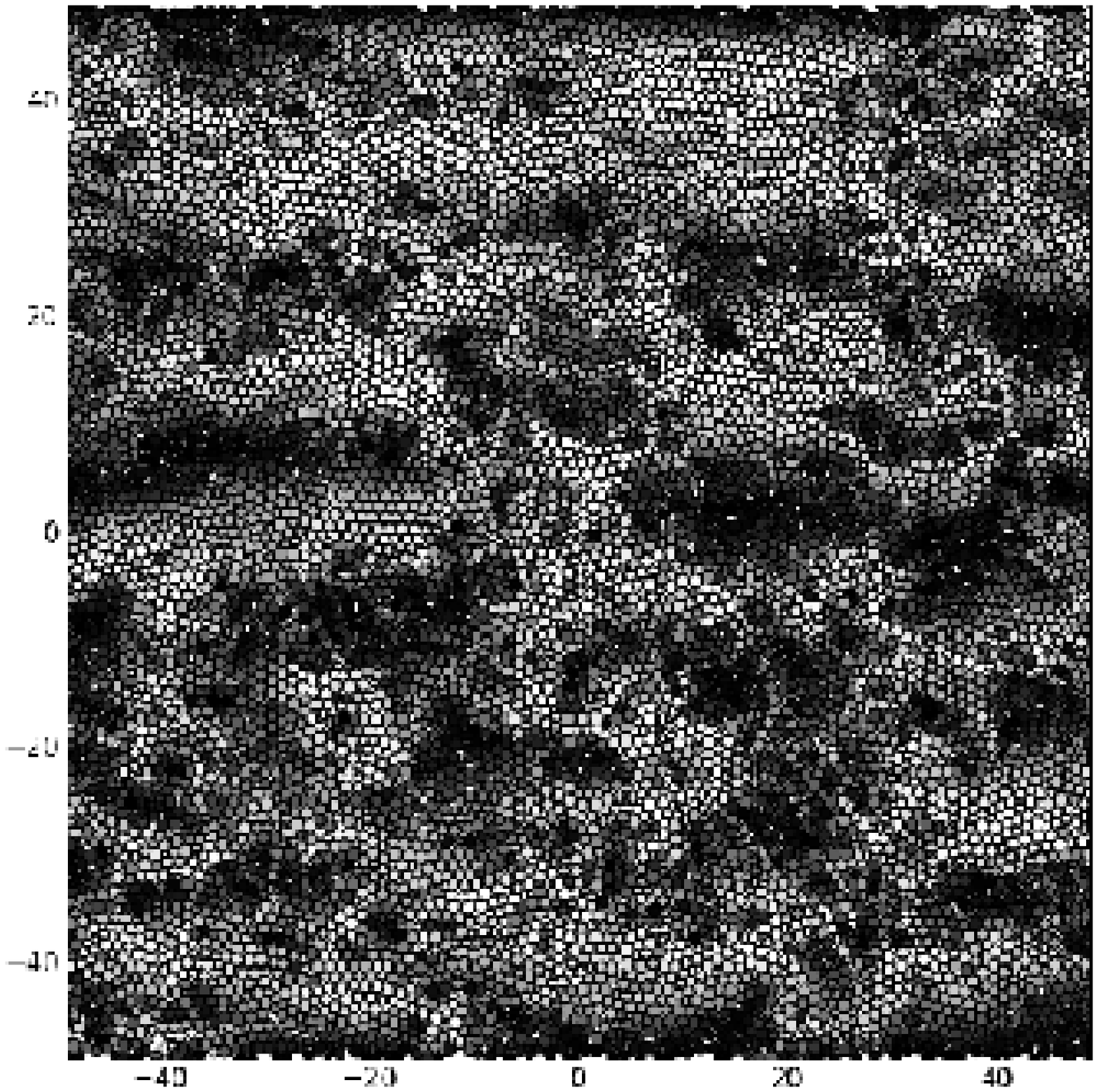}
\includegraphics[width=7cm]{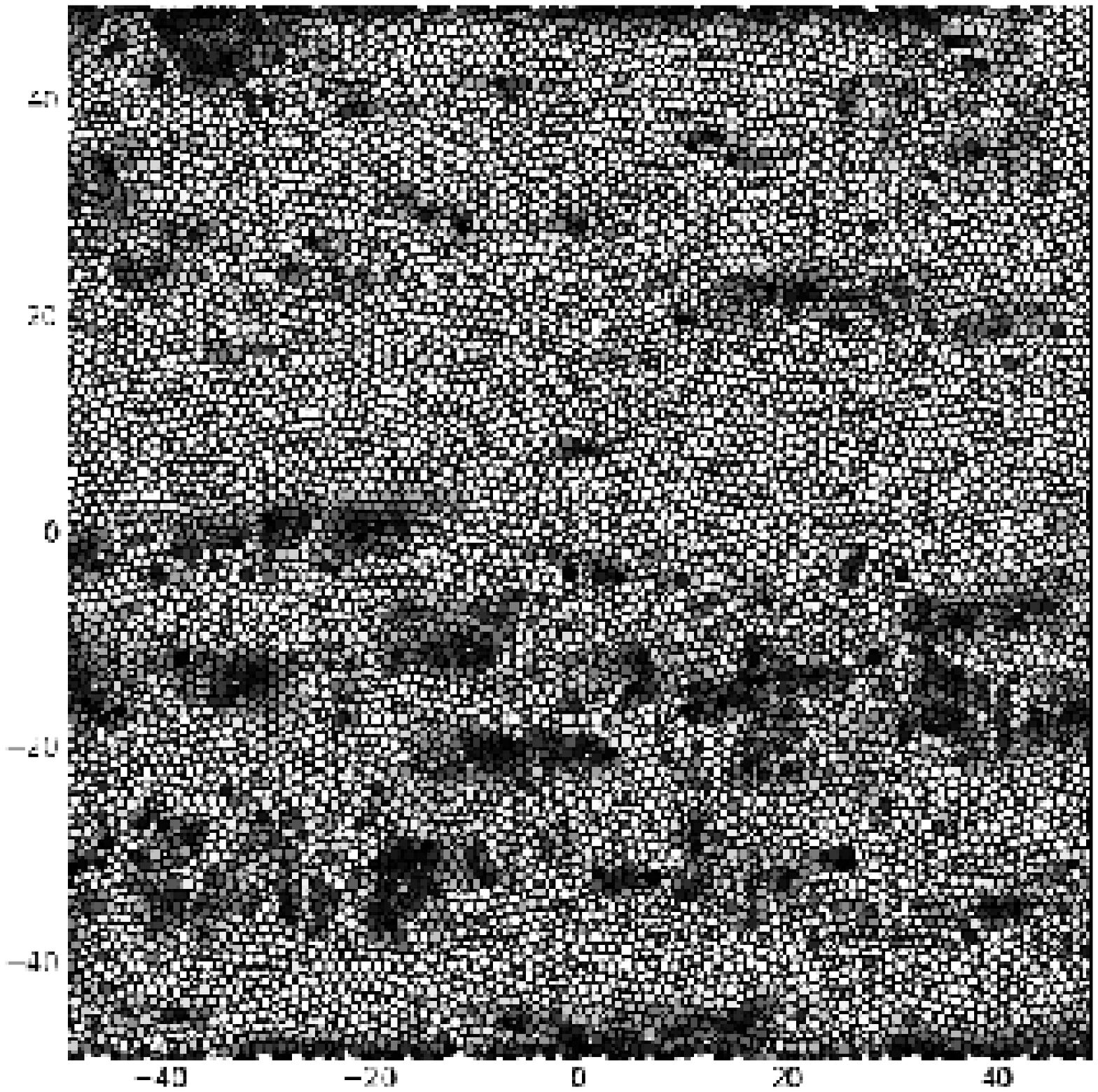}
\caption{\textbf{Top left corner :} $\chi_{4}(\epsilon)$. The red marks correspond to the strain intervals at which the five spatial maps of the self correlation function $Q_{s}^{i}(\epsilon)=\exp{\left(\frac{\Delta r_{i}(\epsilon)^{2}}{2a^{2}}\right)}$ are computed. From top to bottom and from left to right, $\epsilon=2.5\cdot10^{-3}$,$\epsilon=2.5\cdot10^{-3}$,$\epsilon=5\cdot10^{-3}$,$\epsilon=10^{-2}$,$\epsilon=2\cdot10^{-2}$. All figures are obtained on a sample containing 10000 particles under RWBCs and at a shear rate of $\dot\gamma=5\cdot10^{-4}$.}\label{fig:QrnaMapsvsStrain}
\end{center}
\end{figure}

\begin{figure}[!hbtp]
\begin{center}
\includegraphics[width=7cm]{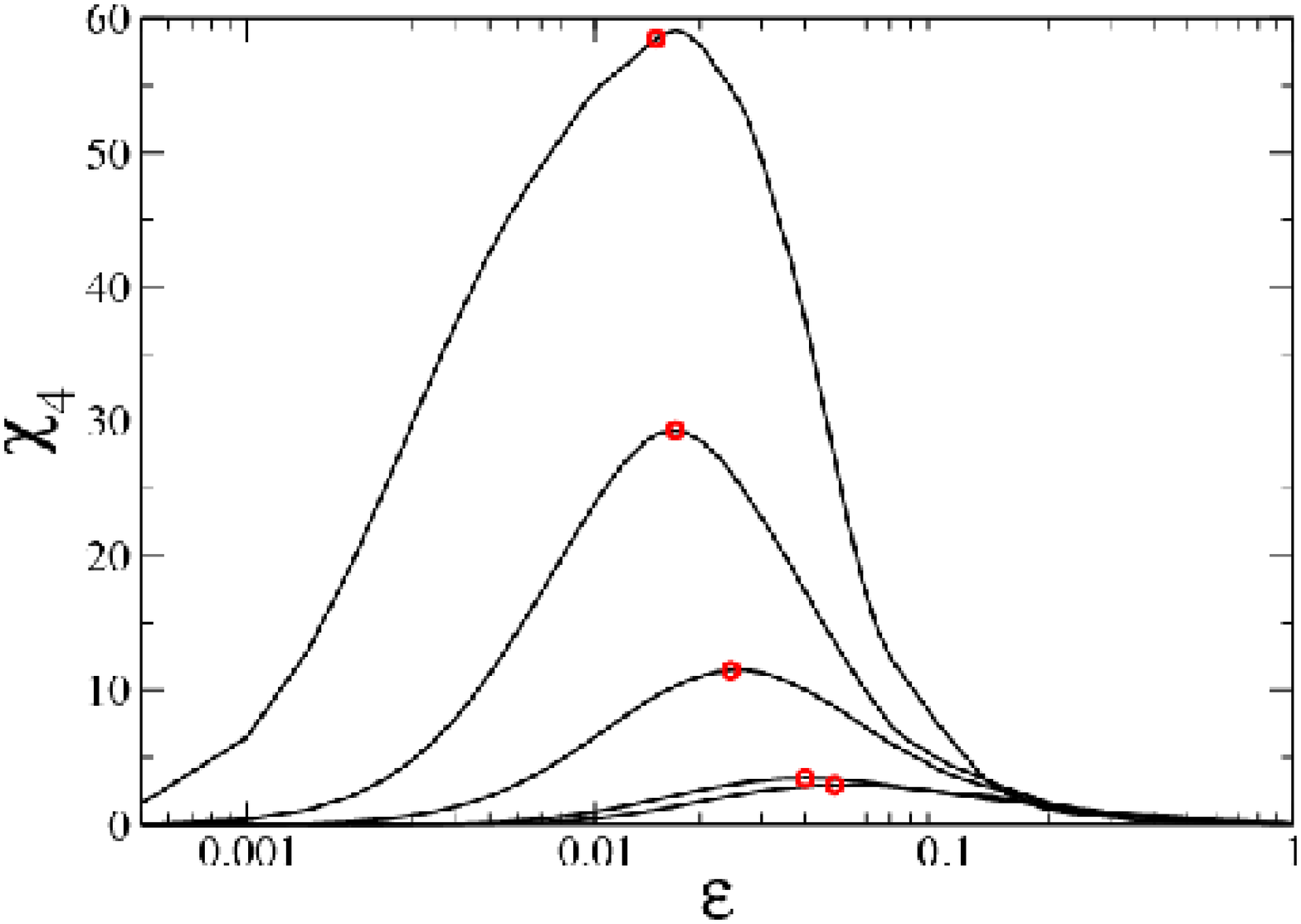} 
\includegraphics[width=7cm]{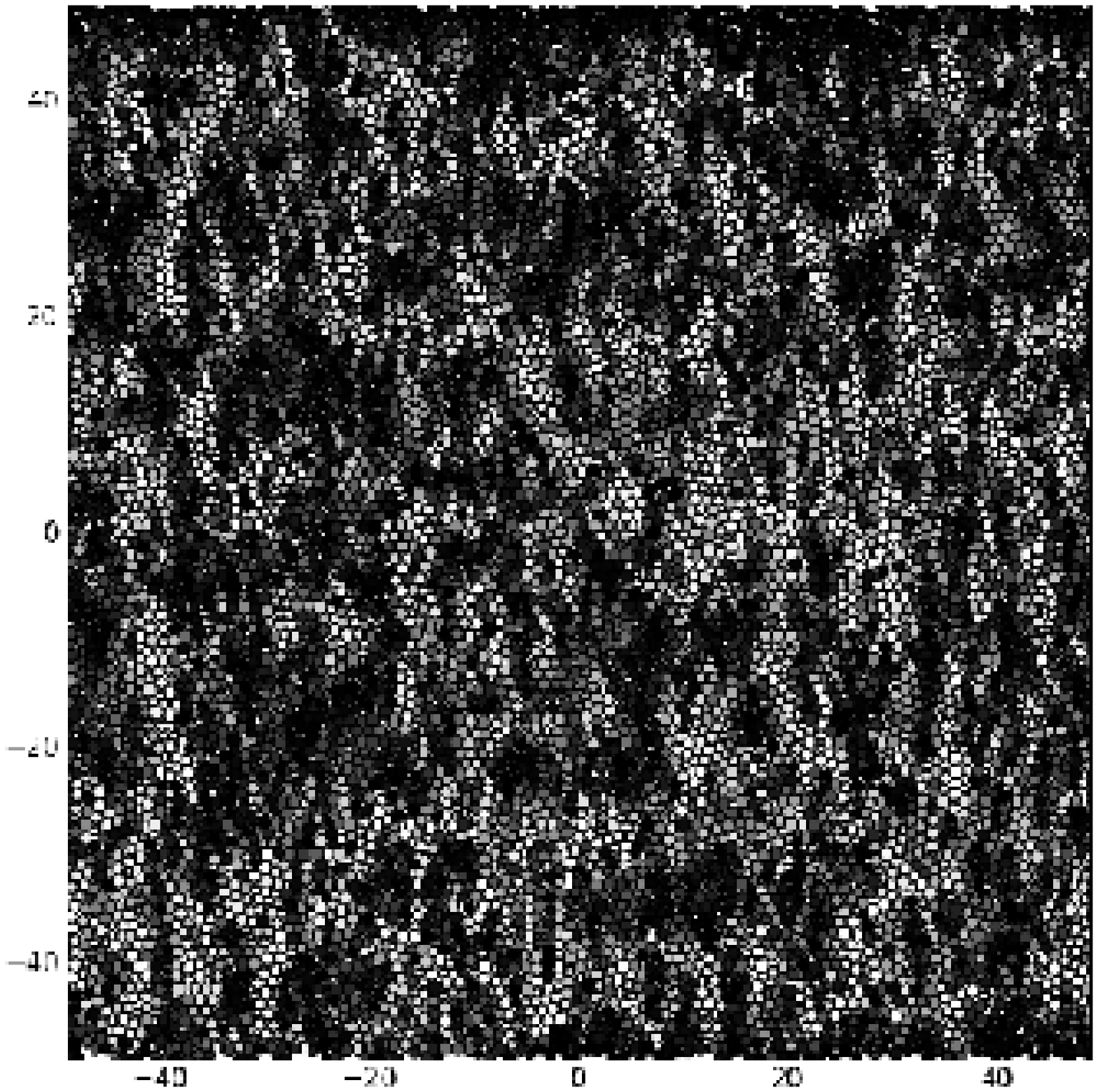}
\includegraphics[width=7cm]{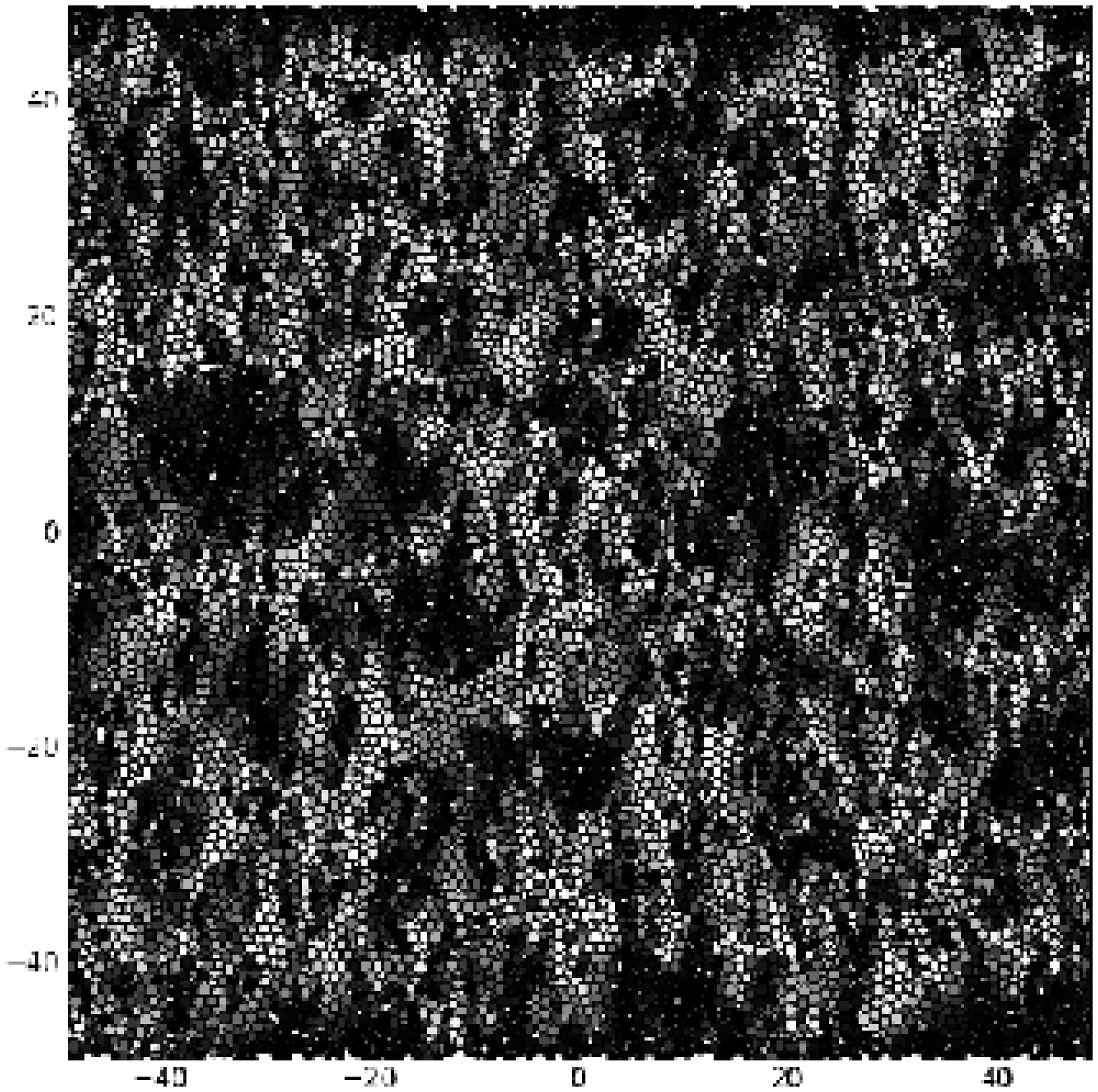}
\includegraphics[width=7cm]{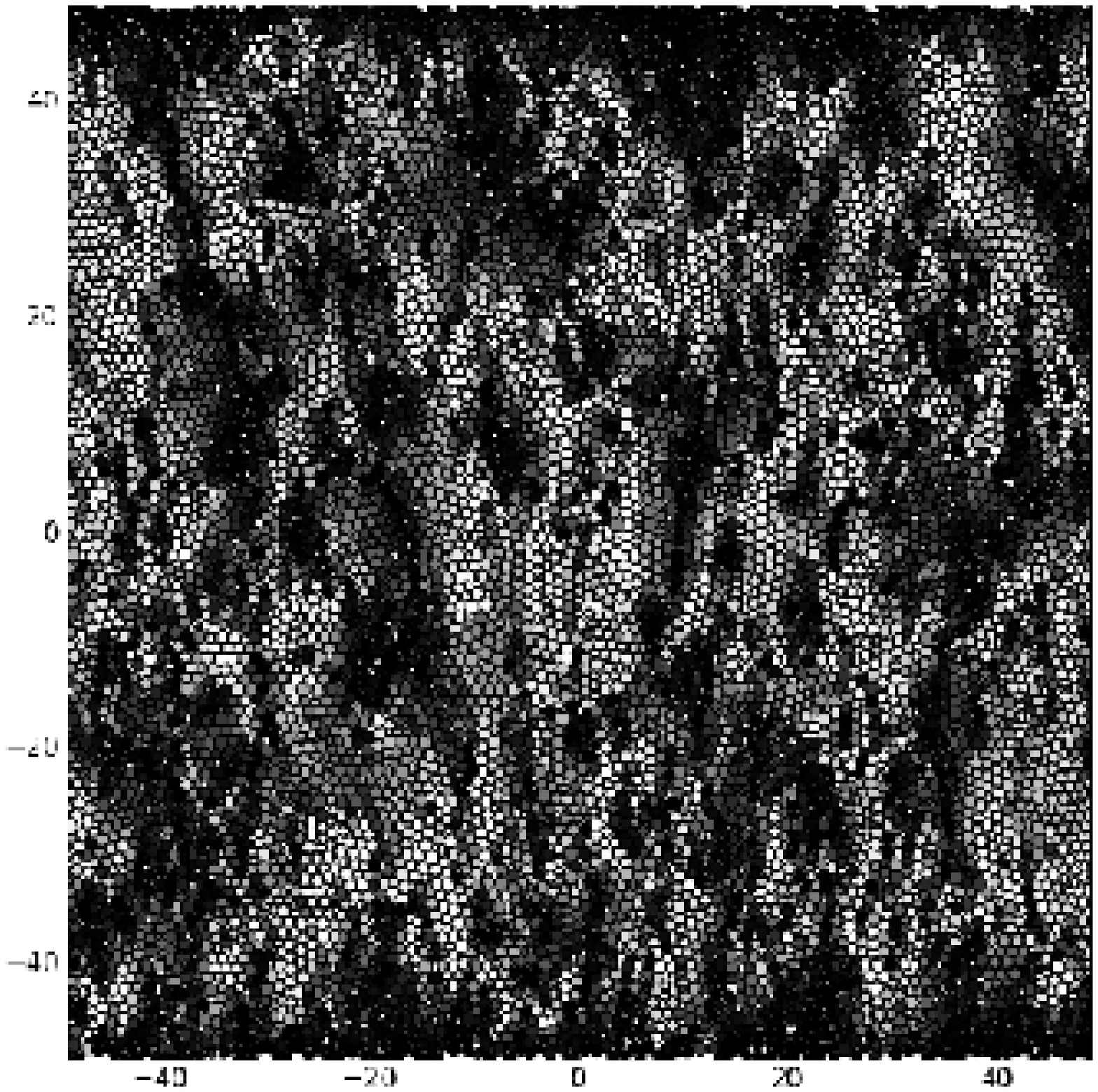}
\includegraphics[width=7cm]{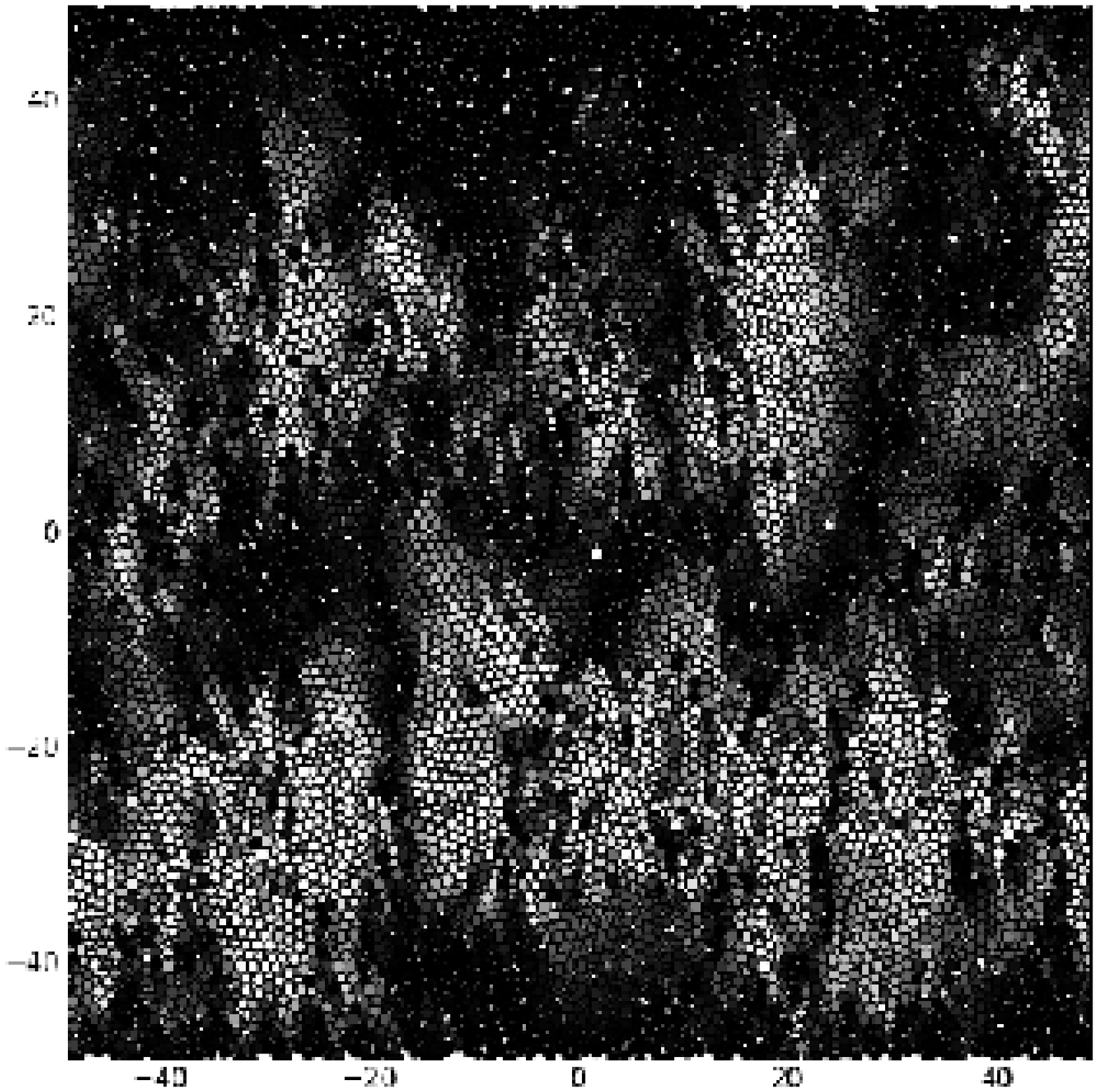}
\includegraphics[width=7cm]{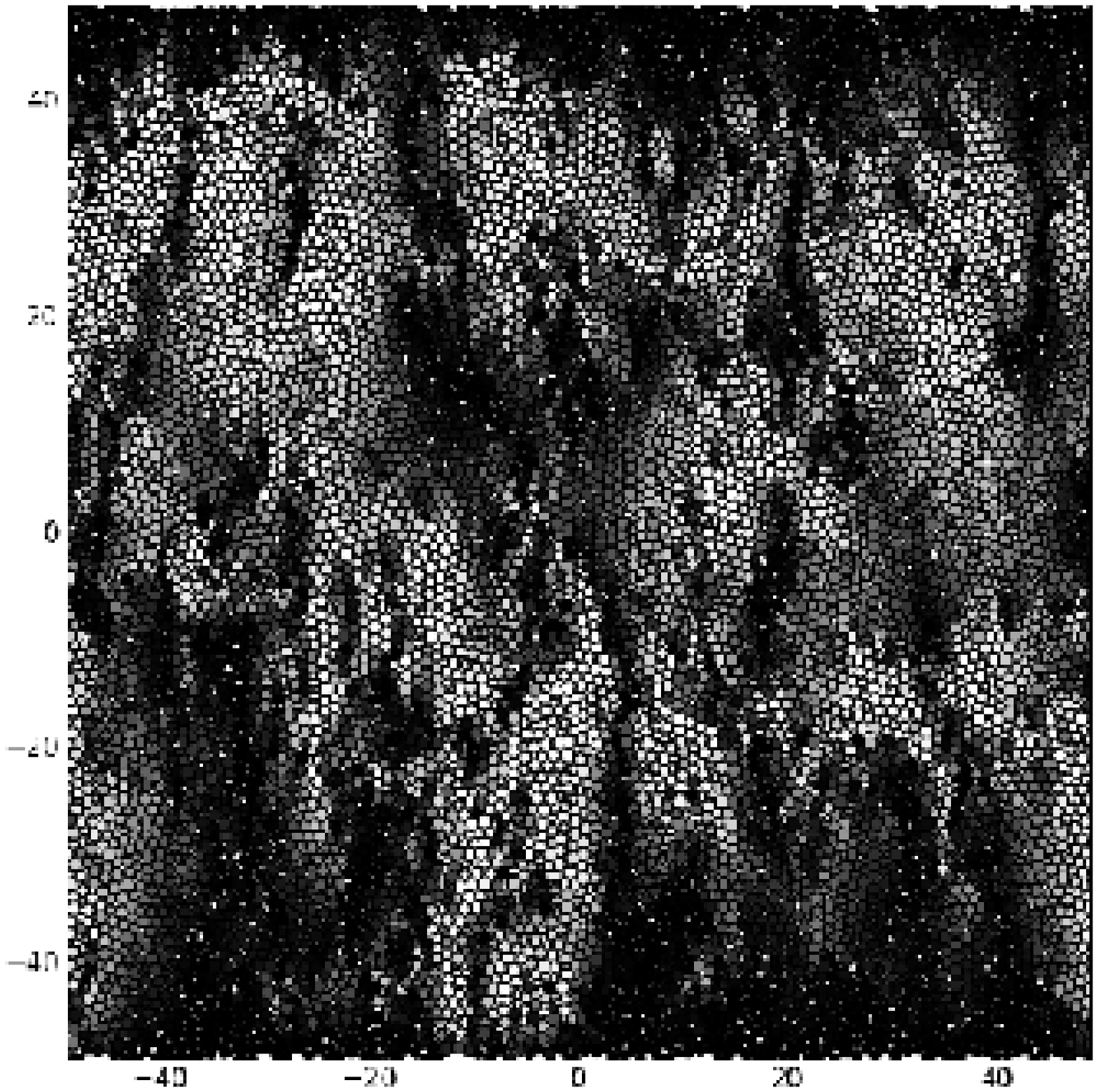}
\caption{\textbf{Top left corner :} $\chi_{4}(\epsilon)$. The red marks correspond to the strain intervals at which the five spatial maps of the self correlation function $Q_{s}^{i}(\epsilon)=\exp{\left(\frac{\Delta y_{i}(\epsilon)^{2}}{2a^{2}}\right)}$ are computed. From top to bottom and from left to right, $\dot\gamma=10^{-3}$,$\dot\gamma=5\cdot10^{-4}$,$\dot\gamma=10^{-4}$,$\dot\gamma=5\cdot10^{-5}$,$\dot\gamma=10^{-5}$. All figures are obtained on a sample containing 10000 particles and for RWBCs.}\label{fig:QyMapsvsStrainRate}
\end{center}
\end{figure}

There are two aspects that we would like to illustrate here. First how does the dynamical heterogeneity build up during the strain of a glassy material, in other words when one looks at spatial maps of $Q_{s}^{i}=\exp{\left(\frac{\Delta y_{i}(\epsilon)^{2}}{2a^{2}}\right)}$ for increasing values of $\epsilon$. Second how the dynamical heterogeneity is affected by the value of the imposed shear rate, in other words when one plots spatial maps of $Q_{s}^{i}$ at $\epsilon_{chi_{4}}^{max}$ but for different shear rates. The next two panels of \ref{fig:QyMapsvsStrain} and \ref{fig:QrnaMapsvsStrain} illustrate respectively the spatial fluctuations of $\exp{\left(\frac{\Delta y_{i}(\epsilon)^{2}}{2a^{2}}\right)}$ and $\exp{\left(\frac{\Delta r_{i}(\epsilon)^{2}}{2a^{2}}\right)}$ for various strain intervals, while the panel \ref{fig:QyMapsvsStrainRate} illustrates the fluctuations of $\exp{\left(\frac{\Delta y_{i}(\epsilon)^{2}}{2a^{2}}\right)}$ at the peak of the four-point correlation function $\chi_{4}$ for different shear rates. Comparing figure \ref{fig:QyMapsvsStrain} and figure \ref{fig:QrnaMapsvsStrain} confirms the already observed fact that the relaxation of the $Q_{s}^{i}$ to zero is faster in the x direction that in the y direction. This anisotropy is here enhanced by the presence of walls but would be visible also under LEBCs. It is striking to see in figure \ref{fig:QyMapsvsStrainRate} the growth of a cooperativity length scale as the shear rate is decrease from $10^{-3}$ to $10^{-5}$. Indeed one sees that the response of the glass to the external shear rates becomes more and more heterogeneous as the shear rate is lowered and that at $\dot\gamma=10^{-5}$, for example, the typical size of the clusters of particles that have moved more than $a=0.1$ (the white particles) seems to reach an important fraction of system size. Note that in all these maps the clusters seem to form anisotropic structures elongated along the y axis. We attribute this apparent anisotropy to the formation of vortices.

Let us finally conclude this article by giving a simple physical argument that allows to understand the scaling of this cooperative length scale with shear rate. The argument goes as follows. Assume that the regions that are prone to fail plastically (we call them the triggering points) under shear are homogeneously distributed in the glass and that the density per unit strain $\rho_{te}$ (te stands for triggering events) of these points is a constant that is independent of strain rate. Then during a time interval t there are $\dot\gamma \times t \times \rho_{te}$ triggering points that are excited and one can define the average distance $l_{te}$ between these points to be $l_{te}=\left(\rho_{te}\dot\gamma t\right)^{-1/d}$, where d is the dimension. These points by definition are triggering a quadrupolar event that as we have discussed in the previous section can induce further plastic rearrangements in its vicinity. More specifically a plastic rearrangement induces a quadrupolar redistribution of the stress in its surrounding and one expects an increased probability of having a new plastic event where the stress is increased, i.e. along the vertical and horizontal axis. The simplest hypothesis concerning the propagation of the events is that it occurs through a diffusive process\footnote{One could make the model more quantitative by mapping it to the problem of diffusion on a grid. One can estimate the diffusion coefficient from atomic considerations as $D\sim\xi^{2}/\tau$, where $\xi$ is the optimal distance between successive plastic rearrangements and $\tau$ is the duration of plastic event. From figure \ref{fig:Frames1e-5} one can estimate $\xi$ to be of the order of $\xi\sim20-30\sigma$ and one can identify $\tau$ with the duration of a plastic event $\tau\sim\tau_{e}\sim10-100$. This gives $1\lesssim D\lesssim100$}. One can then define a new length scale $l_{D}=\sqrt{Dt}$, where D is the diffusion coefficient. Now to extract a characteristic length scale we identify these two length scales, $l_{te}=l_{D}$ which yields a characteristic time $t_{c}$,
\begin{equation}
t_{c}=\rho^{\frac{-1}{d/2+1}}D^{\frac{-1}{1+2/d}}\dot\gamma^{\frac{-1}{d/2+1}}\mbox{ ,}
\end{equation}
and a corresponding length scale scaling as,
\begin{equation}
l_{c}=\rho^{\frac{-1}{d+2}}D^{\frac{1}{d+2}}\dot\gamma^{\frac{-1}{d+2}}\mbox{ .}
\end{equation}
In two dimensions this corresponds to the scalings $t_{c}\propto\dot\gamma^{-1/2}$ and $l_{c}\propto\dot\gamma^{-1/4}$ while in three dimensions one expects $t_{c}\propto\dot\gamma^{-2/5}$ and $l_{c}\propto\dot\gamma^{-1/5}$. These results argue rather well with the observations made earlier on the relaxation time scale and on the growth of a cooperativity length scale. We argue that this length should correspond to the maximum extent of the plastic cascade and therefore to the maximum cooperativity length. Indeed when $l_{D}$ becomes of the order of the distance between triggering events $l_{te}$ each plastic cascade starts to `feel' the presence of the neighboring cascades and its progression is perturbed. Finally we see that the exponent $-1/4$ is in good agreement with what is observed numerically in figure \ref{fig:maxkhi4s} but also with the predictions of the kinetic elastoplastic model introduced in \cite{Bocquet2009} as well as with the prediction of mesoscopic elasto-platic models \cite{Picard2005}.

\section{Conclusion}

The rheology of a model two-dimensional glass was discussed with an emphasis on the complex plastic response and dynamical heterogeneity that builds up in the material under shear. The detailed analysis of the plastic deformation at different shear-rates showed that the glass follows different flow regimes. Below a system size dependent critical shear-rate the mechanical response reaches a quasistatic limit characterized by finite size effects, cascades of plastic rearrangements and yield stress, while at higher shear rates the rheological properties are determined by the externally applied shear-rate. In the later regime we reported on the growth of a cooperativity length scale and discussed the scaling of this length with shear-rate. We proposed a simple model based on the diffusive propagation of plastic events that reproduces this scaling. The glassy system was shown to develop a diverging cooperativity length scale with lowering shear rate and one can ask how this dynamical length scale affects the flow of confined glassy systems. Preliminary results concerning the effect of temperature have shown surprising effects associated with the presence of a small but finite temperature and many further simulation runs at various temperatures and shear rates are needed to apprehend quantitatively the rheology of the glasses. Similarly our results on a two dimensional system call for a generalization to the three dimensional case. Our results have also confirmed at the atomic level many of the predictions of the mesoscopic extremal elasto-plastic models \cite{Baret2002,Picard2005} and quantitative comparison would require an extended amount of simulated material, larger system sizes and lower shear rates. Finally our work is a first step towards a better understanding of the elementary building blocks needed to construct a mesoscopic description of the rheology of glassy materials along with innovative constitutive laws. Another possible extension of this work would be to map the rheology of the structural glasses obtained through the use of atomic scale simulations to a kinetically constrained model, the collaborative nature of the dynamical building blocks presented in this article seems indeed to validate this type of approach.

During the course of this work, I have benefited from discussions with Claus Heussinger, Pinaki Chaudhuri, Lyd\'eric Bocquet, Anne Tanguy and Jean-Louis Barrat. Numerical computations were carried out on P2CHPD (University Lyon) computers.


\onecolumn
\newpage

%
\newpage

%
%
\newpage
\newpage
\newpage

%
%
%
%
\newpage
%
%
%
\newpage
%
\newpage
%
\newpage
%
%
\newpage
%
\newpage
%
%
\newpage
%
\newpage

\end{document}